%% file: solid_electrolyte.tex
\documentclass[review,hidelinks,onefignum,onetabnum]{siamart220329}

\input{solid_electrolyte_shared}
\usepackage{filecontents}

\ifpdf
\hypersetup{
	pdftitle={An Asymptotic Analysis of Space Charge Layers in a Mathematical Model of a Solid Electrolyte},
	pdfauthor={Laura M. Keane and Iain R. Moyles}
}
\fi

\begin{document}
	
	\maketitle
	
	\begin{abstract}
		We review a model for a solid electrolyte derived under thermodynamics principles.
		We non-dimensionalise and scale the model to identify small parameters, where we identify a scaling that controls the width of the space-charge layer in the electrolyte. 
		We present asymptotic analyses and numerical solutions for the one dimensional zero charge flux equilibrium problem. We introduce an auxiliary variable to remove singularities from the domain in order to facilitate robust numerical simulations.
		From the asymptotics we identify three distinct regions: the bulk, boundary layers, and intermediate layers. The boundary and intermediate layers form the space charge layer of the solid electrolyte, which we can further distinguish as strong and weak space-charge-layers respectively. The weak space-charge-layer is characterised by a length, $\lambda$, which is equivalent to the Debye length of a standard liquid electrolyte. The strong space-charge-layer is characterised by a scaled Debye length, which is larger than $\lambda$. We find that both layers exhibit distinct behaviour, we see quadratic behaviour in the strong space-charge-layer and exponential behaviour in the weak space-charge-layer. We find that matching between these two asymptotic regimes is not standard and we implement a pseudo-matching approach to facilitate the transition between the quadratic and exponential behaviours. We demonstrate excellent agreement between asymptotics and simulation.
	\end{abstract}
	
	
	
	\section{Introduction}

	Rechargeable batteries, in particular, lithium-ion batteries (LIBs) have drawn a lot of interest as an alternative, more sustainable energy source over the widely used fossil fuels. LIBs are not only a source of energy, but can also be used to store energy generated by other means. Robust storage can assist with latency and intermittency issues associated to clean energy such as solar, wind, and hydro. LIBs are the most common choice among rechargeable batteries due to their high energy densities \cite{tarascon2001issues} which has resulted in an abundance of research from different fields, both experimental \cite{kennedy2014high,kennedy2017understanding, stokes2017direct} and theoretical \cite{tasaki2003decomposition,huang2020editors,howey2020free}, to optimise the performance of LIBs.  
	
	Mathematical modelling can give insight into observed behaviours, supplement testing, and be used to implement controls in battery management systems.  There are various modelling frameworks used throughout the literature, the most common being continuous models using partial differential equations (PDEs) \cite{newman1962theoretical,newman1975porous, farrell2000primary} and equivalent circuit models \cite{hu2011electro, hu2011linear, marcicki2013design}. PDE models are physics based models which can accurately capture much of the cell dynamics and operation, but can be slow computationally when accounting for all of the electrochemical processes and cell activity. Alternatively, equivalent circuit models treat the cell as a series of resistors and capacitors, and due to their simplicity they are much faster computationally than PDE models. However, they typically cannot capture battery physics at the same resolution as PDE models can \cite{plett2015battery}.  Asymptotic reduction of PDE models can bridge the gap in accuracy and speed. By using asymptotic techniques we can identify the most important processes in the cell and therefore reduce the complexity of the PDE models, increasing their computationally efficiency while retaining the dominant physics. Asymptotics is also used to uncover length scales over which there is significant change in the model. Some examples of an asymptotic approach in the battery community include reduction of porous electrode models \cite{moyles2019asymptotic, sulzer2019faster}
	where the authors retain high accuracy and low computational time in their reduced models of lithium-ion and lead acid batteries respectively. 
	Marquis \etal \cite{marquis2019asymptotic} and Richardson \etal \cite{richardson2020generalised} use asymptotic methods to show how a popular pseudo-two-dimensional model (\cite{doyle1993modeling, fuller1994simulation}) can be reduced to simpler models which attain results in good agreement with the more computationally complex full model.
	Authors have also investigated other features of electrochemical systems via asymptotics, by coupling the electrochemical model with other effects such as mechanical, thermal, and degradation and then proceeding with asymptotic methods, some examples include \cite{foster2017mathematical,hennessy2020asymptotic}, and \cite{planella2022continuum}. For a survey of mathematical modelling and model reduction see the review paper by Planella \etal \cite{planella2022continuum}.

	Solid electrolytes (SEs) have emerged as a topic of interest for further investigation and are becoming increasingly attractive due to their improved safety, lower self discharge, and higher power densities over their liquid counterparts (see \cite{takada2013progress,manthiram2017lithium, zhao2019solid} and references therein). As a result, there has been a plethora of research investigating the use of SEs from experimental perspectives (cf. \cite{liu2011reversible,shen2018unlocking, dixit2021polymorphism}), but less investigation into modelling. Most modelling literature focuses on liquid electrolytes (\cite{newman1962theoretical, newman1975porous, newman2012electrochemical, plett2015battery, doyle1993modeling,fuller1994simulation} to name a few), but some notable exceptions 
	include the work of Li and Monroe (\cite{li2019dendrite, li2021transport}), Mistry and Mukherjee \cite{mistry2020molar}, and Kim \etal \cite{kim2022transport}. In \cite{li2019dendrite} the authors develop an electrochemical mechanical coupling to study dendrite growth in SE. The authors also present a concentrated solution theory model for a SE with two mobile carriers \cite{li2021transport}. 
	Mistry and Mukherjee \cite{mistry2020molar} use non equilibrium thermodynamics to look at the reactions and transport processes in SEs to gain insight into electrodeposition.
	Kim \etal \cite{kim2022transport} look at the transport and mechanical properties of PEO-LLZO composite solid electrolytes using a two dimensional continuum model. 
	However, the resolution of the modelling is not developed to the same level as that of liquid electrolytes. For example, there has been much less investigation into model reduction and less in depth consideration into the underlying mechanism and dynamics of SEs.

	Electrolytes exhibit electric double layers (EDLs), 
	commonly referred to as the Debye layer in liquid electrolytes. This is a layer that forms in the electrolyte at the interface with the electrode. The formation of this layer occurs due to co- and counter-charges being repelled and attracted at the interface \cite{newman2012electrochemical}. It is referred to as a double layer as there are two parallel layers of charge, the Stern layer which is made up of the counter-ions which are attracted and adhere to the electrode surface, and the diffuse layer which is composed of the ions that are attracted to this surface charge but are free to move. When modelling LIBs, assumptions are often made about this electric double layer in order to simplify the model. The Helmholtz model is frequently used \cite{safari2011modeling,smith2017multiphase, moyles2019asymptotic}, which assumes that the Stern layer is much thicker than the diffuse layer, effectively ignoring the diffuse layer.
	The inclusion of diffuse layer effects are more commonly discussed in models of deionization \cite{biesheuvel2011diffuse,singh2018theory,he2018theory}.
	While there exists various models and model reduction incorporating liquid electrolytes, very few also consider an in depth understanding of the electric double layer.

	Simplifications are often made due to the complexity of these EDLs, however, more detailed insights into this double layer are necessary to better understand the dynamics and their implications. For example, authors such as Gross \cite{gross2019modelling} and Sakong \etal \cite{sakong2022structure} indicate that there is insufficient microscopic understanding of these double layers, while Magnussen and Gross \cite{magnussen2019toward} indicate the need for both experimental and theoretical development of our understanding of double layers.
	Swift \etal \cite{swift2021modeling}  highlight that the double layer is an important component of the electrochemical interface as it can control the kinetics and thermodynamics of reactions. Swift \etal also indicate that the success of energy storage materials can often rely on the interface \cite{janek2016solid, manthiram2017lithium, randau2020benchmarking}. Understanding double charge layers may also play a role in understanding the formation of Solid Electrolyte Interphase (see \cite{mozhzhukhina2020direct}, \cite{xu2021competitive} and references therein), a passivation layer
	formed by the electrochemical reduction of the electrolyte at the surface of the electrode \cite{kennedy2016advances}. 

	The investigation into space-charge-layers (SCL), the SE equivalent of electric double layers, is even more sparse than the EDL literature from both experimental and theoretical perspectives. 
	Wu \etal \cite{wu2022solid} highlight that SCL plays an important role in solid-phase reactions and that the physical and chemical properties of SCL can significantly impact the electrochemical performance of materials due to the influence of the SCL on factors such as ion and electron transport. Zhang \etal \cite{zhang2022research} refer to the SCL as one of the most important influences on ion transport by the electrode-electrolyte interface and indicate that understanding the formation, structure, and effect of the SCL is of great significance. 
	In addition, SCL have been shown to contribute to high interfacial resistance, one of the largest hindrances associated with solid state batteries \cite{xia2019practical, liu2022unlocking, ahmad2022chemomechanics}.

	Zoning in on the theoretical literature, there has been some work on modelling SE and SCL from a mathematical point of view, for example, \cite{landstorfer2011advanced,braun2015thermodynamically, swift2021modeling, becker2021statics, katzenmeier2022nature}. However, much of this work is either numerically driven or involves semi analytical solutions. The models tend to be complex which leads to issues in analytics and numerical implementation. Therefore some authors use semi analytical solutions to derive approximate numerical solutions which can add error and can limit the interpretability of the results. The numerically driven models are derived and then simulated without much consideration for the driving mechanisms. 
	Zhang \etal \cite{zhang2022research, zhang2023designing} reiterate this indicating that characterizing the SCL is difficult and oftentimes simulations and calculations need to be employed in parallel to gain insight, they suggest that theoretical simulation calculation cannot fully understand or predict the behaviour.
	By investigating a reduced model for SCL in SEs we are bridging these gaps.

	Numerically solving SE models with SCL has posed challenges in the literature. These difficulties arise as the electrolyte can become exponentially close to fully lithiated or depleted of lithium ions, resulting in singularities in the domain. The numerics struggle with the computations as we near these two limits. As such, many authors avoid direct computation. For example, in \cite{braun2015thermodynamically} the authors use a semi-analytic approach, circumnavigating some of these difficulties. Katzenmeier \cite{katzenmeier2022nature} uses the model derived by \cite{braun2015thermodynamically} for their simulations in COMSOL, employing an empirical equation using Sigmoid functions to compensate for singularities at the boundaries.  In \cite{swift2021modeling} the authors mention that it was difficult to do numerics due to the proximity to the depleted (or zero) limit when charge neutrality is reached.
	The authors use analytic functions to approximate and supplement some of their numerics.  
	Landstorfer \cite{landstorfer2011advanced} also hint at this issue when they indicate that a priori knowledge of the electrode/electrolyte interfaces, which they describe as numerically problematic, was needed to generate the lattice for their numerical solutions.
	However, we want a generalized numerical framework that can be used in standard numerical solving techniques without much pre-processing.

	We develop an improved numerical and analytical assessment of SCL based on the model first derived by Braun, Yada, and Latz \cite{braun2015thermodynamically}. Using this model we carry out a non-dimensionalisation to first identify the important parameters in the model. We use the non-dimensionalisation to inform an asymptotic reduction of the model, enabling us to gain a deeper understanding of the underlying mechanisms. In particular, this approach leads to a deeper understanding of the SCL.
	We note that our work fills certain gaps acknowledged in previous work; Knauth \cite{knauth2009inorganic} indicate that the width of the SCL is proportional to the Debye length. In \cite{yamada2013lithium} the authors find that the SCL is almost twice as thick as the Debye length of a liquid electrolyte. Furthermore, in \cite{braun2015thermodynamically} the authors are also curious about the width of this layer and find that the widths are proportional to 10 times the liquid electrolyte Debye length. In both \cite{li2019dendrite, li2021transport} the authors echo this observation of a SCL or a transition region of about 10 Debye lengths.  In addition Li and Monroe \cite{li2019dendrite} observe the SCL becomes more dilated with rising voltage bias, which contrasts the Debye length contraction typically seen in liquid diffuse layers.  Swift \etal \cite{swift2021modeling} consider a model based on the Poisson-Fermi-Dirac equation to model the SCL. They identify that there are quadratic and exponential regimes in the electrolyte. Our model enables us to quantify the SCL widths more precisely and to explain some of the observations made by these authors (\cite{knauth2009inorganic,yamada2013lithium,braun2015thermodynamically,li2019dendrite,li2021transport}).  We also formalise the ideas of the regimes observed by \cite{swift2021modeling} through our asymptotics, identifying length scales to describe the different regions in the SCL.
	To resolve the numerical difficulties faced by other authors we introduce an auxiliary variable which maps the concentration of ions into another domain to avoid any singularities arising.

	The remainder of this paper is organized as follows: We derive and non dimensionalise the model in section \ref{sec:model_nonD}. 
	In section \ref{sec:num_equi} we present numerical solutions for the one dimensional zero charge flux equilibrium problem. In section \ref{sec:asymptotics} we carry out an asymptotic reduction of the problem. We compare our asymptotic reduction of the model with the numerical results in section \ref{sec:comparison}.
	We discuss the results and conclude the paper in section \ref{sec:discussion}.

	\section{Model and non-dimensionalisation}
	\label{sec:model_nonD}
	We note that in models for liquid electrolytes, both ions diffuse in the solvent and it is usually assumed that the average velocity is zero.
	In SE, typically the anion is assumed to be immobile and so when lithium ions move there are gaps in the lattice that effectively have negative charge from the anion and so this creates a vacancy, this can be modelled chemically as
	$ \ce{LiA
		<=>[\ce{}][\ce{}]
		\underset{}{\ce{ Li^{+} +\text{vacancy} + A^{-}}}}.$
	The vacancies are considered to be massless and chargeless.  Here we will assume that the vacancies are based on Shottky type defects so that each time a cation leaves a spot, it leaves exactly one vacancy behind (\cite{kittel2005introduction}).
	This can be thought of as the fixed anion lattice playing the role of the solvent, as the cations diffuse by hopping between the vacancies. Additionally, because it is typically assumed that only the cation is mobile, the species velocity for anions is zero. The anion immobility can also be justified by the fact that only the lithium intercalates and furthermore, the rigidity of the anion in the solid lattice makes it relatively immobile compared to the lithium. 
	We will consider a model first derived by Braun \etal \cite{braun2015thermodynamically} for a SE under isothermal conditions involving three species, a cation (lithium ions), an anion, and a vacancy.
	Braun \etal derive thermodynamically consistent equations for conservation of mass, charge, momentum, and energy. This leads to the overall model, which has been used by others (for example \cite{becker2021statics, katzenmeier2022nature, katzenmeier2022modeling, sinzig2023finite}), given by
	\begin{subequations}
		\label{Eq:Braun_eqn}
		\begin{align}
		\pderiv{n_c}{t} &= -\frac{1}{m_c} \nabla \cdot \left( \left( 1+ \frac{m_c n_c }{m_a n_a} \right) J_c \right), \label{eq:braun1} \\
		n^F &= - \epsilon_0 \left( 1+\chi \right)\nabla^2 \phi, \label{eq:braun2}\\
		-n^F \nabla \phi &=  \pderiv{(\rho \mathbf{v})}{t}+ \nabla \cdot\left( \rho \mathbf{v} \otimes \mathbf{v} \right)  + \nabla p, \label{eq:braun3}\\
		\frac{J_c}{m_c}&= -M \left( \nabla \mu_c -\frac{m_c}{m_a} \nabla \mu_a - \left(1+  \frac{n_c + n_v}{n_a} \frac{m_c}{m_a}\right) \nabla \mu_v + \left( z_c-\frac{m_c}{m_a} z_a\right) \nabla \phi \right). \label{eq:braun4}
		\end{align}
	\end{subequations}
	For each of the equations the subscripts $c,a,v$ indicate cations, anions, and vacancies respectively. $n_k, m_k$, and $J_k$ represent the number density, ionic mass, and mass flux of species $k$. 
	Time is given by $t$. Charge of species $k$ is denoted $z_k$, $n^F= z_a n_a + z_c n_c$ is the free charge, $\epsilon_0$ ($ 8.854 \times 10^{-12} \text{F}/\text{m}$) is permittivity of free space, $\chi$ is the electric susceptibility (assumed constant), and $\phi$ is the electric potential. We note that here we are referring to the electric potential of the electrolyte, which is distinct from the more commonly measured electric potential with respect to a reference electrode, which is also used in some models (for example, \cite{newman2012electrochemical}). For the details of this distinction and how to go from one to the other, see \cite{richardson2020charge}. We have $p$ representing pressure, the mass density of each species is given by $\rho_k=m_k n_k$, and the total mass density is thus $\rho=\sum \rho_k$. We take $\mathbf{v}$ to be the mass averaged velocity, $\mathbf{v}=\frac{1}{\rho} \sum_k \rho_k \mathbf{v}_k$. $M$ is a mobility term, and $\mu_k$ is chemical potential. 
	%
	The chemical potentials following \cite{braun2015thermodynamically} are given by 
	\begin{subequations}
		\label{eq:incompress_chem_pot}
		\begin{align}
		\mu_c &= \mu_c^* + \frac{1}{n_r} \left(p-p_r\right) + k_BT \log{\left(\frac{n_c}{n_c+n_v}\right)},\\
		\mu_v &= \mu_v^* + \frac{1}{n_r} \left(p-p_r\right) + k_BT \log{\left(\frac{n_v}{n_c+n_v}\right)},\\
		\mu_a &= \mu_a^* +  \frac{1}{n_r} \left(p-p_r\right),
		\end{align}
	\end{subequations}
	where the non ideality is modelled using Margules activity coefficients \cite{margules1895zusammensetzung,gokcen1996gibbs}, which express the excess free energy as a power series of the mole fractions. The subscripts $r$ indicate reference densities and pressures, $k_B$ is the Boltzmann constant, and $K$ is the bulk modulus coming from the assumption of the electrolyte being linearly elastic. $\mu_k^*$ is the reference chemical potential, independent of pressure and composition.  A full derivation of this model is presented in \cref{SM-sec:model_deriv}.
	
	We note that in models for liquid electrolyte it is usually assumed that the average velocity is zero, and therefore, the conservation of momentum equation is implicitly accounted for. However, in considering SEs this is not possible. 
	The vacancies are massless and chargeless, so they do not contribute to the mass-averaged velocity and because we also have stationary anions, zero average velocity would mean lithium does not move. Therefore, we need to retain the momentum equation explicitly. 
	We want to focus on the formation of the SCL and therefore consider lithium-metal electrodes. Therefore, we assume a fixed voltage condition. For the full problem we would have potential and flux boundary conditions along with an initial condition for the lithium ion concentration to close the problem.
	For the purpose of this work we will focus on the one dimensional (1D) problem so we will prescribe the appropriate boundary conditions when we outline the 1D problem.

	We first non dimensionalise the model given by \eqref{Eq:Braun_eqn}. 
	We begin with the full dynamic problem in order to keep things general for the non-dimensionalisation but we will focus on the one dimensional problem for our analyses.
	We choose the following scales
	\begin{align*}
	&x=Lx', \quad t=t_0t', \quad \mu_k =  k_B T \mu_k', \quad \mathbf{v}=\frac{L}{t_0} \mathbf{v}', \quad J_c= \frac{\Delta V q m_c M}{L} J_c', \\
	&n=n_rn', \quad z_k=q z_k',\quad p=p_rp',  \quad \rho = \rho_0 \rho', \quad \phi= \Delta V \phi'
	\end{align*}
	where $L$ is some reference length scale and $q=\frac{F}{N_A}=1.602 \times 10^{-19}$ coulombs is the elementary charge. We chose the velocity scale to be $\frac{L}{t_0}$ in order to balance the two velocity terms of \eqref{eq:braun3}, where $t_0$ is still to be determined.
	We chose $\mu_0$ and the scales for $\phi$ and  $J_c$ in order to balance the flux and potential terms in \eqref{eq:braun4}. 
	We take $\Delta V$ as the potentiostatic hold so that the scaled potential conditions in one dimension become
	\begin{align}
	\phi'(z'=0)&= 1, \notag\\
	\phi'(z'=1) &= 0. \label{eq:non_d_bc1}
	\end{align}
	We also have scaled flux conditions
	\begin{align}
	J_c'(z'=0) &= j_0, \notag\\
	J_c'(z'=1) &= j_1. \label{eq:non_d_bc2}
	\end{align}

	We choose a timescale which leads to sensible balances in both the conservation of mass equation \eqref{eq:braun1} and the conservation of charge equation \eqref{eq:braun2}. 
	The non dimensional form of both equations can be written as
	\begin{equation}\label{eq:find_time}
	\pderiv{n_c'}{t'}=-\frac{t_0 \Delta V  M  q}{n_r  L^2} \nabla' \cdot \left( \left(\frac{m_c n_c'}{m_a n_a'}  + 1\right)J_c' \right) = \frac{1}{z_c} \pderiv{n^{F'}}{t'}, \quad n^{F'}=-
	\frac{\Delta V \epsilon_0 \left(1+\chi\right) }{q n_r L^2}
	\nabla^{2'} \phi'. 
	\end{equation}
	Differentiating the equation for $n^{F'}$ in \eqref{eq:find_time} with respect to time 
	\begin{equation}
	\pderiv{n^{F'}}{t'} = -  \frac{\Delta V \epsilon_0 \left(1+\chi\right) }{q n_r L^2} \nabla^{2'} \pderiv{\phi'}{t'} = -\frac{t_0 \Delta V  M  z_c q}{n_r  L^2} \nabla' \cdot \left(\left(\frac{m_c n_c'}{m_a n_a'}  + 1\right) J_c' \right).
	\end{equation} 
	This suggests that the appropriate scale for time is
	\[ t_0=\frac{\Delta V \epsilon_0 \left(1+\chi\right)  n_r L^2}{\Delta V z_c M q^2 n_r L^2 } 
	= \frac{\epsilon_0 \left(1+\chi\right)}{q^2 M z_c},\]
	which has units of seconds and is a mobility time scale relative to free space. 
	Finally, given these scales we return to the conservation of momentum equation, \eqref{eq:braun3}, which can be written in the following non-dimensional form 
	\begin{equation}
	- \frac{\Delta V q}{k_B T} n^{F'} \nabla' \phi' - \frac{p_r }{n_r  k_B T} \nabla' p' =  \frac{\rho_0 L^2 q^4 M^2 z_c^2 }{n_r  k_B T \epsilon_0^2 (1+\chi)^2} \left(\pderiv{(\rho' \mathbf{v}')}{t'}+ \nabla' \cdot\left( \rho' \mathbf{v}' \otimes \mathbf{v}' \right) \right). 
	\end{equation}
	Due to the choice of the timescale, the non-dimensional coefficient on the right hand side 
	\[\frac{\rho_0 L^2 q^4 M^2 z_c^2 }{n_r  k_B T \epsilon_0^2 (1+\chi)^2} 
	\ll 1, \]
	allowing us to neglect the dynamic and convective terms in the momentum equation.
	
	Overall, we have the following non dimensional problem (where we have dropped the primes)
	\begin{subequations}
		\label{eq:scaled_Braun}
		\begin{align}
		\pderiv{n_c}{t} &= -\frac{\delta^{-1} \lambda^2}{z_c} \nabla \cdot \left( \left( 1+\frac{m_c n_c}{m_a n_a}\right) J_c\right),\\
		n^F &= -\delta^{-1} \lambda^2 \nabla^2 \phi, \label{eq:fullProb_phi_eq}\\
		-\delta^{-1} n^F \nabla \phi &= a^2 \nabla p, \label{eq:fullProb_p_eq}\\
		J_c &= - \nabla \left( \delta \left((\mu_c-\mu_v) - \frac{m_c}{m_a} \left( \mu_a + \frac{n_c+n_v}{n_a} \mu_v \right) \right) + \left( z_c - \frac{m_c}{m_a} \right)\phi  \right),
		\end{align}
	\end{subequations}
	subject to $\phi(0)=1$, $\phi(1)=0$, $J_c(0) = j_0$, and $J_c(1) = j_1$.   
	The chemical potentials are given by
	\begin{equation}	\label{eq:scaled_chem_pot}
	\mu_c = a^2 (p-1) + \log{\left(\frac{n_c}{n_c+n_v}\right)} + \mu_c^*, \quad 
	\mu_v = a^2 (p-1) + \log{\left(\frac{n_v}{n_c+n_v}\right)} + \mu_v^*, \quad
	\mu_a = a^2(p-1) + \mu_a^*,
	\end{equation}
	subject to $n=n_c+n_v+n_a =1$ because we have conservation of total mass (or number density). Finally, we will have some initial concentration of lithium ions
	\begin{equation}\label{eq:IC_nc} 
	n_c(z,t=0)=n_{c0}(z).
	\end{equation}
	We have defined three non dimensional parameters in \eqref{eq:scaled_Braun} and \eqref{eq:scaled_chem_pot}, given by
	\[ a^2= 
	\frac{p_r}{n_r k_BT}, \quad \lambda^2 = 
	\frac{\epsilon_0 \left(1+\chi\right)  k_BT}{q^2 n_r L^2}, \quad \delta = 
	\frac{k_B T}{\Delta V q},\]
	where $a$ denotes a pressure scale relative to the ideal gas pressure, $\lambda$ is a spatial scale similar to the Debye length of liquid electrolytes, and  $\delta$ is a non dimensional parameter related to the potential scale. 
	We note that $\delta$ is the ratio of the thermal voltage to the applied voltage and typically (see for example \cite{braun2015thermodynamically}) $\delta \ll 1$.

	Based on the assumption of Schottky defects as a basis for modelling the vacancies we also have that $n_c+n_v=\nu$, a constant.
	Ignoring the $\mu_k^*$ which will disappear in the gradient we can simplify the expression for the flux by noticing 
	\begin{align}
	\mu_c-\mu_v &= \log{\left( \frac{n_c}{\nu}\right)}-\log{\left( \frac{n_v}{\nu}\right)} = \log{\left( \frac{n_c}{n_v}\right)}, \notag\\
	\mu_a + \frac{\nu}{n_a}\mu_v &= \left( 1+ \frac{\nu}{n_a} \right) a^2 (p-1) + \frac{\nu}{n_a}\log{\left( \frac{n_v}{\nu}\right)},
	\end{align}
	leading to
	\begin{equation}\label{eq:scaled_flux}
	J_c = - \nabla \left( \delta \left(\log{\left( \frac{n_c}{\nu-n_c} \right)} - \frac{m_c}{m_a} \frac{n_a+\nu}{n_a} a^2 p - \frac{m_c}{m_a} \frac{\nu}{n_a} \log{\left(\frac{\nu-n_c}{\nu} \right)}\right) + \left( z_c - \frac{m_c}{m_a}z_a \right) \phi \right).
	\end{equation}
	Before moving on to solutions of the model we will eliminate pressure using \eqref{eq:fullProb_p_eq} and we will rewrite the flux equation in the following form
	\begin{equation}
	J_c = -\left( 1+ \frac{m_c n_c}{m_a n_a}\right) \nabla \left( \delta\log{\left(\frac{n_c}{\nu-n_c}\right)} +  z_c \phi \right),
	\end{equation} 
	so that our final model is
	\begin{subequations}
		\label{eq:final_full}
		\begin{align}
		\pderiv{n_c}{t} &= -\frac{\epsilon^2}{z_c} \nabla \cdot \left( \left( 1+\frac{m_c n_c}{m_a n_a}\right) J_c\right), \label{eq:final_full_nc}\\
		n^F &= -\epsilon^2 \nabla^2 \phi, \label{eq:final_full_phi}\\
		J_c &= -\left( 1+ \frac{m_c n_c}{m_a n_a}\right) \nabla \left( \delta \log{\left(\frac{n_c}{\nu-n_c}\right)} + z_c \phi \right) 
		\label{eq:final_full_jc},
		\end{align}
	\end{subequations}
	with our potential boundary conditions $\phi(0)=1$ and $\phi(1)=0$. 
	We note that $\epsilon^2 = \lambda^2 \delta^{-1}$, and since $\lambda \ll 1$ and $\delta \ll 1$, we will take the distinguished limit that $\epsilon \ll 1$. This is supported by Braun \etal  where parameter values lead to $\lambda=1.5 \times 10^{-3}$ and $\delta \approx 5.88 \times 10^{-3}$, thus $\epsilon \approx 1.96 \times 10^{-2} \ll 1$ \cite{braun2015thermodynamically}. 
	
	To analyse the model given by \eqref{eq:final_full} we concentrate on the one-dimensional case. Physically, we also impose global charge neutrality which leads to the integral constraint 
	\begin{equation}\label{eq:int_constraint_nc}
	\int_0^1 n^F \diff{z}=0,
	\end{equation}
	implying that there is no build up of charge over time via \eqref{eq:final_full_nc}.
	This then imposes a constraint on our flux prescribed at the boundaries, the flux at $z=1$ will be a function of the flux at $z=0$. For simplicity, we take the no flux limit, $J_c=0$, and focus on the steady state, $\deriv{n_c}{t}=0$, reducing \eqref{eq:final_full} to
	\begin{subequations}
		\label{eq:1d_equi_j0_nc}
		\begin{align}
		n^F &= -\epsilon^2 \deriv[2]{\phi}{z}, \label{eq:equi_phi_nc}\\
		c &=  \delta \log{\left(\frac{n_c}{\nu-n_c}\right)} + z_c \phi, \label{eq:equi_nc_nc}
		\end{align}
	\end{subequations}
	where $c$ is an unknown constant of integration. In the full time dependent problem we would have \eqref{eq:final_full_nc} and that integral of the initial condition is preserved.
	When we go to the equilibrium problem we lose this condition and so, mathematically, we need another condition in order to compensate for the additional unknown, $c$.   
	In this limit we see the explicit need for the integral condition as the nullspace constraint used to determine $c$.
	
	Now that we have derived the non dimensional model we will make a substitution in order to simplify the relationship between $n_c$ and $\phi$. We note that the number density of lithium ions, $n_c \in (0,\nu)$, can get exponentially close to its two singular limits of 0 and $\nu$. This will cause issues in this logarithmic term of \eqref{eq:equi_nc_nc}. We introduce an auxiliary variable, $\theta$, defined by
	\begin{equation}\label{eq:auxiliary}
	e^{\theta} = \frac{n_c}{\nu-n_c}
	\end{equation}
	which maps $n_c \in (0,\nu)$ to $\theta \in (-\infty,+\infty)$ to avoid numerical artifacts of near singularity by transforming to a smooth domain.
	Rewriting the problem using the $\theta$ substitution leads to
	\begin{subequations}
		\label{eq:1d_equi_j0}
		\begin{align}
		\epsilon^2 \deriv[2]{\phi}{z}&=-n^F = -\frac{z_c \nu}{1+ e^{-\theta}} - z_a n_a , \label{eq:equi_phi}\\
		c &=  \delta \theta + z_c \phi, \label{eq:equi_nc}
		\end{align}
	\end{subequations}
	The integral constraint \eqref{eq:int_constraint_nc} can be written in terms of $\theta$ as
	\begin{equation}\label{eq:int_constraint}
	\int_0^1 \frac{1}{1+e^{-\theta}} \diff{z} = - \frac{z_a n_a}{z_c \nu}.
	\end{equation}	
	We want to solve \eqref{eq:1d_equi_j0} with \eqref{eq:int_constraint} subject to $\phi(0)=1, \phi(1)=0$.

	\section{Numerical solution}\label{sec:num_equi}
\begin{figure}[htb]
		%
		\begin{subfigure}{0.49\linewidth}
			\includegraphics[width=\linewidth]{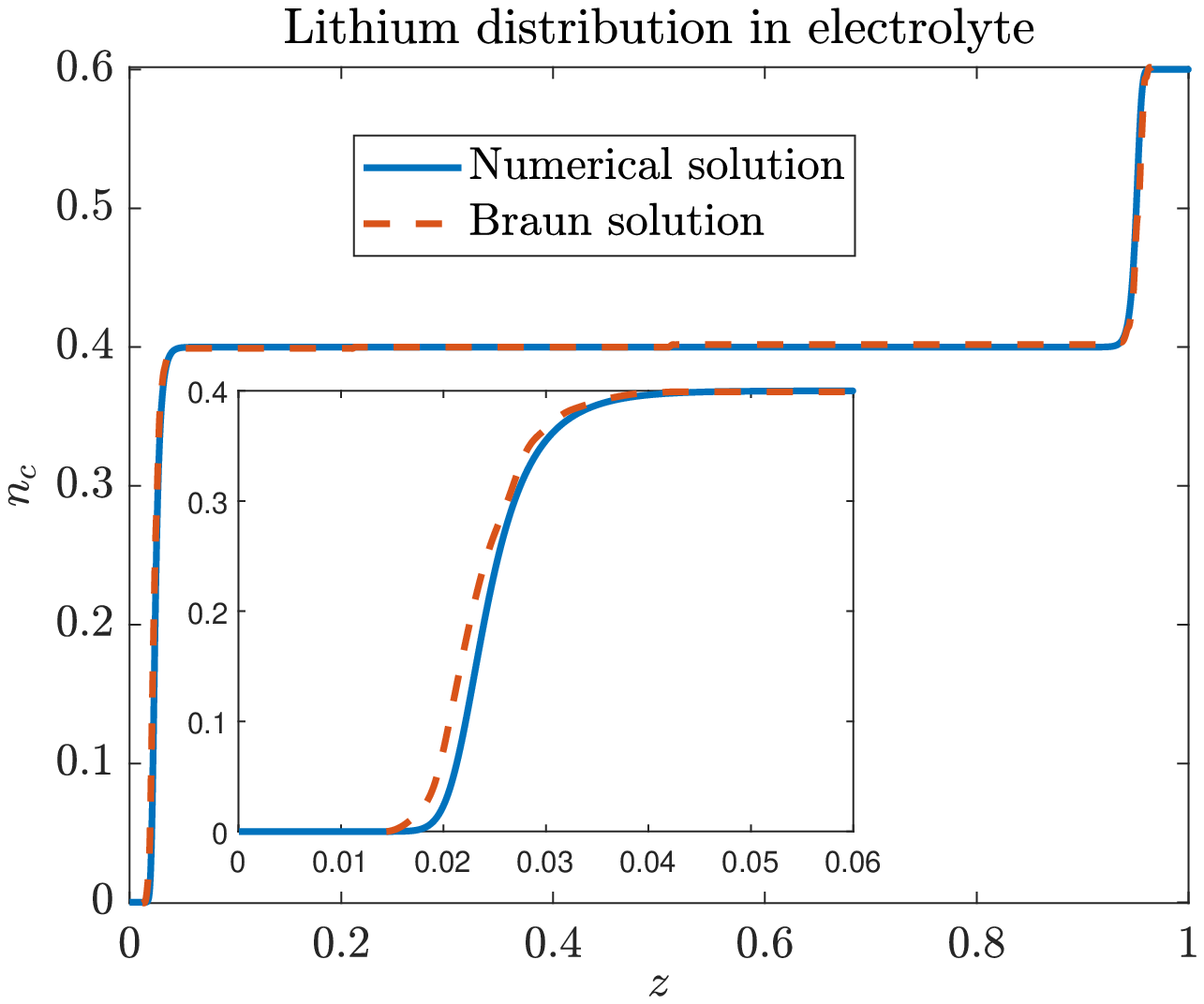}
			\caption{Profile of lithium ions distribution.}
			\label{fig:equi1D_a}
		\end{subfigure}
		\begin{subfigure}{0.49\linewidth}
			\includegraphics[width=\linewidth]{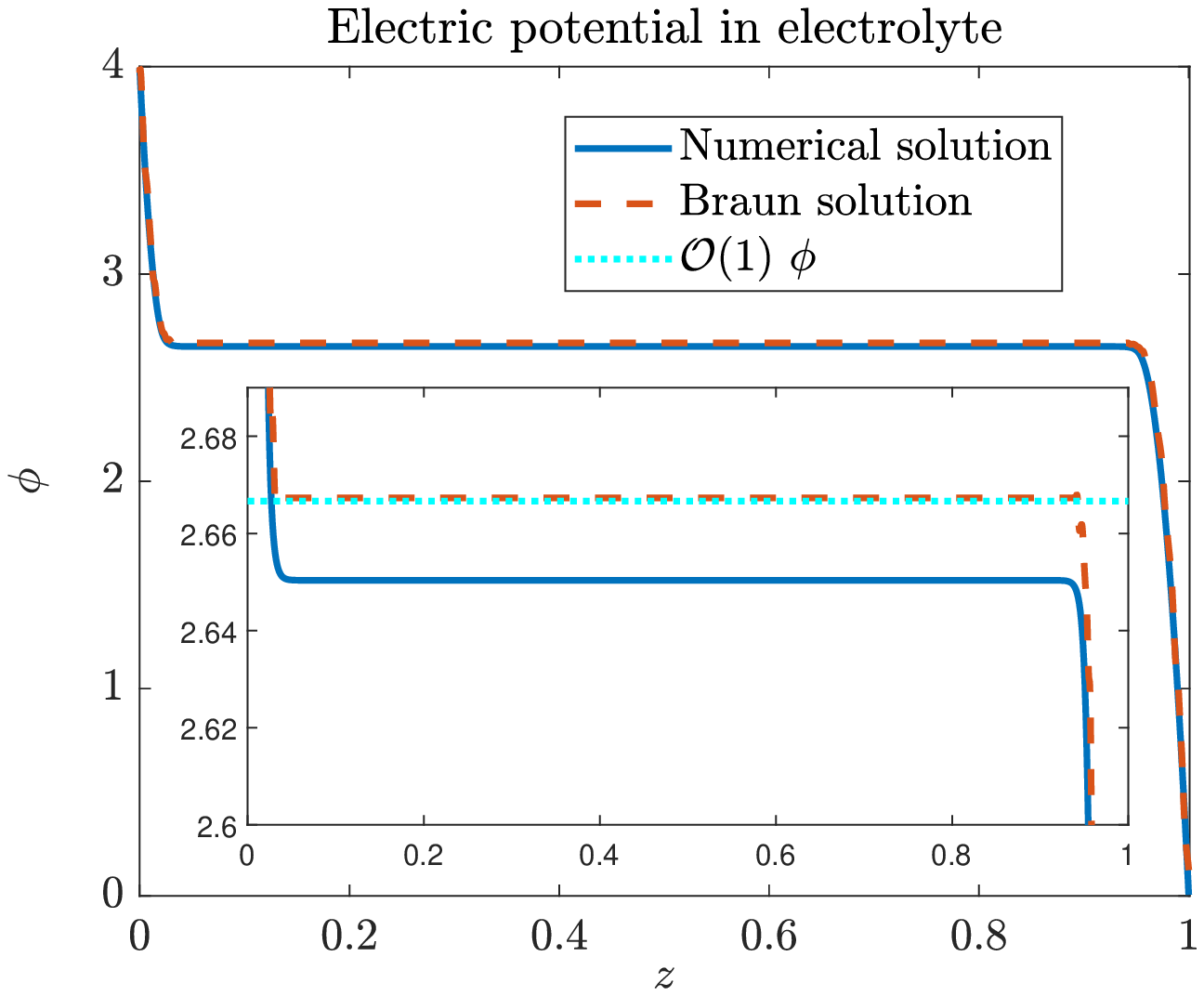}
			\caption{Profile of the electric potential.}
			\label{fig:equi1D_b}
		\end{subfigure}
		\caption{Simulation results for the 1D equilibrium problem   \eqref{eq:1d_equi_j0} with \eqref{eq:int_constraint} subject to $\phi(0)=1$ and $\phi(1)=0$  with parameter values: $z_c=1, \quad z_a=-1, \quad m_c=0.3, \quad m_a=0.7, \quad n_a=0.4,\quad \nu=0.6$	where $n=n_a+\nu=1$, $m=m_a+m_c=1$. We also have $\lambda=1.5 \times 10^{-3}$, $\delta^{-1}=170$, and $\Delta V = 4$. We show our numerical solution in comparison with the solution obtained by \cite{braun2015thermodynamically}. In the inset plot of Figure \ref{fig:equi1D_a} we zoom in on the left boundary to highlight the difference between the two solutions. Similarly in the inset of Figure \ref{fig:equi1D_b} we zoom in on the bulk portion of the solution and include the leading order solution (in light blue) to highlight discrepancies between the solutions.
		}
		\label{fig:equi1D}
	\end{figure}

	We consider the zero flux equilibrium problem,  solving \eqref{eq:1d_equi_j0} with \eqref{eq:int_constraint} subject to $\phi(0)=1, \quad \phi(1)=0$.  We use the same parameter values as used in \cite{braun2015thermodynamically}
	\[z_c=1, \quad z_a=-1, \quad m_c=0.3, \quad m_a=0.7, \quad n_a=0.4,\quad \nu=0.6,\]
	where $n=n_a+\nu=1$ and $m=m_a+m_c=1$ from our non dimensionalisation.
	Numerically we use cell centered finite differences for the derivatives, the midpoint rule for the integral, and Newton's method to solve \eqref{eq:1d_equi_j0}.   We recall that multiple authors have acknowledged difficulties in numerically solving SE models (\cite{katzenmeier2022modeling, swift2021modeling, landstorfer2011advanced}) and to evade these difficulties Braun \etal \cite{braun2015thermodynamically} employ a semi-analytic formulation. 
	In Figure \ref{fig:equi1D} we compare our numerical solutions for the lithium ion and electric potential distributions in the zero flux equilibrium case with the solutions obtained by \cite{braun2015thermodynamically}. We note that the general behaviour agrees between our numerical approach and their semi-analytic approach in both plots of figure \ref{fig:equi1D}, however we highlight some of the discrepancies between the two. 
	On closer inspection of the two solutions we observe some disparity in the boundary layer solutions for the $n_c$ profile, specifically we draw attention to the inset plot of figure \ref{fig:equi1D_a}. The solution of \cite{braun2015thermodynamically} (in red dotted lines) diverges from our numerics (solid blue line) as we move away from the bulk of the solution. We observe that the difference between the two solutions seems to increase the further away we are from the bulk. In addition we notice some difference in the tails of the solutions, the solution in solid blue gets exponentially close to the limiting value of 0 while the red dotted solution appears to terminate much sooner. This likely arises due to the singularities near $0$ and $\nu$.
	We also compare the solutions for the potential profile. Overall we observe similar behaviour, however on inspection of figure \ref{fig:equi1D_b}, and in particular, the inset plot, we notice that the two solutions seem to differ by a constant in the bulk. Considering Equation \eqref{eq:equi_nc} we observe that if we peel off the $\delta \theta $ term and plot the $\frac{c}{z_c}$ part of $\phi$ (shown by the light blue dotted line), \ie neglect the $\mathcal{O}(\delta)$ correction, we seem to capture the plot of \cite{braun2015thermodynamically}. We note that a similar correction also applies to the boundary layers of figure \ref{fig:equi1D_b}. From these observations the semi-analytic approach therefore appears to be a coarse first-order approximation to the problem where as the auxiliary variable is able to remove the singularities and solve the full problem.
	
	In both plots we observe that the profiles of both the lithium ions and the electric potential are constant throughout the bulk of the electrolyte and that there are narrow boundary layers on either side of the bulk. These layers are the SCL, corresponding to the regions in which the solution deviates from electroneutrality. 
	We observe from \eqref{eq:equi_phi} that $\epsilon$ is the small parameter of the boundary layer. 
	Based on the parameter values used by \cite{braun2015thermodynamically} we will have $\epsilon = \sqrt{\lambda^2 \delta^{-1} }= 0.0196 \approx 10 \lambda$, aligning with qualitative observations made by Braun \etal regarding the SCL width \cite{braun2015thermodynamically}. We note that similar comments on these wider space charge layer lengths were indicated in work by other authors \cite{li2019dendrite, li2021transport}. Our scaling therefore provides a quantifiable measure of these double layer widths.

	\section{Asymptotic reduction}
	\label{sec:asymptotics}
	Motivated by the distinct boundary region behaviour observed in our numerical solutions in Figure \ref{fig:equi1D}, we proceed with an asymptotic reduction of the ODE model in order to determine the structure of the layers and to gain a deeper understanding of charge layer thickness. We note that for this one dimensional zero flux problem as presented here it is possible to determine an analytical solution via a first integral approach and that asymptotic methods can also be applied via an integral approach (see the supplemental \cref{SM-sec:appendix})

	We take the 1D zero flux equilibrium problem given by \eqref{eq:1d_equi_j0}, \eqref{eq:int_constraint} and subject to $\phi(0)=1, \quad \phi(1)=0$.
	In equilibrium we note that we can use \eqref{eq:equi_nc} to write $\phi$ in terms of $\theta$, eliminating $\phi$ from the problem. 
	Rewriting the zero flux equilibrium problem in terms of $\theta$ we have
	\begin{subequations}\label{eq:ZeroJ_equi_theta_prob}
		\begin{align}
		\frac{\epsilon^2 \delta}{z_c} \deriv[2]{\theta}{z} &= \frac{\beta_1+\beta_2}{1+ e^{-\theta}} - \beta_1 \label{eq:zero_asym_prob} \\
		\int_0^1 \frac{1}{1+e^{-\theta}} \diff{z} &=\frac{\beta_1}{\beta_1+\beta_2} \label{eq:equi_0_int} \\
		\theta(0)=\frac{c-z_c}{\delta}, & \quad \theta(1)=\frac{c}{\delta},\label{eq:bc_theta_zeroFluxEqui}
		\end{align}
	\end{subequations}
	where we define
	\[ \beta_1 = - z_a n_a, \quad \beta_2=z_a n_a + z_c \nu.\]
	We note here that the boundary conditions for $\theta$ are large, but the problem is sensibly scaled for $\phi$ (where the original boundary conditions are posed) and we have chosen a scaling that preserves $\theta \sim \mathcal{O}(1)$ in the bulk.

	\subsection{Bulk}\label{sec:asy_bulk}
	In the bulk of our solution $z \sim \mathcal{O}(1)$ and \eqref{eq:zero_asym_prob} becomes
	\[ 0 =\frac{\beta_1+\beta_2}{1+ e^{-\theta}} - \beta_1  \]
	with solution
	\begin{equation}\label{eq:theta_0_bulk}
	\theta=\theta_0 := \log{\left( \frac{\beta_1}{\beta_2}\right)}.
	\end{equation}
	We note that all corrections to the bulk are going to be exponentially small and thus there is no formal power series correction of $\mathcal{O}(\delta)$. Now, evidently this bulk solution can satisfy neither boundary condition (and correspondingly the two different prescribed potential values in $\phi$), indicating the need for boundary layer problems at either side of the bulk.
	The corresponding solution for the potential we obtain from \eqref{eq:equi_nc} to be 
	\begin{equation}\label{eq:phi_0_bulk}
	\phi = \frac{c}{z_c} - \delta \frac{\theta_0}{z_c},
	\end{equation}
	where the constant $c$ is, as of yet, unknown and will be determined when we carry out our matching.
	We highlight that $\phi$ does have an $\mathcal{O}(\delta)$ correction which depends on the bulk solution for $\theta$. Equation \eqref{eq:equi_nc} also tells us that $\mathcal{O}(1)$ changes in $\theta$ have $\mathcal{O}(\delta)$ impacts on $\phi$ everywhere in the domain. In the numerical plots in Section \ref{sec:num_equi} we observed that the solutions of \cite{braun2015thermodynamically} capture the $\frac{c}{z_c}$ solution component of $\phi$ only. 
	
	\subsection{Boundary layer}\label{sec:asy_bl} 
	%
	At the boundary near $z=0$ we will scale $z = \epsilon \xl$ and $\theta =  -\delta^{-1} \psibl$ as we expect $\theta$ to be large and negative here from our boundary condition \eqref{eq:bc_theta_zeroFluxEqui}. We use the subscripts $\rm{L}$ and $\rm{BL}$ to indicate the left boundary layer (near $z=0$). Therefore, our problem in this boundary layer can be written as
	\begin{equation}\label{eq:bl_th_at0}
	\deriv[2]{\psibl}{\xl}=z_c \beta_1 - z_c\frac{\beta_1+\beta_2}{1+e^{+\delta^{-1} \psibl}} \approx z_c \beta_1.
	\end{equation}
	If we expand $\psibl$ up to $\mathcal{O}(\delta)$ so that $\theta_{\rm{BL}}$ has an $\mathcal{O}(1)$ component for matching to the bulk,
	\begin{equation}\label{eq:expand_psiL}
	\psibl = \psibl^0+\delta \psibl^1 
	\end{equation}
	then we will have the following problems and boundary conditions
	\begin{subequations}\label{eq:psi0_1_near0_bl}
		\begin{align}
		\mathcal{O}(1): \quad & \deriv[2]{\psibl^0}{\xl} = z_c \beta_1, \quad \psibl^0(0)= z_c -c,\\
		\mathcal{O}(\delta): \quad & \deriv[2]{\psibl^1}{\xl} = 0, \quad \psibl^1(0)=0
		\end{align}
	\end{subequations}
	with solution
	\begin{subequations}\label{eq:psi_sol_lhs}
		\begin{align}
		\mathcal{O}(1): \quad & \psibl^0 = \frac{z_c \beta_1 \xl^2}{2} +  A_0 \xl + z_c -c \\
		\mathcal{O}(\delta): \quad & \psibl^1  = - \frac{A_1}{A_0} \xl,
		\end{align}
	\end{subequations}
	for some unknown constants $A_0$ and  $A_1$.

	We do the same thing for the boundary layer near $z=1$, where we scale $z = 1-\epsilon \xr$ and $\theta = \delta^{-1} \psibr$ as here we expect $\theta$ to be large and positive, with subscripts $\rm{R}$ and $\rm{BR}$ denoting the boundary layer near 1.
	In this case we will have 
	\[ \deriv[2]{\psibr}{\xr} = -z_c \beta_1 + z_c\frac{\beta_1+\beta_2}{1+ e^{-\delta^{-1}\psibr}} \approx z_c \beta_2.\]
	We expand $\psibr$ in the same way as \eqref{eq:expand_psiL} to obtain 
	\begin{subequations}\label{eq:psi_sol_rhs}
		\begin{align}
		\mathcal{O}(1): \quad & \psibr^0 = \frac{z_c \beta_2 \xr^2}{2} + B_0 \xr +c \\
		\mathcal{O}(\delta): \quad & \psibr^1 =  - \frac{B_1}{B_0} \xr
		\end{align}
	\end{subequations}
	for constants $B_0$ and $B_1$.
	
	We note that as a consequence of the integral constraint \eqref{eq:equi_0_int}, from \eqref{eq:fullProb_phi_eq} we must have 
	\begin{equation}
	\lambda^2 \delta^{-1} \int_0^1 \deriv[2]{\phi}{z} \diff{z} = - z_c \int_0^1 n_c \diff{z} - z_a n_a =0, 
	\end{equation}
	implying that 
	\begin{equation} \deriv{\psibl}{\xl}\biggr\rvert_{\left(\xl=0\right)} = \deriv{\psibr}{\xr}\biggr\rvert_{\left(\xr=0\right)}
	\end{equation}
	and thus that
	\[B_0=A_0, \quad B_1= A_1. \]

	To determine the constants $A_0$ and $A_1$ we notice that the full problem \eqref{eq:zero_asym_prob} has a first integral. We would normally find $A_0$ and $A_1$ through the integral condition \eqref{eq:equi_0_int}, but this involves matching together asymptotic solutions which likely introduces error. We can avoid this by noting that we can instead determine a local condition by taking the first integral. Multiplying both sides by $\deriv{\theta}{z}$ and integrating with respect to $z$
	\begin{align}\label{eq:first_int}
	\frac{\epsilon^2 \delta}{z_c} \deriv[2]{\theta}{z} \deriv{\theta}{z} &= \left(\frac{\beta_1+\beta_2}{1+e^{-\theta}}-\beta_1\right)\deriv{\theta}{z} \notag \\
	\frac{1}{2} \frac{\epsilon^2 \delta}{z_c} \left(\deriv{\theta}{z}\right)^2 &= \int \left(\frac{\beta_1+\beta_2}{1+e^{-\theta}}-\beta_1\right)\diff{\theta} + z_c f_0\notag \\ & = \left(\beta_1+\beta_2\right) \log{\left(1+e^{-\theta}\right)} -\beta_1 \theta + f_0.
	\end{align}
	Based on the bulk value we know that $\deriv{\theta}{z}=0$ when $\theta=\theta_0$, therefore we find
	\begin{equation}\label{eq:f0}
	f_0=\beta_1 \log{\left(\frac{\beta_1}{\beta_2}\right)} - \left(\beta_1+\beta_2\right) \log{\left(\frac{\beta_1+\beta_2}{\beta_2}\right)}.
	\end{equation}
	Then we can impose a local condition at $z=0$: from our boundary condition we know $\theta(0)=\frac{c-z_c}{\delta}$, our boundary layer solution says that $\deriv{\theta}{z}= -\frac{1}{\epsilon \delta} \left( z_c \beta_1 \frac{z}{\epsilon} + A_0 - \delta \frac{A_1}{A_0}\right)$. Therefore we must have that
	\begin{align}
	\frac{\epsilon^2 \delta}{2 z_c} \left(-\frac{1}{\epsilon \delta} \left( z_c \beta_1 \frac{z}{\epsilon} + A_0 - \delta \frac{A_1}{A_0}\right)\right)^2\rvert_{z=0} =  &\left(\beta_1+\beta_2\right) \log{\left(1+e^{-\frac{c-z_c}{\delta}}\right)} -\beta_1 \left(\frac{c-z_c}{\delta}\right) \notag \\ & + \beta_1 \log{\left(\frac{\beta_1}{\beta_2}\right)} - \left(\beta_1+\beta_2\right) \log{\left(\frac{\beta_1+\beta_2}{\beta_2}\right)}.
	\end{align}
	Expanding and matching $\mathcal{O}(\delta^{-1})$ and $\mathcal{O}(1)$ terms, noting that $e^{-\frac{c-z_c}{\delta}}$ is asymptotically small, we find that
	\begin{equation}\label{eq:A0sol}
	A_0=-\sqrt{2 z_c \beta_1 (z_c-c)},
	\end{equation}
	and
	\begin{equation}\label{eq:A1_sol_firstInt}
	A_1=z_c \left(\left(\beta_1+\beta_2\right)\log{\left(\frac{\beta_1+\beta_2}{\beta_2}\right)}- \beta_1\log{\left(\frac{\beta_1}{\beta_2}\right)}\right).
	\end{equation}

	In order to determine the value of the remaining unknown $c$ we will have to match between the different regions. 
	We note that we will also need $\theta \sim \mathcal{O}(1)$ towards the bulk to match there, but the boundary layer solution on the left \eqref{eq:psi_sol_lhs} cannot reach the bulk because the quadratic doesn't plateau and begins to decrease before the bulk value. 
	We consider again equation \eqref{eq:zero_asym_prob} noting that there are two possible scalings which retain the derivative when $z \sim \epsilon$
	\[ \theta \sim \delta^{-1} \text{ or } x \sim \sqrt{\delta}.\]
	Therefore, we must have some intermediate layer where $\theta \sim \mathcal{O}(1)$ and $x \sim \sqrt{\delta}$ to facilitate where $\theta$ transitions from large to $\mathcal{O}(1)$ and to ensure that we have continuity between the bulk and boundary layers. 

	\subsection{Intermediate layer}\label{sec:asy_il} 
	
	To investigate our intermediate layer solution we will start with the left side. We rescale
	\[ \xl = \xil  + \sqrt{\delta} \yl,  \]
	where $\xil$ is the point of continuity between the boundary and the intermediate layer solutions. 
	In this region we will denote the solution for $\theta$ by 
	\[\theta_{\text{IL}}(\yl)=\theta( \xil+\sqrt{\delta} \yl), \]
	where the subscript ${\text{IL}}$ represents the left intermediate solution.
	This leads to the following equation for $\theta$
	\begin{equation}\label{eq:theta_IL_left_eq}
	\deriv[2]{\theta_{\text{IL}}}{\yl}= -z_c\beta_1+ z_c \frac{\beta_1+\beta_2}{1+ e^{-\theta_{\text{IL}}}}.
	\end{equation}
	Before going further we acknowledge a couple of things.
	We note that this intermediate layer is not needed to resolve the leading order potential problem. Recalling that $z_c \phi + \delta \theta =c$, this intermediate layer will have its strongest effect at $\mathcal{O}(\delta)$ in $\phi$, whereas it has $\mathcal{O}(1)$ effects in $\theta$. In other words, $ \phi_{\rm{IL}} = \frac{c}{z_c} - \frac{\delta}{z_c} \theta_{\rm{IL}} $ so the intermediate layer behaviour is a correction to $\phi$. 
	We will assume that the intermediate $\theta$ is also close to its bulk value
	\[\theta_{\text{IL}} = \theta_0 + \thl, \]
	where $\thl \ll 1$ is the intermediate solution correction to the bulk value.
	Substituting this expression for $\theta_{\text{IL}}$ into \eqref{eq:theta_IL_left_eq} we have
	\begin{align}\label{eq:xi_left_prob}
	\deriv[2]{\thl}{\yl} &= -z_c \beta_1+ z_c\frac{\beta_1+\beta_2}{1+e^{-\theta_0-\thl}}  \notag \\   &\approx -z_c \beta_1+ z_c \left(\beta_1+\beta_2\right) \left( \frac{1}{1+e^{-\theta_0}} + \frac{e^{-\theta_0}}{\left(1+e^{-\theta_0}\right)^2} \thl  \right) \notag \\
	&= -z_c \beta_1+ z_c \left(\beta_1+\beta_2\right)\left( \frac{\beta_1}{\beta_1 + \beta_2} + \frac{\beta_1 \beta_2}{\left(\beta_1+\beta_2\right)^2} \thl  \right) \notag \\
	&= \frac{\beta_1 \beta_2}{\nu} \thl
	\end{align}
	where we have Taylor expanded about $\thl=0$.
	Solving \eqref{eq:xi_left_prob} yields
	\begin{equation}\label{eq:xi_sol_leftIL}
	\thl = \left(\theta_1-\theta_0\right) e^{-\sqrt{\frac{\beta_1 \beta_2}{\nu}}\yl} =  \left(\theta_1-\theta_0\right) e^{-\sqrt{\frac{\beta_1 \beta_2}{\delta \nu}} \left(\xl-\xil \right)}=  
	\left(\theta_1-\theta_0\right) e^{-a \frac{\left(z-z_L \right)}{\epsilon \sqrt{\delta}}},
	\end{equation}
	for some constant $\theta_1$ with $z_L=\epsilon \xil$, and where we define $ a=\sqrt{\frac{z_c \beta_1 \beta_2}{\beta_1+\beta_2}}=\sqrt{\frac{\beta_1 \beta_2}{ \nu}}$. We note that $\thl \to 0$ as $\yl \to \infty$, thus $\theta_{\text{IL}} \to \theta_0$  as required.

	For the intermediate layer near $z=1$ we will get analogous expressions when we scale $ \xr = \xir + \sqrt{\delta}\yr $ and let $\theta_{\text{IR}} = \theta_0 + \thr$ to find the following form for the intermediate layer solution for $\theta$ which we denote as $\theta_{\text{IR}}$
	\begin{equation}\label{eq:theta_sol_rightIL}
	\theta_{\text{IR}} = \theta_0 + \left( \theta_2-\theta_0 \right) e^{-\sqrt{\frac{\beta_1 \beta_2 }{\delta \nu}}\left(\xr-\xir\right)}= \theta_0 + \left( \theta_2-\theta_0 \right) e^{-a \frac{\left(z_R-z\right)}{\epsilon \sqrt{\delta}}},
	\end{equation}
	for some constant $\theta_2$ with $z_R=1-\epsilon \xir$.

	\subsection{Matching}\label{sec:asy_matching} 
	Naturally, continuity and differentiability between the boundary and intermediate solutions at $z_L$ and $z_R$ would furnish the remaining conditions to solve the problem. However, the intermediate layer solution is derived from a far-field expansion near the bulk $\theta=\theta_0$ and therefore these two layers need not agree at $z_L$ and $z_R$. The monotonicity of the intermediate layer provides an mechanism for continuity, but differentiability can, in general, not be satisfied. Instead, we will perform a pesudo-matching where we minimize the error in differentiability between the two layers.
	Of course, this is not a classic boundary layer as we need two solutions going to infinity to match. There is some other layer where the full non-linear problem must be realized, but we are just seeking an approximation of these layer locations so we will proceed with this minimisation and pseudo matching instead.

	We start with the left hand side. Here we want to match our solution for $\psibl$ in the boundary layer with our solution for $\theta_{\text{IL}}$ in the intermediate layer.
	Enforcing continuity at $z_L$ requires
	$\theta_{\text{BL}} (z_L) =-\delta^{-1} \psibl (z_L) =  \theta_{\text{IL}} (z_L)$, so we will have
	\begin{align}\label{eq:match_lhs}
	- \delta^{-1} \left(\frac{z_c \beta_1}{2} \left(\frac{z_L}{\epsilon}\right)^2 +  A_0 \frac{z_L}{\epsilon} + z_c -c  - \delta \frac{A_1}{A_0} \frac{z_L}{\epsilon}\right)  &=\theta_1, 
	\end{align}
	which we match term by term in orders of $\delta$.
	We define $F_1$ to be the difference in the derivatives, given by
	\begin{align}
	F_1 &= \deriv{\theta_{\text{BL}}}{z}\rvert_{z_L}-\deriv{\theta_{\text{IL}}}{z}\rvert_{z_L} \notag \\
	&= -\frac{1}{\delta \epsilon} \left( z_c \beta_1 \frac{z_L}{\epsilon} +A_0 - \delta \frac{A_1}{A_0} \right) + \frac{a}{\sqrt{\delta} \epsilon}\left(\theta_1-\theta_0\right)
	\label{eq:F1}.
	\end{align}
	We can then substitute $\theta_1$ from the continuity condition \eqref{eq:match_lhs} into \eqref{eq:F1}. Normally, we would match to enforce differentiability so that $F_1=0$.  There will be scenarios where we can have differentiability but we will also have scenarios where this is not possible, in which case we determine the value of $z_L$ that minimises $|F_1|$. We note that
	\begin{equation}
	F_1'' = -\frac{ a z_c \beta_1}{\lambda^3} <0,
	\end{equation}
	so $F_1$ is convex.
	We have similar matching conditions for the right hand side, for continuity $\delta^{-1} \psibr\left(z_R\right) =  \theta_{\text{IR}} (z_R)$ so
	\begin{align}\label{eq:match_rhs}	
	\delta^{-1} \left(\frac{z_c \beta_2}{2} \left(\frac{1-z_R}{\epsilon}\right)^2 +  A_0 \frac{1-z_R}{\epsilon} + c  - \delta \frac{A_1}{A_0} \frac{1-z_R}{\epsilon}\right)  &=\theta_2
	\end{align}
	and we define $F_2$ as 
	\begin{align}
	F_2 &= \deriv{\theta_{\text{BR}}}{z}\rvert_{z_R}-\deriv{\theta_{\text{IR}}}{z}\rvert_{z_R} \notag \\
	&= -\frac{1}{\delta \epsilon} \left( z_c \beta_2 \left( \frac{1-z_R}{\epsilon}\right) +A_0 - \delta \frac{A_1}{A_0} \right) -\frac{a}{\sqrt{\delta} \epsilon}\left(\theta_2-\theta_0\right), \label{eq:F2}
	\end{align}
	which we note is also convex.
	
	We note that as we have $F_i''<0$, if $F_i\left(z_i^*\right)>0$, where $z_i^*$ is the critical point, then the parabola will have a root and we can enforce differentiability.
	In this case we can solve for the root by setting $F_i\left(z_i\right)=0$, 
	we will take the root closer to the intermediate layer (larger $z_i$). Otherwise, we will take the value of $z_i$ that minimises $|F_i|$, \ie $z_i^*$.
	We find that $F_1'=0$ if
	\begin{equation}\label{eq:zl_sol}
	z_L^* = -\epsilon \left( \frac{A_0}{z_c \beta_1}+ \frac{\sqrt{\delta}}{a} -  \frac{\delta A_1}{z_c \beta_1 A_0}\right), 
	\end{equation}
	and $F_2'=0$ for
	\begin{equation}\label{eq:zr_sol}
	z_R^*=1-\epsilon \left(-\frac{A_0}{z_c \beta_2} - \frac{\sqrt{\delta}}{a} + \frac{\delta A_1}{ z_c \beta_2 A_0}\right).
	\end{equation}
	
	Given the parameters obtained from our matching using the first integral (\eqref{eq:A0sol} for $A_0$ and \eqref{eq:A1_sol_firstInt} for $A_1$) and the values of $z_L^*$ from \eqref{eq:zl_sol} and $ z_R^*$ from \eqref{eq:zr_sol} we will have 
	\begin{align}
	F_1\left(z_L^*\right) 
	&= \frac{z_c \beta_2}{2 \epsilon \sqrt{\delta}} \sqrt{\frac{\beta_1+\beta_2}{z_c \beta_1 \beta_2}} \left( \frac{\beta_1}{\beta_2} - 2 \log{\left(1+ \frac{\beta_1}{\beta_2}\right)}\right)
	\end{align}
	We note that we can determine whether we have differentiability or whether we instead minimize $F_i$ based on the values of $\beta_i$
	\begin{equation}
	\frac{\beta_1}{\beta_2}-2\log{\left(1+\frac{\beta_1}{\beta_2}\right)}=0 \text{ when } \frac{\beta_1}{\beta_2}=2.51286 =\beta^*.
	\end{equation}
	Therefore we have that $F_1(z_L^*)<0$ for $\frac{\beta_1}{\beta_2}<\beta^*$ and $F_1(z_L^*)>0$ otherwise, \ie $\beta^*$ is the critical value, since $F_1$ is convex then if $F_1(z_L^*)<0$ then there are no roots and differentiability is not permissible.
	Similarly, for $F_2$ we find
	\begin{equation}
	F_2 \left(z_R^*\right)= \frac{z_c \beta_1}{2 \epsilon \sqrt{\delta} }\sqrt{\frac{\beta_1+\beta_2}{z_c \beta_1 \beta_2}} \left(\frac{\beta_2}{\beta_1} - 2 \log{\left(1+ \frac{\beta_2}{\beta_1}\right)}\right),
	\end{equation}
	and we will have $F_2>0$ when $\frac{\beta_2}{\beta_1} - 2 \log{\left(1+ \frac{\beta_2}{\beta_1}\right)}>0$ which occurs when $\frac{\beta_2}{\beta_1}> \beta^*$.

	We can substitute these value for $z_L$ and $z_R$ obtained from $F(z_i)=0$ or from the minimisation into equations \eqref{eq:match_lhs} and \eqref{eq:match_rhs} to determine $\theta_1$ and $\theta_2$.
	If we consider, as an example, the scenario where both $F_i(z_i^*)<0$ and we take $z_L=z_L^*, z_R=z_R^*$.
	Using \eqref{eq:zl_sol} for $z_L$ in the continuity condition \eqref{eq:match_lhs} 
	at $\mathcal{O}(1)$ yields
	\begin{equation}\label{eq:th1sol}
	\theta_1= - \frac{\beta_1+\beta_2}{2 \beta_2} - \frac{A_1}{z_c \beta_1}. 
	\end{equation}
	noting that we only take the leading order solution for $\theta$.
	Similarly, using \eqref{eq:zr_sol} for $z_R$ we find
	\begin{equation}\label{eq:th2sol}
	\theta_2 = \frac{\beta_1+\beta_2}{2 \beta_1}+ \frac{A_1}{z_c \beta_2}. 
	\end{equation}

	We highlight that $\xil \neq \xir$, \ie the boundary layers on either side have different widths and therefore the problem is not quite symmetric. As highlighted by Braun \etal \cite{braun2015thermodynamically}, this is due to the fixed anion lattice since the free charge density, $ n^F\in [-n_a, \nu-n_a]$, is only symmetric if $\nu=2 n_a$.
	Substituting $z_{R}$ and $ \theta_2$ into the continuity condition \eqref{eq:match_rhs}, using our value of $A_0$ \eqref{eq:A0sol}, and matching at orders of $\delta$ we find that 
	\begin{equation}\label{eq:c_value}
	c=\frac{\beta_1}{\nu},
	\end{equation}
	completing our solution.

	\subsection{SCL Summary}
	We summarise the characterisation of the space-charge-layers in terms of the solution for the auxiliary variable, $\theta$, as shown in table \ref{table:scl_summary}.
	\begin{table}[htb]
		\caption{Summary of the solution obtained in each region of the solid electrolyte.}
		\centering
		\begin{tabular}{||p{3.67cm}||p{9cm}||} 
			\hline
			\multicolumn{2}{|c|}{Summary of the solutions obtained in each of the regions} \\
			\hline
			Region & Solution for $\theta$ \\ [0.5ex] 
			\hline\hline
			Left boundary layer & $\theta_{\rm{BL}}= -\delta^{-1} \psibl = - \delta^{-1} \left(  \frac{z_c \beta_1 \xl^2}{2} +  A_0 \xl + z_c -c - \delta \frac{A_1}{A_0} \xl\right)$\\
			\hline
			Left intermediate layer & $\theta_{\text{IL}} = \theta_0 +   \left(\theta_1-\theta_0\right) e^{-a \frac{\left(z-z_L \right)}{\epsilon \sqrt{\delta}}}$\\
			\hline
			Bulk &  $\theta_0= \log{\left( \frac{\beta_1}{\beta_2}\right)}$\\
			\hline
			Right intermediate layer &  $\theta_{\text{IR}} = \theta_0 + \left( \theta_2-\theta_0 \right) e^{-a \frac{\left(z_R-z\right)}{\epsilon \sqrt{\delta}}}$ \\
			\hline
			Right boundary layer &  $\theta_{\rm{BR}} = \delta^{-1} \psibr = \delta^{-1} \left(  \frac{z_c \beta_2 \xr^2}{2} +  A_0 \xr + c - \delta \frac{A_1}{A_0} \xr\right)$\\
			[1ex] 
			\hline\hline
		\end{tabular}
		\label{table:scl_summary}
	\end{table}
	We reiterate that the left boundary then meets the left intermediate layer at $z_L$, the right intermediate layer joins the right boundary layer at $z_R$, and the bulk solution connects the two intermediate layer solutions.
	We have defined $ a=\sqrt{\frac{\beta_1 \beta_2}{ \nu}}$ and we have the determined the constants $A_0, A_1, z_L, z_R, \theta_1, \theta_2$ and $c$ via matching to be given by \eqref{eq:A0sol}, \eqref{eq:A1_sol_firstInt}, \eqref{eq:zl_sol}, \eqref{eq:zr_sol}, \eqref{eq:th1sol}, \eqref{eq:th2sol}, and \eqref{eq:c_value} respectively.
	
	Having completed our asymptotic solution to the problem we highlight that the boundary layer solutions are of width $\epsilon$ and the intermediate layer is of width $\lambda$. On investigation of these solution profiles we observe that there is rapid change in the boundary layer solutions, which we will refer to as a strong SCL as a result. The intermediate layer solutions involve a rapidly decaying exponential and facilitate the transition between the region of fully lithiated/depleted of cations and the constant bulk concentration, therefore we will refer to this region as the weak SCL. From our SCL summary in table \ref{table:scl_summary} we note that the strong SCL exhibits quadratic behaviour while the weak SCL exhibits exponential behaviour, agreeing with the observations made by \cite{swift2021modeling}.

	\section{Comparison of asymptotics and numerics}
	\label{sec:comparison}
	
\begin{figure}[htb]
		\begin{subfigure}{0.47\linewidth}
			\includegraphics[width=\linewidth]{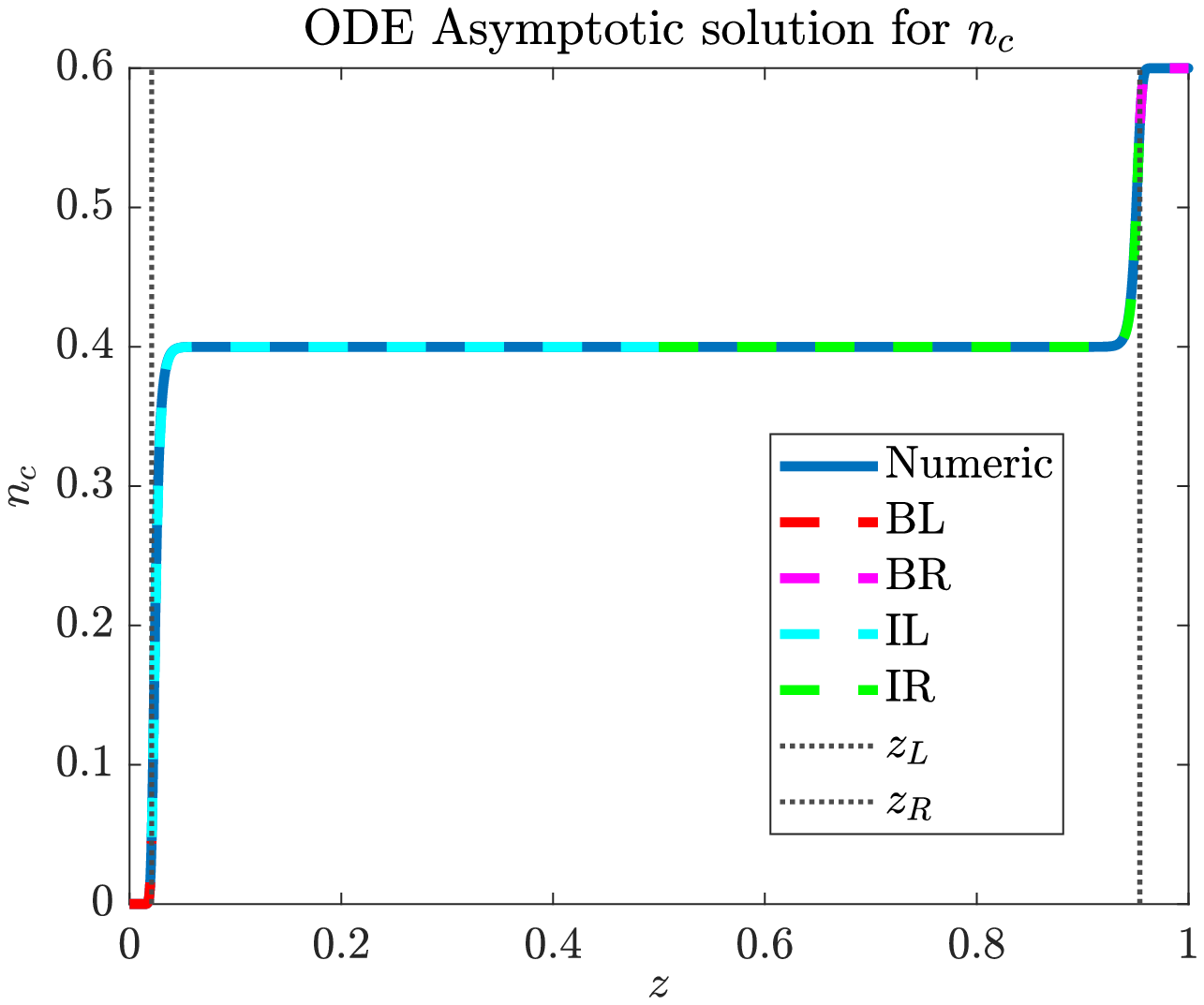}
			\caption{Lithium concentration asymptotics versus numerics}
			\label{fig:asym_v_numerics_a}
		\end{subfigure}
		\begin{subfigure}{0.47\linewidth}
			\includegraphics[width=\linewidth]{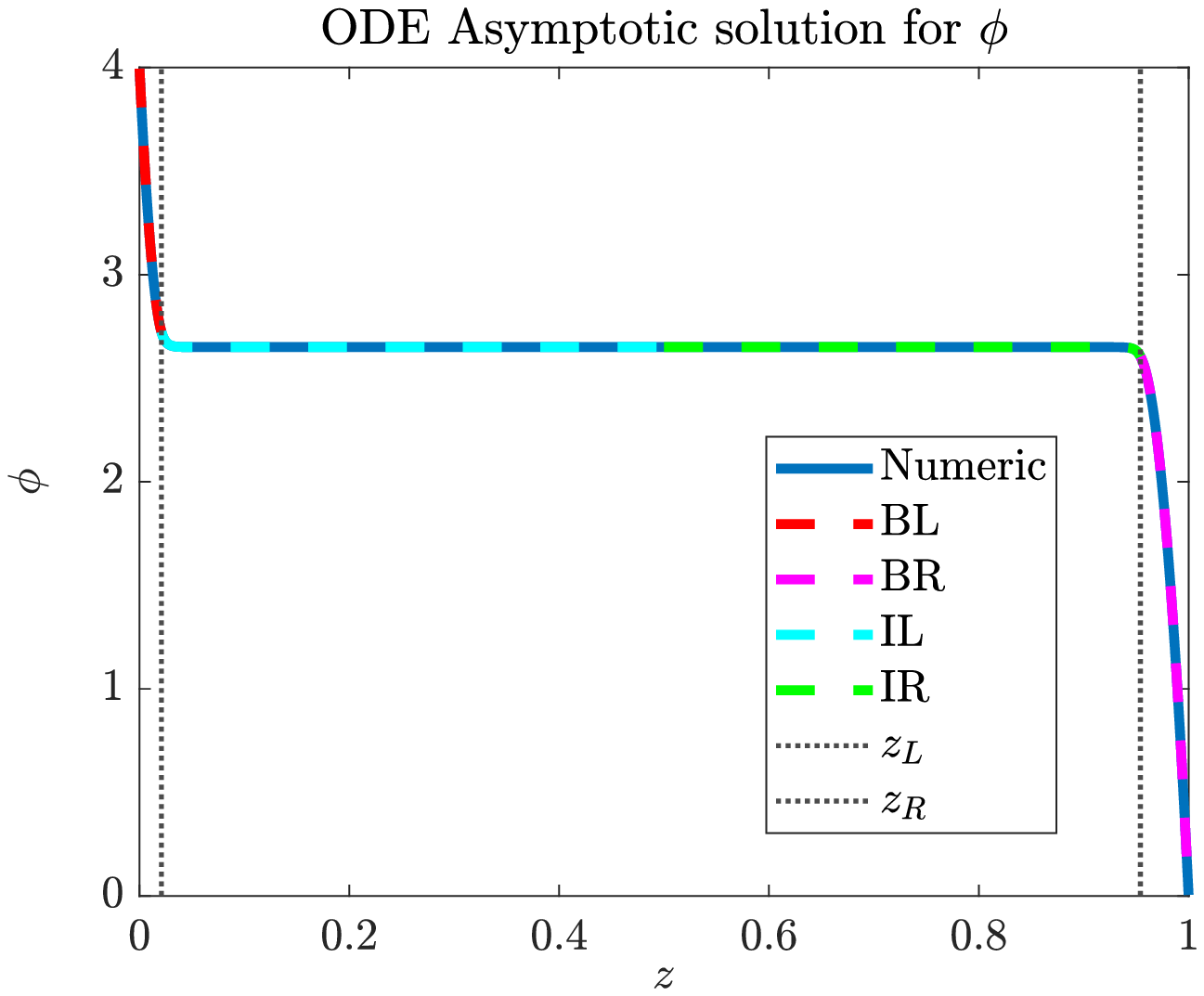}
			\caption{Electric potential asymptotics versus numerics}
			\label{fig:asym_v_numerics_b}
		\end{subfigure}
		\caption{Asymptotic solution from section \ref{sec:asymptotics} versus numerical solution from section \ref{sec:num_equi} for both the lithium concentration and electric potential. Parameter values are as follows: $z_c=1, \quad z_a=-1, \quad m_c=0.3, \quad m_a=0.7, \quad n_a=0.4,\quad \nu=0.6$, with $\frac{\beta_1}{\beta_2}=2 < \beta^*, \quad \frac{\beta_2}{\beta_1}=\frac{1}{2} < \beta^*$. We also have $\lambda=1.5 \times 10^{-3}$, $\delta^{-1}=170$, and $\Delta V = 4$. BL, BR refer to the left and right boundary layer solutions respectfully, similarly IL, IR refer to the left and right intermediate layer solutions respectfully. $z_L, z_R$ refer to the leading order transition points from the boundary to the intermediate layers.}
		\label{fig:asym_v_numerics}
	\end{figure}

	We show the asymptotic solution for both the lithium concentration and the electric potential in comparison with the results obtained in section \ref{sec:num_equi}. For each of these plots we show the solution for $n_c$, obtained from the solution for $\theta$ via the auxiliary relation \eqref{eq:auxiliary}. We note that for these parameter values 
	\[z_c=1, \quad z_a=-1, \quad m_c=0.3, \quad m_a=0.7, \quad n_a=0.4,\quad \nu=0.6,\]
	with $ \beta_1 = - z_a n_a, \quad \beta_2=z_a n_a + z_c \nu$, we have 
	\[ \frac{\beta_1}{\beta_2}=2 < \beta^*, \quad \frac{\beta_2}{\beta_1}=\frac{1}{2} < \beta^*,\]
	so neither $F_1$ nor $F_2$ has a root and thus we minimise both $|F_i|$.
	Plotting the asymptotic solutions versus the numerics for this set of parameters in Figure \ref{fig:asym_v_numerics} we see excellent agreement between our asymptotic regimes and the numerical solutions. The vertical lines in both plots of Figure \ref{fig:asym_v_numerics} represent the leading order point where the solution transitions between the boundary and intermediate layers.  In both cases we can see how the intermediate layer solutions merge seamlessly into the bulk.

\begin{figure}[htb]
		%
		\begin{subfigure}{0.47\linewidth}
			\includegraphics[width=\linewidth]{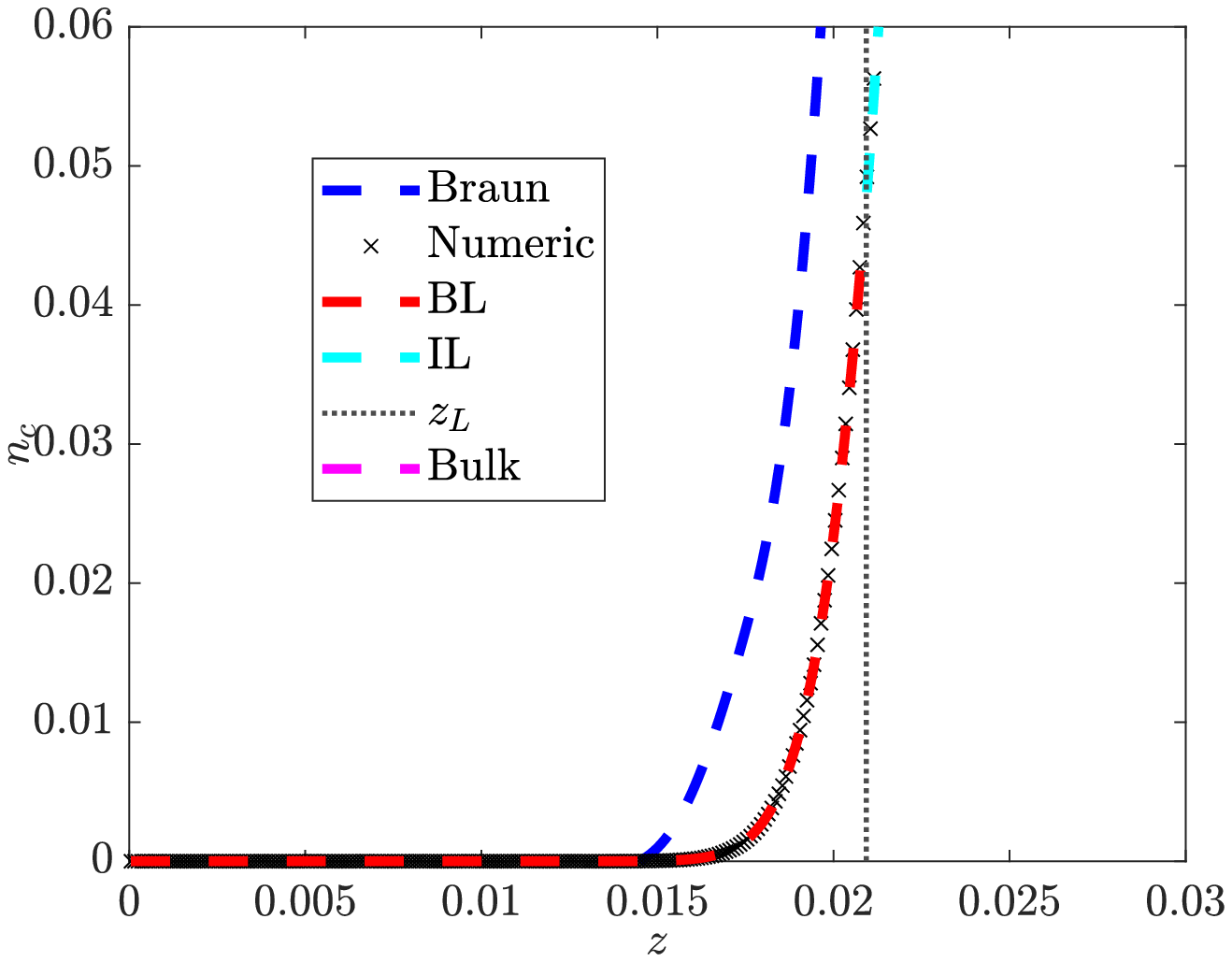}
			\caption{Lithium concentration in Boundary layer}
			\label{fig:asym_zeroF_multiPlot_a}
		\end{subfigure}
		\begin{subfigure}{0.47\linewidth}
			\includegraphics[width=\linewidth]{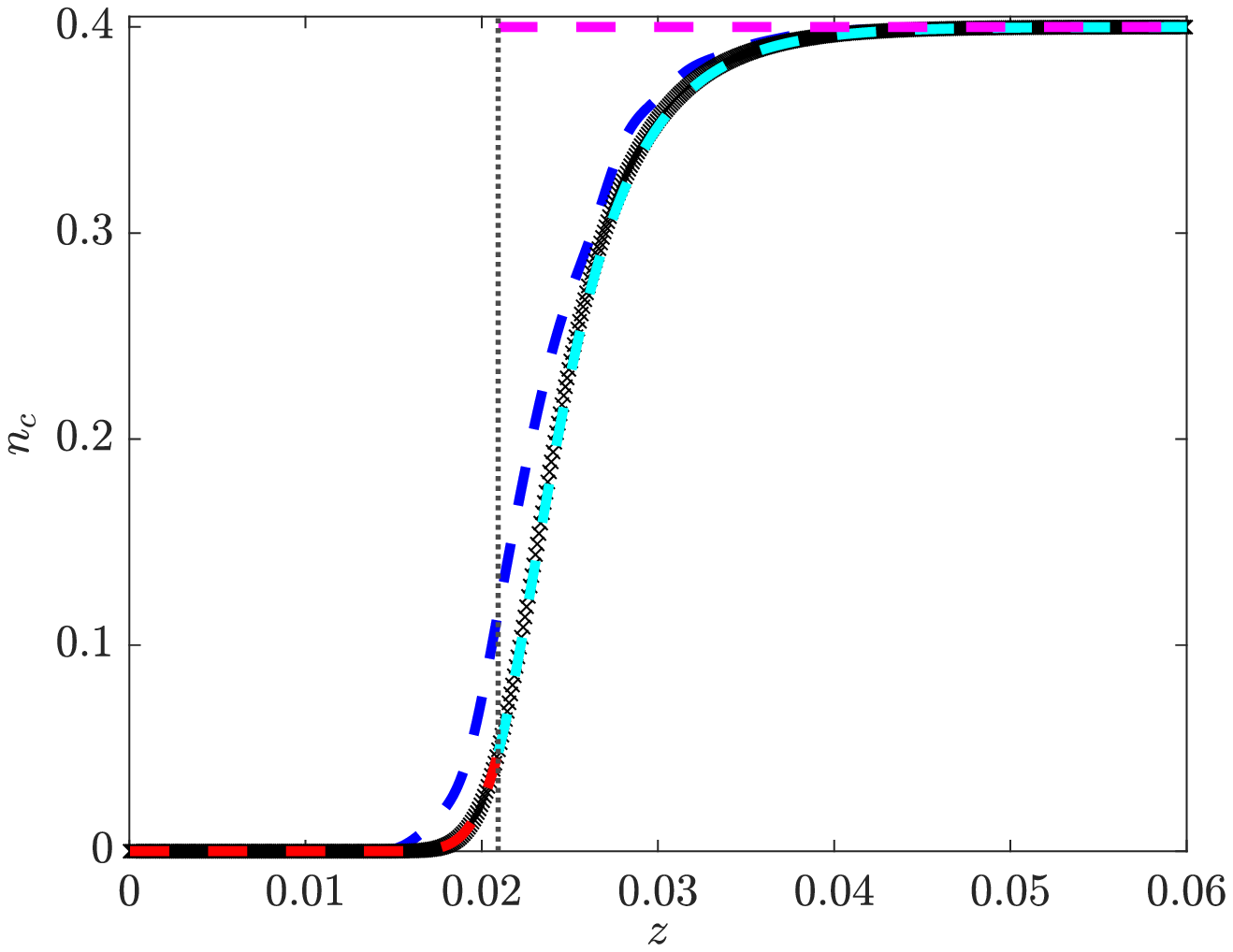}
			\caption{Lithium concentration in Intermediate layer}
			\label{fig:asym_zeroF_multiPlot_b}
		\end{subfigure}
		\centering
		\begin{subfigure}{0.47\linewidth}
			\includegraphics[width=\linewidth]{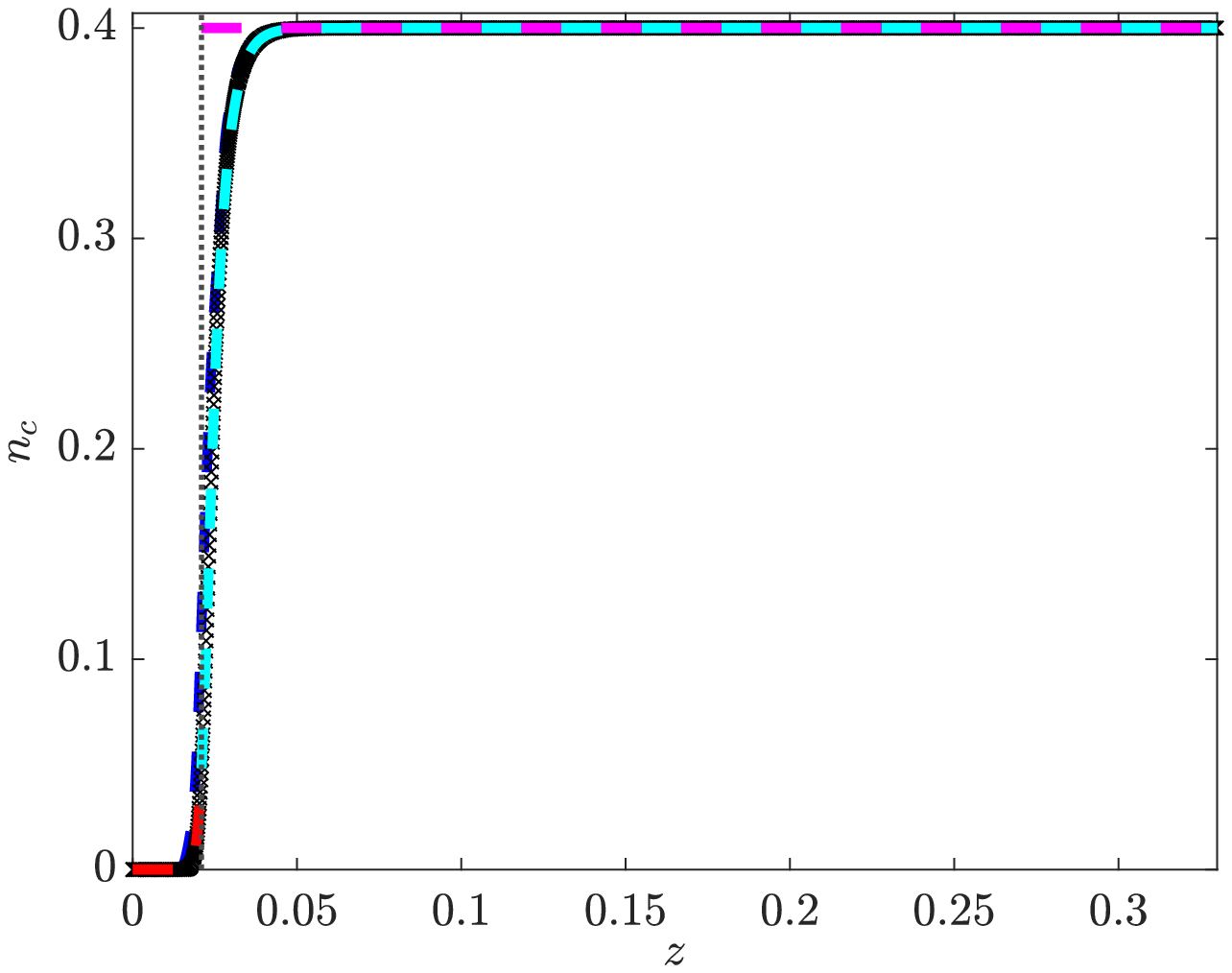}
			\caption{Lithium concentration in bulk}
			\label{fig:asym_zeroF_multiPlot_c}
		\end{subfigure}
		\caption{Considering figure \ref{fig:asym_v_numerics_a}, we zoom in on each of the regions near $z=0$ of the asymptotic versus numerical solution for equilibrium profile for $n_c$ to show the agreement of the asymptotics with the numerics.}
		\label{fig:asym_zeroF_multiPlot}
	\end{figure}

	We highlight the agreement of the various asymptotic regimes with the numerical solution for the left hand side in Figure \ref{fig:asym_zeroF_multiPlot}. We first plot the solution obtained by \cite{braun2015thermodynamically} in dashed blue lines for comparative purposes and we plot the numerical solution of solving the full model (\eqref{eq:1d_equi_j0} with \eqref{eq:int_constraint} subject to $\phi(0)=1, \phi(1)=0$) in black markers. Figure \ref{fig:asym_zeroF_multiPlot_a} shows our boundary layer solution \eqref{eq:psi_sol_lhs} in red and the vertical grey line shows our transition point $z_L$ given by \eqref{eq:zl_sol}. We see fantastic agreement between the asymptotics and numerics in this region. As before, in section \ref{sec:num_equi}, we observe the discrepancy of the semi-analytic solution in the boundary layer. Figure \ref{fig:asym_zeroF_multiPlot_b} shows our intermediate layer solution given by $\theta_0$ plus the intermediate correction term \eqref{eq:xi_sol_leftIL} in teal. We see that the solution agrees very well with the numerical solution, with the agreement worsening as we move further from the bulk as is expected. Again, we note that the asymptotics outperforms the semi-analytic solution of \cite{braun2015thermodynamically} in this region. We show the bulk solution with $\theta$ given by \eqref{eq:theta_0_bulk} in pink in Figure \ref{fig:asym_zeroF_multiPlot_c}. Plots near $z=1$ are similar.
	We provide some examples of the results for scenarios where either $F_1$ or $F_2$ have a root and the other is minimised in the supplemental \cref{SM-sec:betas}.

\begin{figure}[htb]
		%
		\includegraphics[width=\linewidth]{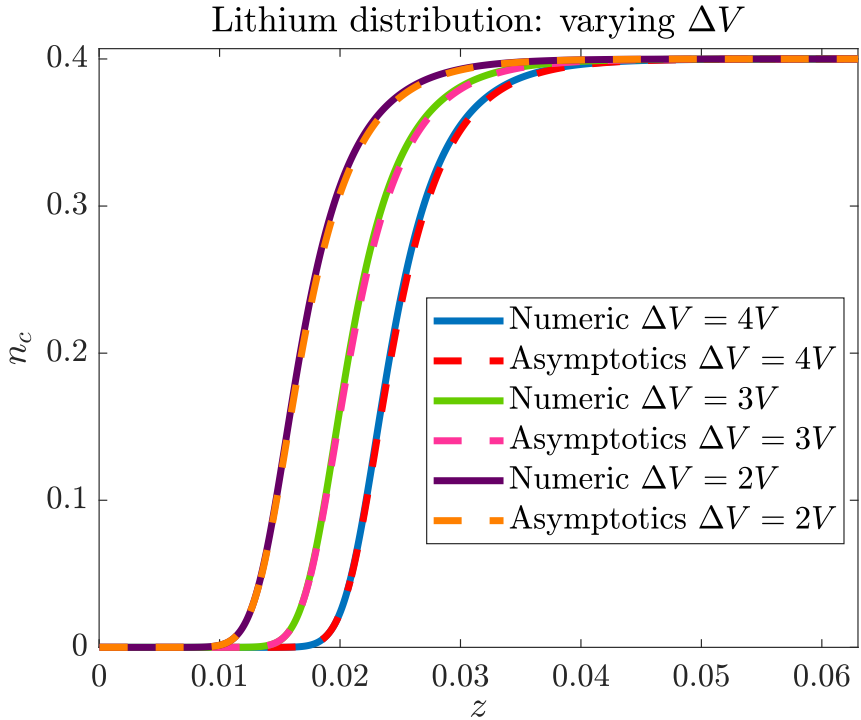}
		\caption{Solution profile of the number density of the lithium ions for both the asymptotics and the numerics shown for different applied voltages. We have fixed $z_c=1, \quad z_a=-1, \quad m_c=0.3, \quad m_a=0.7, \quad n_a=0.4,\quad \nu=0.6$, $\lambda=1.5 \times 10^{-3}$ as before. We provide numerical/asymptotic solutions for $\Delta V=4\rm{V}, 3\rm{V}, 2\rm{V}$ shown in solid blue/dotted red, solid green/dotted pink, and solid purple/dotted orange respectively.}
		\label{fig:equiVolt}
	\end{figure}

	Braun \etal \cite{braun2015thermodynamically} also present some additional results pertaining to the lithium concentration in various scenarios in order to investigate the SCL. The first is concerned with varying the applied potential difference $\Delta V$. The authors show separately the concentration near the positive and negative electrodes. For comparison purposes we simply include the comparisons at the negative electrode side, noting that similar results are obtained on the positive electrode side. This is shown in figure \ref{fig:equiVolt}, where we show numerical and asymptotic results for three different values of $\Delta V$. As noted by \cite{braun2015thermodynamically}, increasing $\Delta V$ increases the width of the boundary layers, but does not impact the region of transition between the boundary layer and the bulk. That is, a larger $\Delta V$ results in wider regions which are either depleted or saturated with cations. This agrees with our description of the strong SCL having a width $\epsilon = \lambda \sqrt{\delta^{-1}}$, that is from our scaling we know that $\Delta V$ affects $\delta$ and that the IL thickness is $\lambda$, thus explaining why $\Delta V$ does not impact this width. We further note that this also resolves the observation of \cite{li2019dendrite} regarding dilation of the SCL for an increase in voltage bias.  
	
\begin{figure}[htb]
		%
		\includegraphics[width=\linewidth]{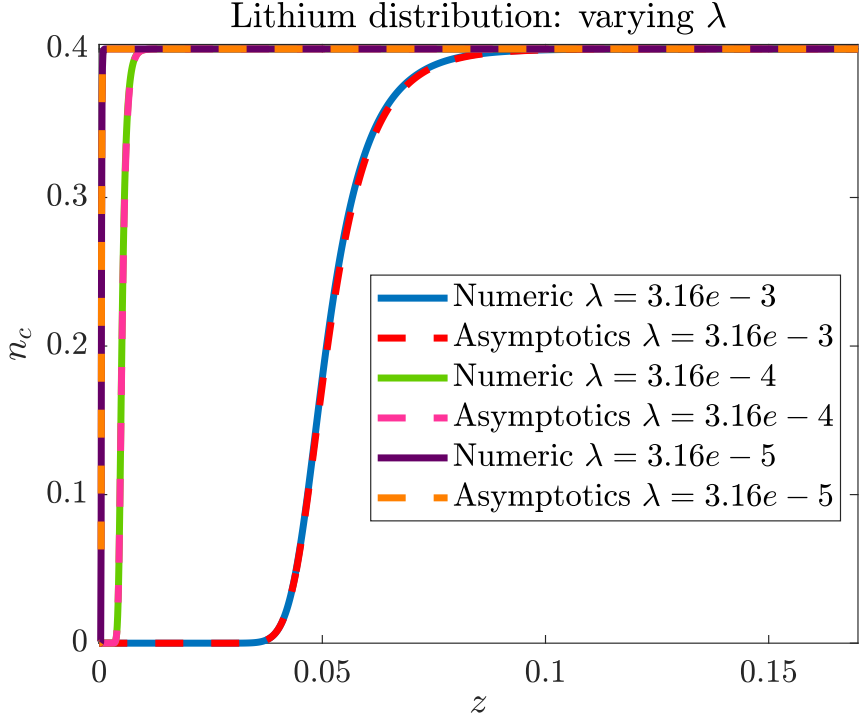}
		\caption{Solution profile of the number density of the lithium ions for both the asymptotics and the numerics shown for different $\lambda$. We have fixed $z_c=1, \quad z_a=-1, \quad m_c=0.3, \quad m_a=0.7, \quad n_a=0.4,\quad \nu=0.6$, $\Delta V=4V$ as before. We provide numerical/asymptotic solutions for $\lambda=3.16 \times 10^{-3}, 3.16 \times 10^{-4}, 3.16 \times 10^{-5}$ shown in solid blue/dotted red, solid green/dotted pink, and solid purple/dotted orange respectively.}
		\label{fig:equiLambda}
	\end{figure}

	The authors also investigate the concentration profile for various values of the parameter $\lambda$. We provide a similar analysis in figure \ref{fig:equiLambda}, comparing the numerical and asymptotic solutions for three variations of the parameter $\lambda$.
	In this case increasing $\lambda$ will increase the width of both the boundary layer and the region transitioning between the boundary and bulk as observed by \cite{braun2015thermodynamically}. Again, this observation is in agreement with our description of both the strong and weak SCL having dependence on the parameter $\lambda$.

	We reiterate that the observed space charge layers widths can be determined and explained by our asymptotic approach and our representation of strong and weak Debye layers, with $\epsilon$ representing the width of the strong SCL, and $\lambda$ representing the width of the weak SCL.

	\section{Discussion and Conclusions} \label{sec:discussion}
	Mathematical modelling is a valuable tool in gaining understanding into the behaviour of a system.  
	We have noted the increased interest in the use of solid electrolytes, in addition to the need for a deeper understanding of double charge layer dynamics.
	The limited literature, and in particular, the scarcity of mathematical models studying both solid electrolytes and electric double layers motivated us to carry out this asymptotic analysis of space charge layers in solid electrolyte. 
	
	Overall, we have presented a non-dimensional model for a SE derived from non-equilibrium thermodynamics. 
	In our non-dimensionalisation of the model we uncover the true length scale of the boundary layer in comparison with previous literature.
	We used asymptotics to reduce the model, revealing three important regions in the SE - the bulk, the boundary layer of width $\epsilon$, and the intermediate layers of width $\lambda$. The boundary and the intermediate layers together form the SCL of the SE.
	
	By exploring this reduced model for SCL in SE we have determined the existence of two distinct regions in the double layer - strong and weak double layers. We have observed, based on our asymptotic solutions, that the strong SCL exhibits quadratic behaviour while the weak SCL exhibits exponential behaviour, which is in agreement with the findings of Swift \etal \cite{swift2021modeling}. We have explicitly determined a length scale to characterise both of these regimes within the SCL, thus addressing the observations of other authors regarding the length of the SCL compared to EDLs in liquid electrolytes (\cite{knauth2009inorganic,yamada2013lithium,braun2015thermodynamically,li2019dendrite,li2021transport}).
	
	In addition, by introducing an auxiliary variable into the model we were able to address many of the numerical issues faced by other authors (\cite{braun2015thermodynamically, katzenmeier2022modeling, katzenmeier2022nature, swift2021modeling}). The use of the auxiliary variable enabled us to transform the problem to a smooth domain whereby we could avoid numerical difficulties caused by the proximity to singularities in the true domain.

	We have presented results for a zero flux, one dimensional problem. While these results can give insights into the behaviours occurring in full battery cells, extending the model to consider non zero flux conditions and a two dimensional version enables us to better model real problems with prescribed flux conditions and to connect to a fuller battery model with Butler Volmer type conditions. Future work will aim to extend the analysis presented here to those scenarios. 
	
	A potential drawback of this model is the lack of consideration of coulombic interactions between the vacancies, De Klerk and Wagemaker \cite{de2018space} find that these interactions can play a significant role in the impact of SCL and the effects in solid state batteries.

	The behaviour we observe in the boundary and intermediate regions can have implications in further modelling and analysis of lithium-ion batteries. For example, for the Butler Volmer boundary condition many models use a potential difference based on the bulk electric potential or use a jump across the electrolyte to determine the change in potential. If just the bulk is used then the differences between EDLs and SCLs cannot be realized because those effects are ignored. With our model we can pinpoint precisely what the potential difference should look like. While these differences may be negligible, they are still worth further investigation. This also forms part of our future work.

	In conclusion, our numerical framework more robustly computes charge and electric potential in SCLs and our asymptotic analysis has elucidated the double-layer structure.
	
	\section*{Acknowledgments}
	L.M.K. acknowledges the financial support of an NSERC Vanier Canada Graduate scholarship Grant No. 434051. I.R.M. acknowledges the Natural Sciences and Engineering Research Council of Canada Discovery Grant 2019-06337.
	
	\bibliographystyle{siamplain}
	\bibliography{References}
	\makeatletter\@input{yy.tex}\makeatother

\end{document}


\maketitle
	
	\section{Solid Electrolyte: model derivation}\label{sec:model_deriv}
	We present a derivation of the model first derived by Braun \etal \cite{braun2015thermodynamically} for a solid electrolyte with cations $c$, stationary anions $a$, and massless vacancies $v$.
	The conservation equation for each species is given by the following equations
	\begin{subequations}
		\begin{align}
		\pderiv{\rho_{c}}{t} + \nabla \cdot (\rho_{c} \mathbf{v}_{c}) &=0,  \label{eq:cons_mass_c}  \\
		\vspace{3mm}
		\pderiv{\rho_{a}}{t} + \nabla \cdot (\rho_{a} \mathbf{v}_{a}) &=0, \label{eq:cons_mass_a} \\
		\vspace{3mm}
		\pderiv{n_{v}}{t} + \nabla \cdot (n_{v} \mathbf{v}_{v}) &=0, \label{eq:cons_mass_v}
		\end{align}
	\end{subequations}
	where $\rho$ is mass density, $m$ is mass, $n$ is number density, $\mathbf{v}$ is velocity, and $\rho_{\alpha}=m_{\alpha} n_{\alpha}$. We conserve mass for both the cations and anions, but as the vacancies are massless, we instead conserve the number density of the vacancies.  Due to the stationarity of the anion, $\mathbf{v}_a=0$, we can reduce \cref{eq:cons_mass_a} to
	\begin{equation}\label{eq:cons_mass_anion}
	\pderiv{\rho_{a}}{t}=0.
	\end{equation}
	We define the total density by $\rho = \sum_k \rho_k$ and mass averaged velocity by
	$ \mathbf{v} = \frac{1}{\rho} \sum_k \rho_k \mathbf{v}_k$, for $k=v,a,c$.
	The diffusive fluxes are given by $J_c=\rho_c(\mathbf{v}_c-\mathbf{v}), \quad J_a=\rho_a(\mathbf{v}_a-\mathbf{v})$, and because of massless vacancies, $J_v=0$. Therefore, because of total conservation of mass, $\sum_k J_k =0$, which then implies that $J_a=-J_c$. We can also express $J_v$ in terms of $J_c$. If we consider writing the flux equation for each species in terms of number densities, 
	\begin{subequations}
		\label{eq:mass_Eqs_density}		\begin{align}
		&\pderiv{n_{c}}{t} + \nabla \cdot (n_{c} \mathbf{v}_c) = 0  \\
		&\pderiv{n_{a}}{t}  = 0, \text{ since } \mathbf{v}_a=0\\
		&\pderiv{n_{v}}{t} + \nabla \cdot (n_{v} \mathbf{v}_v) = 0. 
		\end{align}
	\end{subequations}
	Adding all three equations 
	\begin{equation}
	\pderiv{n}{t} + \nabla \cdot \left( n_c \mathbf{v}_c + n_v \mathbf{v}_v \right) =0,
	\end{equation}
	where $n=n_c+n_a+n_v$. We must have a global conservation of mass,  therefore $n$ is constant, so we must have 
	\begin{equation}
	\nabla \cdot \left( n_c \mathbf{v}_c + n_v \mathbf{v}_v \right) =0,
	\end{equation}
	thus $n_c \mathbf{v}_c + n_v \mathbf{v}_v= \nabla \times B$ for some $B$ which we will set to zero.
	Writing the cation and vacancy flux equations from $J_k=\rho_k(\mathbf{v}_k-\mathbf{v})$ in terms of the number density 
	\begin{equation}
	\frac{J_v}{m_v}+n_v \mathbf{v}=n_v \mathbf{v}_v, \qquad \frac{J_c}{m_c}+ n_c \mathbf{v} = n_c \mathbf{v}_c,
	\end{equation} 
	adding these together and using $n_c \mathbf{v}_c + n_v \mathbf{v}_v=0$ we will have the expression
	\begin{equation}\label{eq:flux_vandc}
	\frac{J_v}{m_v} +\frac{J_c}{m_c}+(n_v+n_c)\mathbf{v}=0.
	\end{equation}
	Similarly, the anion flux $J_a=\rho_a(\mathbf{v}_a-\mathbf{v})$, can be rewritten as
	\begin{equation}\label{eq:flux_ja}
	\frac{J_a}{m_a} = -n_a \mathbf{v} 
	\end{equation}
	since $\mathbf{v}_a=0$. 
	Rewriting $\mathbf{v}$ in \cref{eq:flux_vandc} using \cref{eq:flux_ja}, and rearranging with $J_a=-J_c$, we find that we can also express $J_v$ in terms of $J_c$ as $\frac{J_v}{m_v}=-\kappa \frac{J_c}{m_c}$, where $\kappa$ is defined as \begin{equation}\label{eq:kappa_Def}
	\kappa= 1+  \left(\frac{n_c + n_v}{n_a}\right)  \left(\frac{m_c}{m_a}\right).
	\end{equation}
	Conservation of energy from non-equilibrium thermodynamics leads to the entropy production equation
	\begin{equation}\label{eq:entropy}
	\xi = -\frac{1}{T} \sum_{k} (\nabla \mu_{k} + z_{k} \nabla \phi ) \cdot \frac{J_{k}}{m_{k}},
	\end{equation}
	where $T$ is temperature, $z_k$ is species charge, $\mu$ is chemical potential, and $\phi$ is the electric potential. This can be derived by considering conservation of mass, momentum, and energy (kinetic, potential and thermodynamic), for more detail see chapter 2 of \cite{taylor1993multicomponent}.
	We assume that the fluxes depend linearly on the thermodynamic driving forces,  $\nabla \left(\mu_k+z_k \phi \right)$ (see \cite{onsager1931reciprocal,prigogine1963introduction,de2013non,rajagopal2004thermomechanical}), which allows us to write 
	\begin{gather}
	\begin{bmatrix}
	\frac{J_c}{m_c} \\  \frac{J_a}{m_a} \\ \frac{J_v}{m_v} 
	\end{bmatrix}
	=
	\begin{bmatrix}
	\alpha_1 & \beta_1 & \gamma_1 \\
	\alpha_2 & \beta_2 & \gamma_2 \\
	\alpha_3 & \beta_3 & \gamma_3 
	\end{bmatrix}
	\begin{bmatrix}
	\nabla \mu_c + z_c  \nabla \phi \\
	\nabla \mu_a + z_a  \nabla \phi  \\
	\nabla \mu_v  
	\end{bmatrix}
	=
	-\begin{bmatrix}
	-\alpha_1 & -\beta_1 & -\gamma_1 \\
	\frac{m_c}{m_a}\alpha_1 & 	\frac{m_c}{m_a}\beta_1 & 	\frac{m_c}{m_a}\gamma_1 \\
	\kappa \alpha_1 & \kappa \beta_1 & \kappa \gamma_1 
	\end{bmatrix}
	\begin{bmatrix}
	\nabla \mu_c + z_c  \nabla \phi \\
	\nabla \mu_a + z_a  \nabla \phi  \\
	\nabla \mu_v  
	\end{bmatrix}
	\end{gather}
	where we have used $J_a=-J_c$ and $\frac{J_v}{m_v}=- \kappa \frac{J_c}{m_c}$.
	We determine the coefficients $\alpha_1, \beta_1, \text{ and } \gamma_1$ from the following; physically, we need entropy production to be positive so the matrix must be positive semi definite, \cite{onsager1931reciprocal} tells us that the matrix should be symmetric, and the Curie-Prigogine principle (also referred to as \textit{Curie's law}) says that thermodynamic forces of different tensorial components are uncoupled \cite{de2013non}. These conditions tell us that $\alpha_1=-M<0$, $\beta_1=-\frac{m_c}{m_a} \alpha_1$, and $\gamma_1=-\kappa \alpha_1$, where $M$ is a free positive constant, leading to the reduced expression for the flux equation
	\begin{equation}\label{eq:reduced_flux}
	\frac{J_c}{m_c}= -M \left( \nabla \mu_c -\frac{m_c}{m_a} \nabla \mu_a -\kappa \nabla \mu_v + \left(z_c-\frac{m_c}{m_a}z_a \right) \nabla \phi \right).
	\end{equation}
	
	For the chemical potentials we note that only $n_c$ and $n_v$ mix, so we will have
	\begin{align}\label{eq:chem_potentials}
	\mu_c &= \mu_{c0}(T,p) + k_BT \log{\left(\frac{n_c}{n_c+n_v} \gamma_c \right)}\\
	\mu_v &= \mu_{v0}(T,p)  + k_BT \log{\left(\frac{n_v}{n_c+n_v} \gamma_v\right)}\\
	\mu_a &= \mu_{a0}(T,p). 
	\end{align}
	The authors in \cite{braun2015thermodynamically} use Margules activity coefficients \cite{margules1895zusammensetzung,gokcen1996gibbs} to model the non-ideality. This means that the excess free energy is expressed as a power series of the mole fractions which yields
	\begin{subequations}
		\label{eq:margules}
		\begin{align}
		\ln{\gamma_c} &= \left(A_{12} + 2(A_{21}-A_{12})x_c \right) x_v^2 = A x_v^2, \\
		\ln{\gamma_v} &= \left(A_{21} + 2(A_{12}-A_{21})x_v \right) x_c^2 = A x_c^2, 
		\end{align}
	\end{subequations}
	where $x_i = \frac{n_i}{n_c+n_v}$ and we have taken $A_{12}=A_{21}=A$ .
	For an equation of state we have 
	\begin{equation}\label{eq:ofState}
	p=p_r + K \left(\frac{n}{n_r}-1 \right),
	\end{equation}
	where $p$ is pressure, $K$ is the bulk modulus based on the assumption that the electrolyte is an elastic medium, and the subscript $r$ indicates a reference variable. We take the Helmholtz free energy, given by 
	\begin{equation}\label{Eq:helmholtz}
	\psi =\psi(T,V)=U-ST = -pV + \sum_k \mu_k n_k V,
	\end{equation}
	where $U$ is the internal energy, $V$ is volume, and $S$ is entropy.  If we define the chemical potentials as the partial derivative of the free energy per mass $\mu_k = \pderiv{\rho \bar{\psi}}{n_k}$, where $\bar{\psi}=\frac{\psi}{m}$,
	The chemical potentials can then be written as
	\begin{subequations}
		\begin{align}
		\mu_c &= \mu_c^* + \frac{K}{n_r} \log{\left(\frac{p-p_r}{K}+1 \right)} + k_BT \log{\left(\frac{n_c}{n_c+n_v}\right)} + \frac{A n_v^2}{\left(n_c+n_v\right)^2} \notag\\
		\mu_v &= \mu_v^* + \frac{K}{n_r} \log{\left(\frac{p-p_r}{K}+1 \right)} + k_BT \log{\left(\frac{n_v}{n_c+n_v}\right)} + \frac{ A n_c^2}{\left(n_c+n_v\right)^2}\\
		\mu_a &= \mu_a^* +  \frac{K}{n_r} \log{\left(\frac{p-p_r}{K}+1 \right)} \notag
		\end{align}
	\end{subequations}
	We take the incompressibility limit $K \to \infty \text{ in } p  =p_r+K \left(\frac{n}{n_r}-1\right)$, which means we must have $n=n_r$ to prevent infinite pressure. Taking this limit in the chemical potentials we obtain
	\begin{subequations}
		\begin{align}
		\mu_c &= \mu_c^* + \frac{K}{n_r} \left(\frac{p-p_r}{K}\right) + k_BT \log{\left(\frac{n_c}{n_c+n_v}\right)}+ \frac{A n_v^2}{\left(n_c+n_v\right)^2}, \notag\\
		\mu_v &= \mu_v^* + \frac{K}{n_r} \left(\frac{p-p_r}{K} \right) + k_BT \log{\left(\frac{n_v}{n_c+n_v}\right)} +  \frac{A n_c^2}{\left(n_c+n_v\right)^2},\\
		\mu_a &= \mu_a^* +  \frac{K}{n_r} \left(\frac{p-p_r}{K} \right), \notag
		\end{align}
	\end{subequations}
	with $\mu^*$ denoting the reference chemical potential.
	From here we set $A=0$, assuming ideality.
	
	We will also have conservation of charge from Gauss' law
	\begin{equation}\label{eq:cons_of_charge}
	\nabla \cdot (\epsilon_0 (1 + \chi ) E ) = n^F,
	\end{equation}
	where $n^F= z_c n_c + z_a n_a$ is our free charge, $\epsilon_0$ is the permittivity of free space, $\chi$ is the electric susceptibility, and $E$ is the electric field with $E=-\nabla \phi$.
	Conservation of momentum takes the general form
	\begin{equation}\label{Eq:cons_momentum}
	-n^F \nabla \phi =  \pderiv{(\rho \mathbf{v})}{t}+ \nabla \cdot\left( \rho \mathbf{v} \otimes \mathbf{v} \right)  + \nabla p.
	\end{equation}
	Finally, we can rewrite the conservation of mass equation using the relationship between the diffusive fluxes and noting that $\rho_c \mathbf{v}_c = \rho \mathbf{v} = \frac{\rho}{(\rho-\rho_c)}J_c $ resulting in
	\begin{equation}\label{eq:cons_mass_final}
	\pderiv{n_c}{t} = -\frac{1}{m_c} \nabla \cdot \left( \frac{m_c n_c + m_a n_a}{m_a n_a} J_c \right).
	\end{equation}
	Therefore, the model can be defined by the following set of equations, as first derived by \cite{braun2015thermodynamically}
	\begin{subequations}
		\label{eq:braun_model}
		\begin{align}
		\pderiv{n_c}{t} &= -\frac{1}{m_c} \nabla \cdot \left( \left( 1+ \frac{m_c n_c }{m_a n_a} \right) J_c \right), \label{eq:braun_mass} \\
		n^F &= - \epsilon_0 \left( 1+\chi \right)\nabla^2 \phi, \label{eq:braun_charge}\\
		-n^F \nabla \phi &=  \pderiv{(\rho \mathbf{v})}{t}+ \nabla \cdot\left( \rho \mathbf{v} \otimes \mathbf{v} \right)  + \nabla p, \label{eq:braun_momentum}\\
		\frac{J_c}{m_c}&= -M \left( \nabla \mu_c -\frac{m_c}{m_a} \nabla \mu_a - \left(1+  \frac{n_c + n_v}{n_a} \frac{m_c}{m_a}\right) \nabla \mu_v + \left( z_c-\frac{m_c}{m_a} z_a\right) \nabla \phi \right). \label{eq:braun_flux}
		\end{align}
	\end{subequations}
	with chemical potentials
	\begin{subequations}
		\label{eq:braun_incompress_chem_pot}
		\begin{align}
		\mu_c &= \mu_c^* + \frac{1}{n_r} \left(p-p_r\right) + k_BT \log{\left(\frac{n_c}{n_c+n_v}\right)},\\
		\mu_v &= \mu_v^* + \frac{1}{n_r} \left(p-p_r\right) + k_BT \log{\left(\frac{n_v}{n_c+n_v}\right)},\\
		\mu_a &= \mu_a^* +  \frac{1}{n_r} \left(p-p_r\right),
		\end{align}
	\end{subequations}
	corresponding to \cref{main-Eq:Braun_eqn} and \cref{main-eq:incompress_chem_pot} in the main text.

	\section{Additional examples of ODE asymptotic results}
	\label{sec:betas}
	
\begin{figure}
		\centering
		\includegraphics[width=\linewidth]{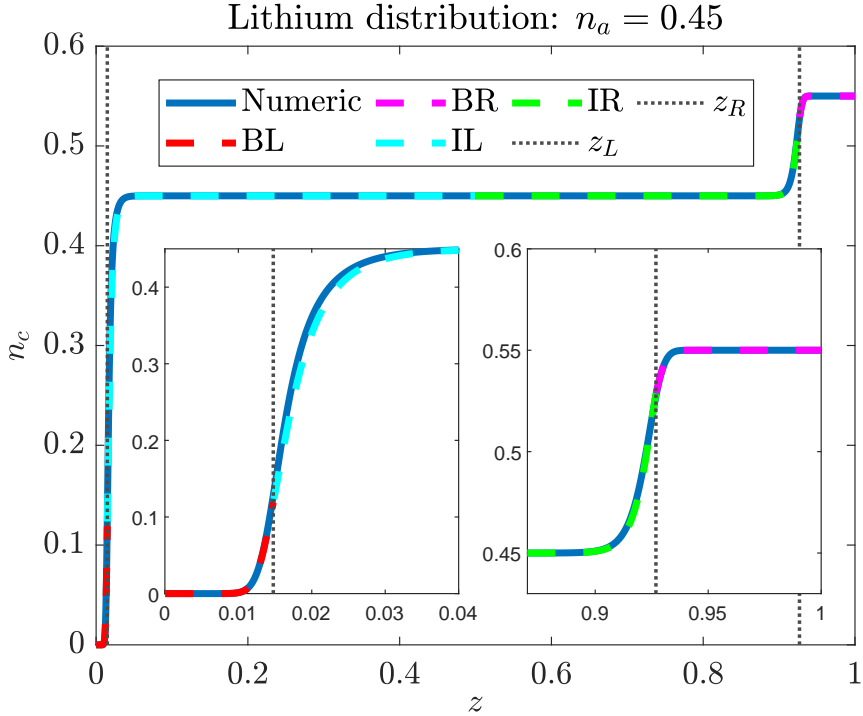}
		\caption{Asymptotic versus numerical solution for $n_c$ in the scenario where $F_1>0, F_2<0$: $z_a=-1, n_a=0.45, \beta_1=0.45, \beta_2=0.1$. We also have $\lambda=1.5 \times 10^{-3}$, $\delta^{-1}=170$, and $\Delta V = 4$. We include insets to show a close up of the agreement at the two boundaries.  BL, BR refer to the left and right boundary layer solutions respectfully, similarly IL, IR refer to the left and right intermediate layer solutions respectfully. $z_L, z_R$ refer to the leading order transition points from the boundary to the intermediate layers. }
		\label{fig:case2beta}
	\end{figure}

\begin{figure}
		\centering
		\includegraphics[width=\linewidth]{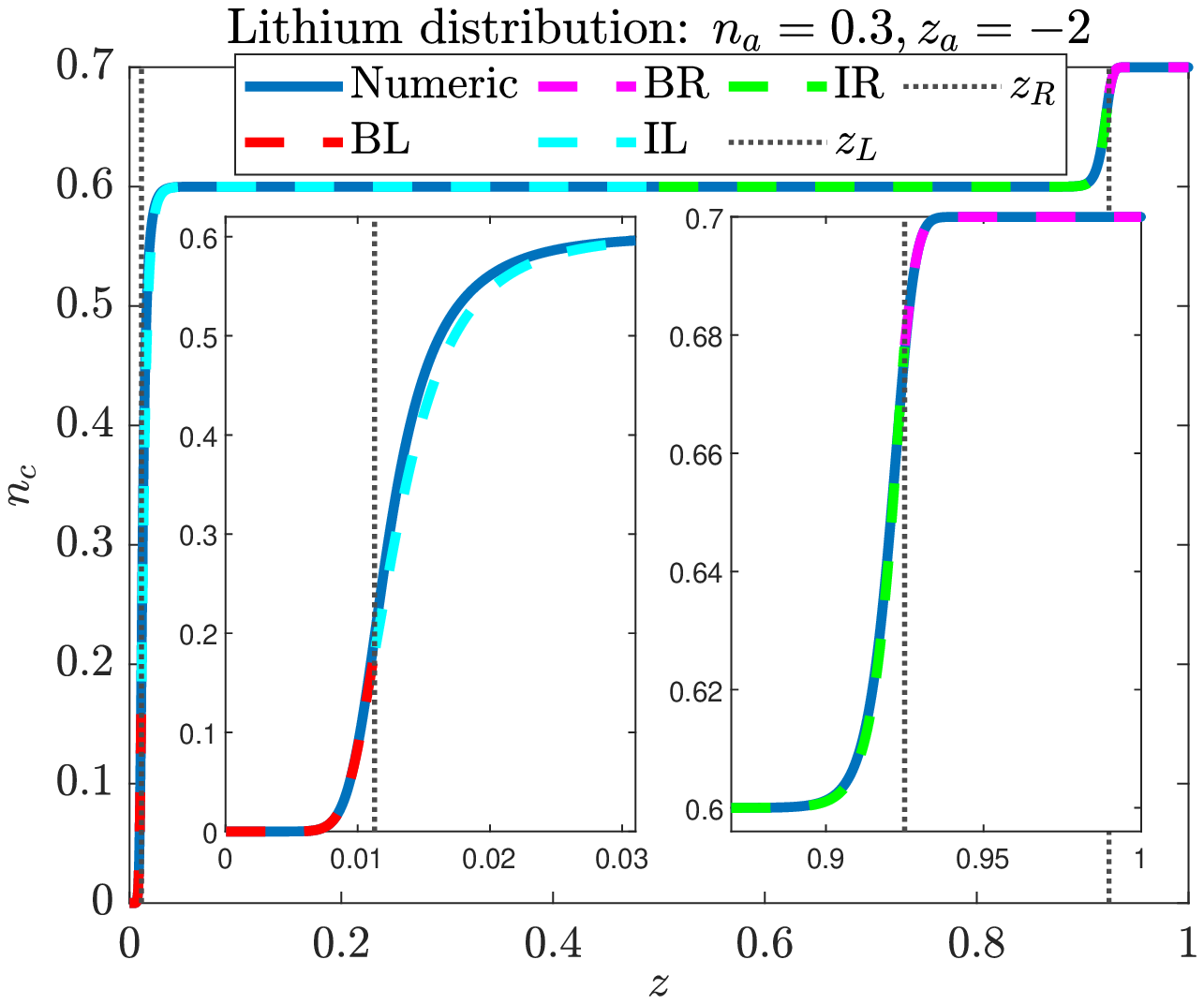}
		\caption{Asymptotic versus numerical solution for $n_c$ in the scenario where $F_1>0, F_2<0$: $z_a=-2, n_a=0.3, \beta_1=0.6, \beta_2=0.1$. We also have $\lambda=1.5 \times 10^{-3}$, $\delta^{-1}=170$, and $\Delta V = 4$. We include insets to show a close up of the agreement at the two boundaries.  BL, BR refer to the left and right boundary layer solutions respectfully, similarly IL, IR refer to the left and right intermediate layer solutions respectfully. $z_L, z_R$ refer to the leading order transition points from the boundary to the intermediate layers.  }
		\label{fig:case2betaiain}
	\end{figure}

\begin{figure}
		\centering
		\includegraphics[width=\linewidth]{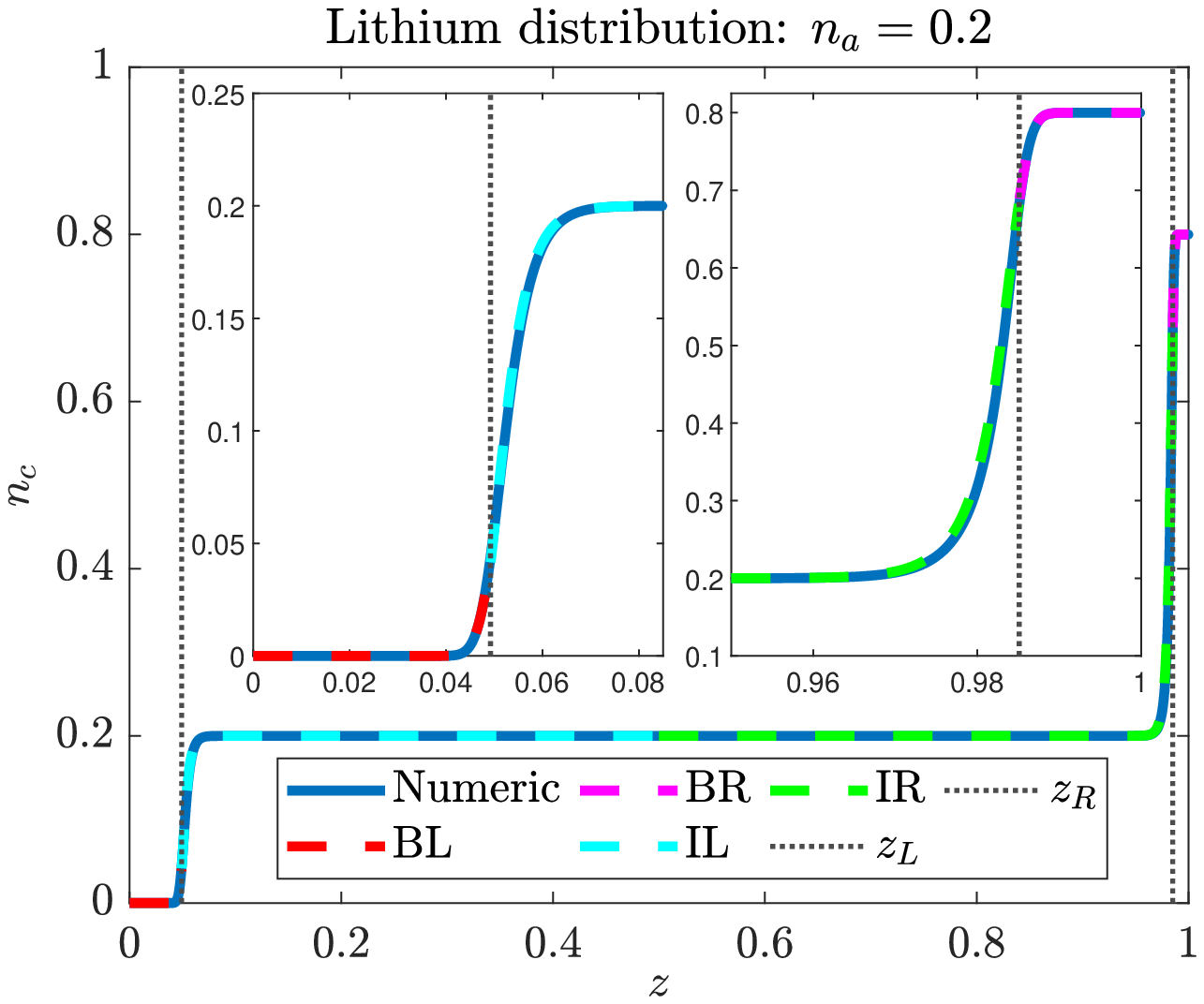}
		\caption{Asymptotic versus numerical solution for $n_c$ in the scenario where $F_2>0, F_1<0$. We take $z_a=-1, z_c=1, \beta_1=0.2, \beta_2=0.6$.  We also have $\lambda=1.5 \times 10^{-3}$, $\delta^{-1}=170$, and $\Delta V = 4$. We include insets to show a close up of the agreement at the two boundaries.  BL, BR refer to the left and right boundary layer solutions respectfully, similarly IL, IR refer to the left and right intermediate layer solutions respectfully. $z_L, z_R$ refer to the leading order transition points from the boundary to the intermediate layers.}
		\label{fig:case3beta}
	\end{figure}

	We also provide examples of the solution obtained via the ODE asymptotics for cases where have a root for one of $F_i$, $i=1,2$.   
	We first outline the case where 
	\[ \frac{\beta_1}{\beta_2}> \beta^*, \quad \frac{\beta_2}{\beta_1}< \beta^*,\]
	in other words, this is the scenario where $F_1$ has a root, but $F_2$ does not and therefore we use the value of $z_R^*$ to minimise $|F_2|$. The results for the lithium concentration profile from both asymptotic and numerical approaches can be seen in \cref{fig:case2beta} and \cref{fig:case2betaiain}.
	We need to ensure we retain $\beta_i>0$ so we take $\beta_1=0.45, \beta_2=0.1$ by choosing $z_a=-1, n_a=0.45$ with the results shown in \cref{fig:case2beta}. We also include an example in \cref{fig:case2betaiain} where we take $z_a=-2, n_a=0.3$ to obtain $\beta_1=0.6, \beta_2=0.1$. In both cases we see similar agreement between the asymptotics and the numerics as was observed in the case presented in \cref{main-fig:asym_v_numerics_a} in the main text. In the insets of both plots we have zoomed in on the two boundary layers to further demonstrate the agreement.
	We also show an example of the scenario where $F_1$ is minimised and $F_2$ has a root, corresponding to
	\[ \frac{\beta_1}{\beta_2}< \beta^*, \quad \frac{\beta_2}{\beta_1}>\beta^*.\]
	To demonstrate this we take $\beta_1=0.2, \beta_2=0.6$ as shown in \cref{fig:case3beta}. Again, we zoom in on the boundaries as seen in the inset plots and we observe good agreement between the numerics and asymptotics.

	\section{Integral approach} 
	\label{sec:appendix}

	We observe that the one dimensional equilibrium problem, written in terms of $\theta$, given by 
	\begin{subequations}\label{eq:ZeroJ_equi_theta_prob_intSol}
		\begin{align}
		\frac{\epsilon^2 \delta}{z_c} \deriv[2]{\theta}{z} &= \frac{\beta_1+\beta_2}{1+ e^{-\theta}}  -\beta_1      \label{eq:zero_asym_prob_intSol} \\
		\int_0^1 \frac{1}{1+e^{-\theta}} \diff{z} & = \frac{\beta_1}{\beta_1+\beta_2} \\
		\theta(0)=\frac{c-z_c}{\delta}, & \quad \theta(1)=\frac{c}{\delta}. 
		\end{align}
	\end{subequations}
	has a first integral. The resulting problem is separable, therefore we can write down an analytical solution which will hold across the whole domain. We present the analytical solution and we also perform an asymptotic reduction of the first integral problem, which reiterates the various regimes we described in \cref{main-sec:asymptotics} of the main text, in this section.
	We note that we introduce the ODE asymptotic solution in the main body of this work over this analytical solution as we wanted to present an approach that could be generalised. In the case of this problem it is not clear that a first integral could be obtained in the non zero flux scenario, whereas we can carry out a similar analysis of the ordinary differential equations of the non zero flux problem in the same manner as shown for the zero flux case.

	\subsection{Analytical solution}
\begin{figure}[htb]
		%
		\begin{subfigure}{0.325\linewidth}
			\includegraphics[width=\linewidth]{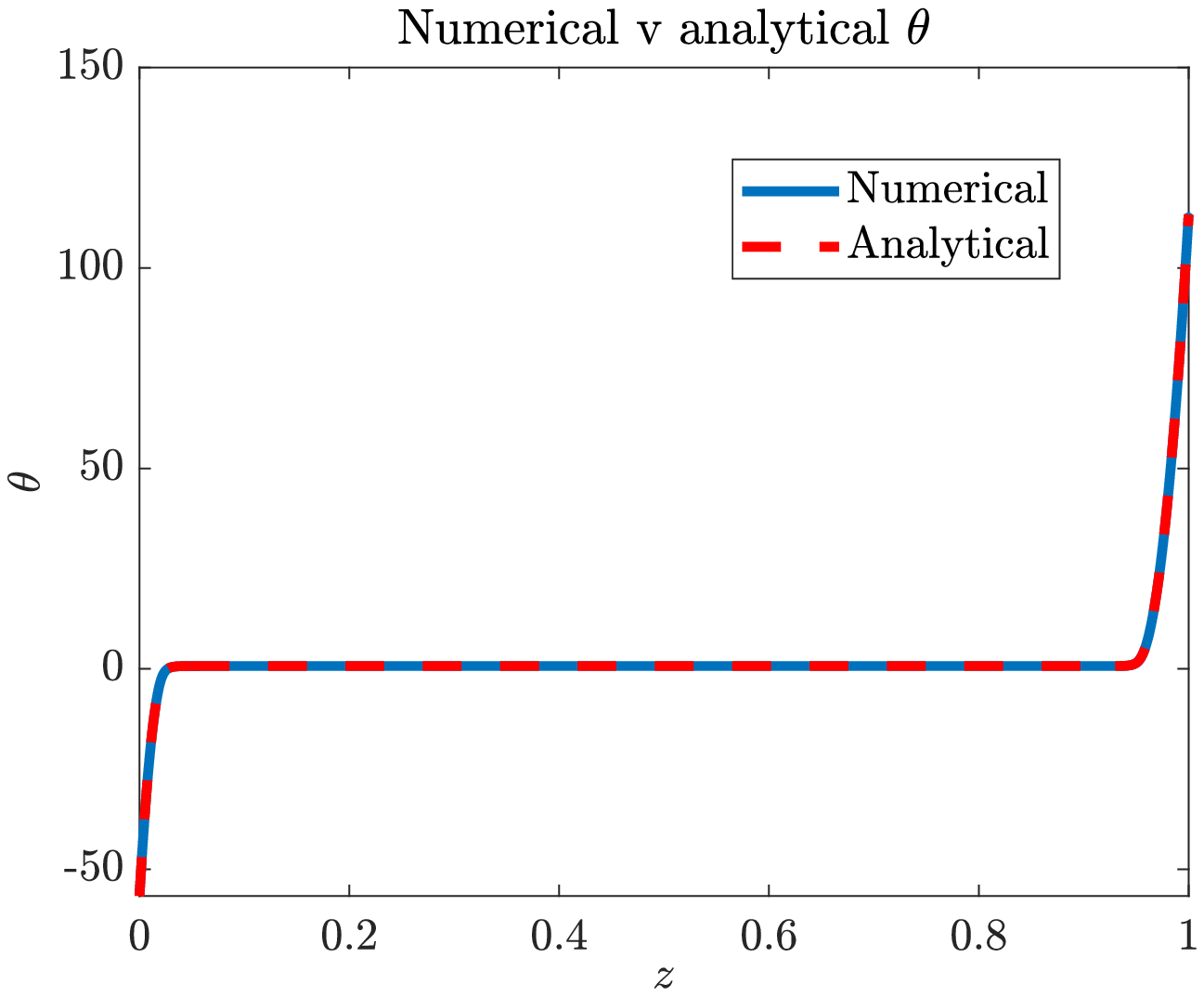}
			\caption{Numerical versus analytical $\theta$ solution}
			\label{fig:analytical_sol_a}
		\end{subfigure}
		\begin{subfigure}{0.325\linewidth}
			\includegraphics[width=\linewidth]{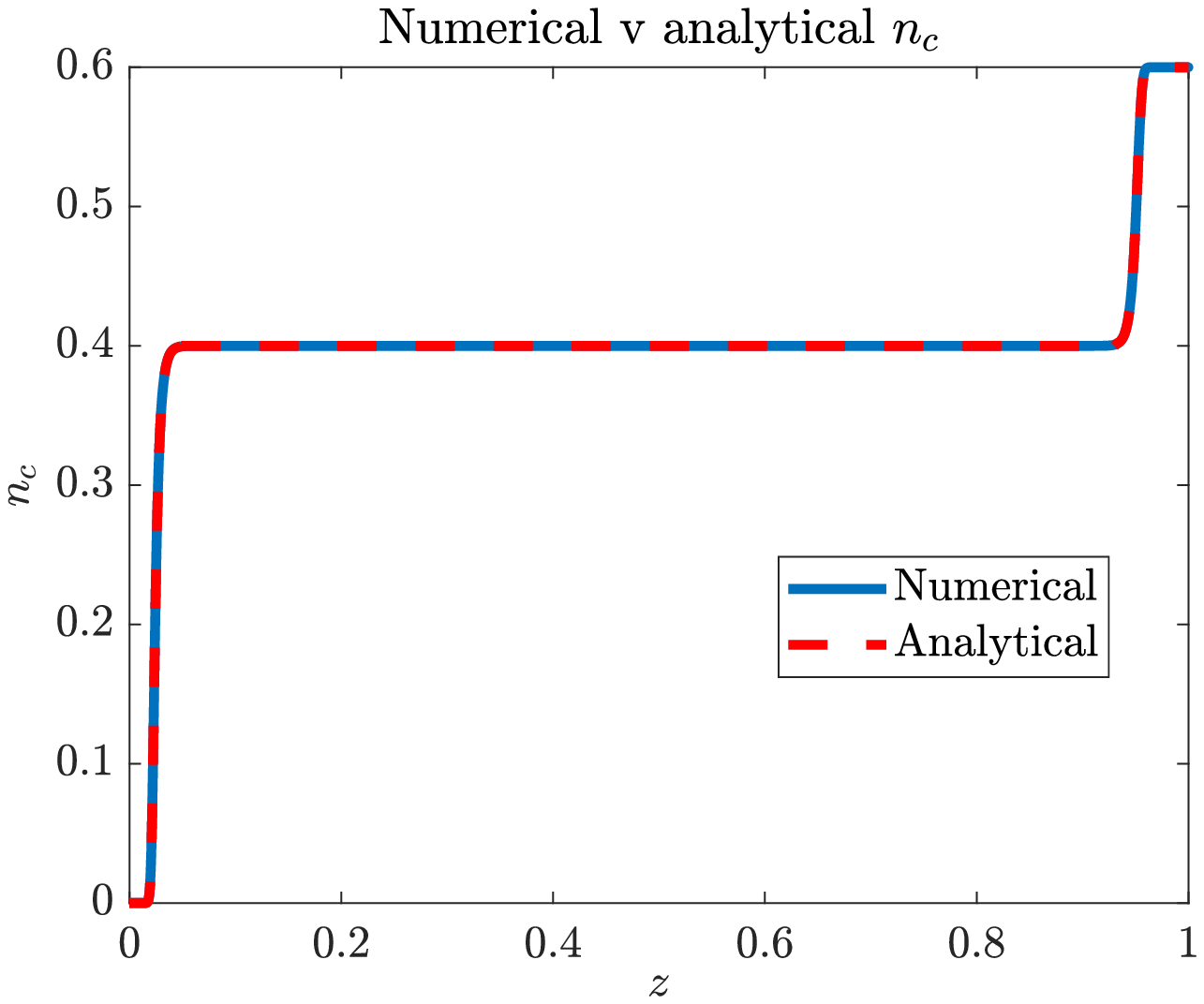}
			\caption{Numerical versus analytical $n_c$ solution}
			\label{fig:analytical_sol_b}
		\end{subfigure}
		\begin{subfigure}{0.325\linewidth}
			\includegraphics[width=\linewidth]{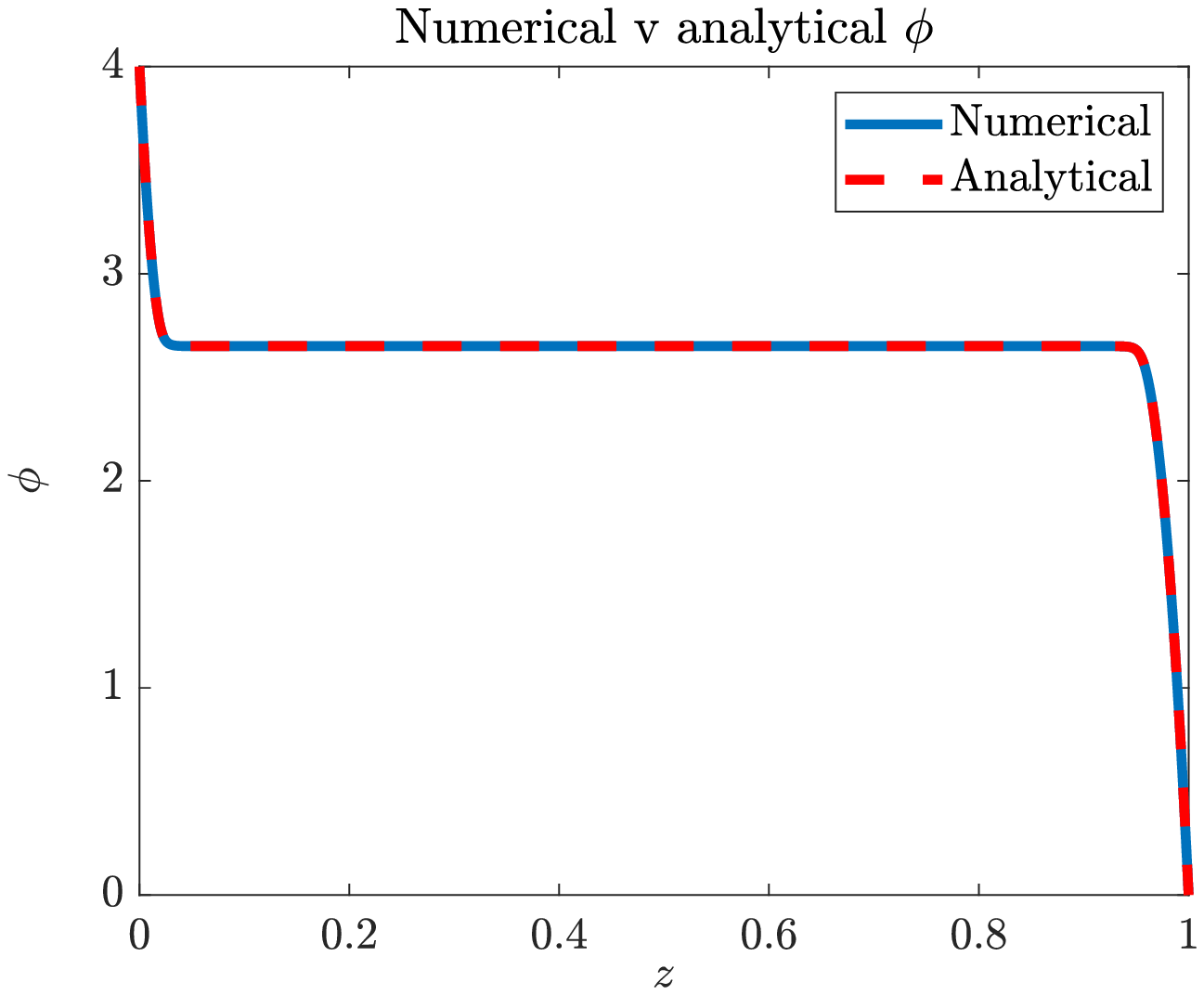}
			\caption{Numerical versus analytical $\phi$ solution}
			\label{fig:analytical_sol_c}
		\end{subfigure}
		\caption{Analytical versus numerical solutions
			for $\theta$, the lithium concentration $n_c$, and electric potential $\phi$. Parameter values are as follows: $z_c=1, \quad z_a=-1, \quad m_c=0.3, \quad m_a=0.7, \quad n_a=0.4,\quad \nu=0.6$, with $\frac{\beta_1}{\beta_2}=2 < \beta^*, \quad \frac{\beta_2}{\beta_1}=\frac{1}{2} < \beta^*$. We also have $\lambda=1.5 \times 10^{-3}$, $\delta=170$, and $\Delta V = 4$. }
		\label{fig:analytical_sol}
	\end{figure}

	We multiply \cref{eq:zero_asym_prob_intSol} by $\deriv{\theta}{z}$ 
	and note that $\deriv{}{z} \log{(1+e^{\theta})} =  (1+e^{-\theta})^{-1} \deriv{\theta}{z}$ so that we have
	\begin{align}\label{eq:integral_deriv_Eq}
	\frac{\delta \epsilon^2}{2 z_c} \deriv{}{z} \left(\deriv{\theta}{z} \right)^2 &= \deriv{}{z} \left(  \left(\beta_1+\beta_2\right)  \log{(1+e^{\theta})} - \beta_1 \theta \right). 
	\end{align}
	We solve for an expression for $\theta$ using separation of variables and integrating, defining $F(\theta)$ by
	\begin{equation}\label{eq:int_eq_theta}
	F(\theta)= \int_{\frac{c-z_c}{\delta}}^{\theta} \frac{\diff{u}}{\sqrt{\left(\beta_1+\beta_2\right) \log{(1+e^{u})} - \beta_1 u - C_0}} = \sqrt{\frac{2 z_c}{\delta \epsilon^2}} z,
	\end{equation}
	where we take the positive root so that $\theta$ increases with increasing $z$, and $C_0$ is some constant of integration.
	As a consequence of the integral constraint $ \deriv{\theta}{z}(z=0) = \deriv{\theta}{z}(z=1)$ and from our boundary conditions $\theta(0)= \frac{c-z_c}{\delta}, \quad \theta(1)=\frac{c}{\delta}$, applying these to \cref{eq:integral_deriv_Eq} we will have
	\[ \left(\beta_1+\beta_2\right)\log{(1+e^{\frac{c-z_c}{\delta}})}- \beta_1\left(\frac{c-z_c}{\delta} \right)  = \left(\beta_1+\beta_2\right) \log{(1+e^{\frac{c}{\delta}})} -\beta_1\left(\frac{c}{\delta} \right),  \]
	which we can use to solve for $c$ exactly  
	\begin{equation}
	c = \delta \log{\left( \frac{1-e^{z_c \beta_1/\delta \left(\beta_1+\beta_2\right)}}{e^{-z_c \beta_2/\delta \left(\beta_1+\beta_2\right)}-1}\right)} = \frac{z_c \beta_1}{\beta_1+\beta_2} + \delta \log{\left( \frac{1-e^{-z_c \beta_1/\delta \left(\beta_1+\beta_2\right)}}{1-e^{-z_c \beta_2/\delta \left(\beta_1+\beta_2\right)}}\right)}.
	\end{equation}
	Noting that $\delta \ll 1$ so that 
	\begin{equation}
	c   \approx \frac{z_c \beta_1}{\beta_1+\beta_2} = \frac{\beta_1}{\nu}
	\end{equation}
	as before in \cref{main-eq:c_value} of \cref{main-sec:asy_matching} of the main text. That is, any corrections to $c$ beyond leading order are exponentially small.
	To find $C_0$ we recall that $\theta(1)= \frac{c}{\delta}$, so that $F(\frac{c}{\delta})= \sqrt{\frac{2 z_c}{\delta \epsilon^2}}$ to close the problem. 
	Practically determining $C_0$ can be difficult because the bulk value is near a singularity of the integral, however we can exploit this to approximate $C_0$ by using the fact that $\delta \epsilon^2 \ll 1$ so that there is a bulk value for $\theta$ when $\frac{\beta_1+\beta_2}{1+e^{-\theta}} = \beta_1$, corresponding to
	\begin{align}
	\theta 
	= \log{\left(\frac{\beta_1}{\beta_2}\right)},
	\end{align}
	which is what we find from the ODE approach.
	Now, in the bulk the derivative in \cref{eq:integral_deriv_Eq} is approximately zero, therefore we approximate $C_0$ by
	\begin{align}\label{eq:c0_analytical_approx}
	\left(\beta_1+\beta_2\right) \log{\left(\frac{\beta_1+\beta_2}{\beta_2}\right)} - \beta_1 \log{\left(\frac{\beta_1}{\beta_2}\right)} &=C_0,
	\end{align}
	completing the analytical solution to the zero flux equilibrium problem. We note that here the constant $z_c C_0$ is equivalent to $A_1$ from the ODE approach to the asymptotics in \cref{main-sec:asymptotics} of the main text, given by \cref{main-eq:A1_sol_firstInt}. In \cref{fig:analytical_sol_a} we plot the analytical solution for $\theta$ in comparison with the numerical solution obtained in \cref{main-sec:num_equi} of the main text. Using the solution for $\theta$ we also compute the analytical solutions for $n_c$ and $\phi$. We present these solutions, with their numerical counterparts in \cref{fig:analytical_sol_b} and \cref{fig:analytical_sol_c}. In all three figures we have excellent agreement between the analytics and the numerics.
	
	\subsection{Integral asymptotics}
	Having a full analytic solution as outlined in the last section can be useful, however there are drawbacks. For example, we do not gain insight into the behaviour in the different regimes of the solution as we do from the asymptotic approach. 
	We note that using the approximation \cref{eq:c0_analytical_approx} for $C_0$ introduces a singularity in the integral at $\theta = \theta_0 = \log{\left( \frac{\beta_1}{\beta_2}\right)}$ which leads us to consider splitting the integral. In addition, from our boundary conditions in \cref{eq:ZeroJ_equi_theta_prob_intSol} we see that $\theta \sim \mathcal{O}(\delta^{-1})$ at the two limits, therefore we will have four cases for an asymptotic approximation to the solution. We will outline the first two cases, and then briefly overview the solutions for the last two, which are symmetric to the first two. We further note that this section echoes that of \cref{main-sec:asymptotics} in the main text. 
	
	\subsubsection{Case 1: $\theta < \log{\left(\frac{\beta_1}{\beta_2}\right)}, \quad \theta \sim \mathcal{O}(\delta^{-1})$}
	The boundary value $c-z_c<0$ so $\theta$ must be large and negative. We scale $u = -\delta^{-1}w, \quad \diff{u}=-\delta^{-1} \diff{w}$. With this change of variable, when $u=\frac{c-z_c}{\delta}, \quad w=z_c-c$ and when $u=\theta, \quad w = -\delta \theta = \psi$ with $\theta = -\delta^{-1} \psi$.
	Using this substitution in \cref{eq:int_eq_theta}, noting that $\log{\left(1+e^{-\delta^{-1} w}  \right)}\approx 0$
	\begin{align}\label{eq:case1intprob}
	-\delta^{-1} \int_{z_c-c}^{-\delta \theta} \frac{\diff{w}}{\sqrt{  \left(\beta_1+\beta_2\right)\log{(1+e^{-\delta^{-1}w})} +\beta_1\delta^{-1} w - C_0 }}  \notag
	& \approx 
	\sqrt{\delta^{-1}}  \int_{-\delta \theta}^{z_c-c} \frac{\diff{w}}{\sqrt{ \beta_1 w-\delta C_0}}, \notag \\
	&= \sqrt{\frac{2z_c}{\delta}} \frac{z}{\epsilon},
	\end{align}
	where we have just set the simplified integral equal to the right hand side of \cref{eq:int_eq_theta} in the last line.
	Simplifying and solving \cref{eq:case1intprob} yields
	\begin{align}
	\theta = -\delta^{-1} \left( \frac{z_c\beta_1 }{2 } \frac{z^2}{\epsilon^2} - \sqrt{2 z_c \left( \beta_1(z_c-c)-\delta C_0\right) }\frac{z}{\epsilon}+ z_c-c\right).
	\end{align}
	We can the expand about $\delta \ll 1$ to get 
	\begin{equation}
	\theta \sim -\delta^{-1} \left( \frac{z_c\beta_1 }{2} \frac{z^2}{\epsilon^2}-\sqrt{2 z_c \beta_1(z_c-c)}\frac{z}{\epsilon} +  z_c-c   \right) - \frac{z_c \delta C_0}{\sqrt{2 z_c \beta_1(z_c-c)}}\frac{z}{\epsilon} + \mathcal{O}(\delta^{3/2})
	\end{equation}
	highlighting our $\mathcal{O}(1)$, $\mathcal{O}(\delta)$ solutions 
	\begin{align}\label{eq:int_asy_case1_o1}
	\mathcal{O}(1):
	\psi_0 &= \frac{z_c \beta_1}{2} \left(\frac{z}{\epsilon} \right)^2 -\sqrt{2(z_c-c)z_c \beta_1}\frac{z}{\epsilon} + z_c-c,\\
	\mathcal{O}(\delta):  
	\psi_1 &=  
	\frac{z_c C_0}{\sqrt{2 z_c \beta_1 (z_c-c)}} \frac{z}{\epsilon},
	\end{align}
	with $\theta= -\delta^{-1} \left(\psi_0 + \delta \psi_1 \right)$.
	We note that this expression \cref{eq:int_asy_case1_o1} for $\psi_0$ matches the solution we recover from the ODE approach \cref{main-eq:psi_sol_lhs} with $A_0$ given as \cref{main-eq:A0sol}. We can see that we will lose asymptotic consistency when $\frac{z}{\epsilon} \sim \delta^{-1}$. 
	This region is valid up to some $\hat{z_1} \quad (\hat{\psi} = \psi(\hat{z_1}))$ where $\theta$ transitions to $\mathcal{O}(1)$, and as this is an $\mathcal{O}(1)$ value we would expect $\hat{\psi} = 0 + \delta \hat{\psi_1} \quad (\hat{\theta}=\hat{\theta_1})$.
	
	\subsubsection{Case 2: $\theta < \log{\left(\frac{\beta_1}{\beta_2}\right)}, \theta \sim \mathcal{O}(1)$}
	We want to exploit the singularity when $\theta=\theta_0=\log{\left(\frac{\beta_1}{\beta_2}\right)}$. We will define $u=\theta_0-y$ and we will integrate from the point where the previous solution terminates, which we call $\hat{\theta_1}$.
	This will give us
	\begin{equation}\label{eq:case2int}
	\int_{\hat{\theta_1}}^{\theta} \frac{\diff{u}}{\sqrt{  (\beta_1+\beta_2) \log{(1+e^{u})} -\beta_1 u -C_0 }} = \sqrt{\frac{2 z_c}{\delta \epsilon^2}} \left(z-\hat{z_1}\right).
	\end{equation}
	If $u=\theta_0- y$, $\diff{u}=- \diff{y}$, then when $u=\hat{\theta_1}, y=\theta_0-\hat{\theta_1}$, and $u=\theta, y=\theta_0-\theta$.	
	With this we can expand about the singularity 
	\begin{align}\label{eq:singularity_expand}
	\left(\beta_1+\beta_2 \right) \log{(1+e^u)} -\beta_1 u -  C_0  &= \left(\beta_1+\beta_2 \right)  \log{(1+e^{\theta_0- y})} -\beta_1 \left(\theta_0-y\right) -  C_0  \notag \\
	& \approx  \frac{\beta_1 \beta_2}{2 \left(\beta_1+\beta_2 \right) } y^2
	\end{align}
	since $ \left(\beta_1+\beta_2 \right)  \log{\left(1+e^{\theta_0}\right)} 
	-\beta_1 \theta_0  -C_0 = 0 $ as it corresponds to the bulk solution, $\frac{\left(\beta_1+\beta_2 \right)  e^{\theta_0}}{1+e^{\theta_0}} -\beta_1 = \beta_1 -\beta_1 =0$, and $\frac{\left(\beta_1+\beta_2 \right)   y^2 e^{\theta_0}}{2 \left(1+e^{\theta_0}\right)^2} = \frac{\beta_1 \beta_2}{2 \left(\beta_1+\beta_2 \right) } y^2$.
	Substituting back into the left hand side of \cref{eq:case2int} we then have
	\begin{equation}
	\int_{\hat{\theta_1}}^{\theta} \frac{\diff{u}}{\sqrt{ \left(\beta_1+\beta_2 \right) \log{(1+e^{u})} -\beta_1 u -C_0 }} = \sqrt{\frac{2 \left(\beta_1+\beta_2 \right) }{\beta_1 \beta_2}} \int_{\theta_0-\theta}^{\theta_0-\hat{\theta_1}} \frac{\diff{y}}{y}
	= \sqrt{\frac{2 \left(\beta_1+\beta_2 \right) }{\beta_1 \beta_2}} \log{\left( \frac{\theta_0-\hat{\theta_1}}{\theta_0-\theta}\right)}.
	\end{equation}
	Therefore,
	\begin{equation}
	\sqrt{\frac{2 \left(\beta_1+\beta_2 \right) }{\beta_1 \beta_2}} \log{\left( \frac{\theta_0-\hat{\theta_1}}{\theta_0-\theta}\right)} = \sqrt{\frac{2 z_c}{\delta \epsilon^2}} \left( z-\hat{z_1}\right),
	\end{equation}
	which we rearrange to obtain the following solution for $\theta$
	\begin{equation}\label{eq:int_asy_case2}
	\theta = \theta_0 - \left(  \theta_0-\hat{\theta_1}\right) e^{-\sqrt{\frac{ z_c\beta_1 \beta_2}{\left(\beta_1+\beta_2 \right)  \delta \epsilon^2}} \left( z - \hat{z_1} \right)}. 
	\end{equation}
	We note that because of the singularity at $\theta=\theta_0$ this will go to $\theta_0$ as  $z \to \infty$. 
	We observe that this also agrees with the intermediate solution found in \cref{main-eq:theta_sol_leftIL}.

	\subsubsection{Left matching}\label{section:left_match_int}
	We can now look to determine $\hat{z_1}$ and $\hat{\theta_1}$. Ideally we would impose both continuity and differentiability at $\hat{z_1}$, however as the boundary layer is concerned with $\theta \sim \delta^{-1}$ and the intermediate layer is based on $\theta \sim \theta_0$ there may be no natural single point for matching. We impose continuity 
	\begin{equation}\label{eq:contin_lhs}
	\theta \left( \hat{z_1} \right) = - \delta^{-1} \left( \frac{z_c \beta_1}{2} \left(\frac{\hat{z_1}}{\epsilon}\right)^2 - \sqrt{2 z_c \beta_1\left(z_c-c\right) - 2 z_c\delta C_0 }\left(\frac{\hat{z_1}}{\epsilon}\right)+z_c-c\right)
	= \hat{\theta_1}.  \end{equation}
	We define 
	\begin{align}
	F_1 &= \deriv{\theta}{z}\rvert_{\hat{z} \text {case 1}}-\deriv{\theta}{z}\rvert_{\hat{z} \text {case 2}} \notag \\
	&= -\frac{1}{\delta \epsilon} \left( z_c \beta_1 \left( \frac{\hat{z_1}}{\epsilon}\right) - \sqrt{2 z_c \beta_1\left(z_c-c\right) - 2 z_c\delta C_0 }\right) - \frac{a}{\sqrt{\delta} \epsilon}\left(\theta_0-\hat{\theta_1}\right), \label{eq:F1_def}
	\end{align}
	where $a=\sqrt{\frac{z_c \beta_1 \beta_2}{\beta_1+\beta_2}}$. 
	Ideally we would choose $\hat{z_1}$ so that $F_1=0$, but instead we will minimise $\left|F_1\right|$. Note that $F_1'' = -\frac{ a z_c \beta_1}{\lambda^3} <0$ so $F_1$ is convex.
	Using the continuity constraint \cref{eq:contin_lhs} in $F1$ \cref{eq:F1_def} and setting this equal to zero
	\begin{equation}
	F_1' = - \frac{z_c \beta_1}{\delta \epsilon^2} - \frac{a z_c \beta_1 \hat{z_1}}{\delta^{3/2} \epsilon^3} + \frac{a \sqrt{2z_c\beta_1\left(z_c-c\right)-2z_c\delta C_0}}{\delta^{3/2} \epsilon^2} =0,
	\end{equation}
	is satisfied for 
	\begin{equation}
	\hat{z_1} =  \frac{\left(a \sqrt{2z_c\beta_1\left(z_c-c\right)-2z_c\delta C_0} - z_c \beta_1 \sqrt{\delta}\right)}{a z_c \beta_1 } \epsilon. 
	\end{equation}
	This gives us that
	\begin{equation}
	\theta\left(\hat{z_1}\right) = - \left( \frac{\beta_1+\beta_2}{2 \beta_2} + \frac{C_0}{\beta_1} \right).
	\end{equation}

	\subsubsection{Case 3: $\theta>\log{\left(\frac{\beta_1}{\beta_2}\right)}, \quad \theta \sim \mathcal{O}(\delta^{-1})$}
	We now consider the symmetric cases on the other side of the singularity. On this side we redefine the integral as
	\begin{equation}\label{eq:case3int}
	\int_{\hat{\theta}}^{\frac{c}{\delta}} \frac{\diff{u}}{\sqrt{\left(\beta_1+\beta_2\right)\log{(1+e^{u})} -\beta_1 u - C_0}} = \sqrt{\frac{2z_c}{\delta \epsilon^2}} \left(1-z\right) = \sqrt{\frac{2 z_c}{\delta \epsilon^2}} \zeta,
	\end{equation}
	where $\zeta=1-z$. As before we scale $u=\delta^{-1}w, \quad \diff{u}=\delta^{-1} \diff{w}, \quad \theta =\delta^{-1} \xi$. We proceed with the same step as taken in Case 1 to find the following solution
	\begin{equation}
	\theta = \delta^{-1} \left( \frac{z_c \beta_2}{2} \left(\frac{\zeta }{\epsilon}\right)^2 - \sqrt{2 z_c \beta_2 c - 2 z_c C_0 \delta} \left(\frac{\zeta}{\epsilon}\right)+c\right).
	\end{equation}
	As before, we expand about $\delta \ll 1$ to obtain
	\begin{equation}
	\theta \sim \delta^{-1} \left( \frac{z_c\beta_2 }{2} \frac{\zeta^2}{\epsilon^2}-\sqrt{2 z_c \beta_2 c}\frac{\zeta}{\epsilon} +  c   \right) +\frac{z_c \delta C_0}{\sqrt{2 z_c \beta_2 c }}\frac{\zeta}{\epsilon} + \mathcal{O}(\delta^{3/2})
	\end{equation}
	again, highlighting our $\mathcal{O}(1)$, $\mathcal{O}(\delta)$ solutions
	\begin{subequations}
		\begin{align}
		\mathcal{O}(1): \quad &\xi_0 = \frac{z_c \beta_2}{2} \left(\frac{\zeta}{\epsilon} \right)^2 - \sqrt{2 z_c \beta_2 c}
		\left( \frac{\zeta}{\epsilon}\right)+c,    \\
		\mathcal{O}(\delta): \quad &\xi_1 = \frac{C_0}{\sqrt{2     z_c \beta_2 c}} \frac{\zeta}{\epsilon},
		\end{align}
	\end{subequations}
	corresponding to \cref{main-eq:psi_sol_rhs} from the ODE approach.
	As on the other side, this will be valid up to some $\hat{\zeta}$ where $ \xi(\hat{\zeta})= \hat{\xi} = 0+ \delta \hat{\xi}_1$ .

	\subsubsection{Case 4: $\theta>\log{\left(\frac{\beta_1}{\beta_2}\right)}, \quad \theta \sim \mathcal{O}(1)$}

	We will split the integral based on where the solution is valid. $\theta$ in case 3 will be valid to some $\theta=\hat{\theta_2}$ and 
	\begin{equation}
	\int_{\theta}^{\hat{\theta_2}} \frac{\diff{u}}{\sqrt{\left(\beta_1+\beta_2\right) \log{\left( 1+e^u\right)-\beta_1 u -C_0}}} = \sqrt{\frac{2 z_c}{\delta \epsilon^2}} \left( \zeta - \hat{\zeta}\right)
	\end{equation}
	We let $u=\theta_0 + y, \quad \diff{u}= \diff{y}$ so that when $u=\theta$, $y=\theta-\theta_0$ and when $u=\hat{\theta_2}$, then $y=\hat{\theta_2}-\theta_0$. We proceed as in \cref{eq:singularity_expand}, expanding about the singularity and taking the leading order to obtain
	\begin{align}
	\int_{\theta}^{\hat{\theta_2}} \frac{\diff{u}}{\sqrt{\left(\beta_1+\beta_2\right) \log{\left( 1+e^u\right)-\beta_1 u -C_0}}} &= \sqrt{\frac{2 \left(\beta_1+\beta_2\right)}{ \beta_1 \beta_2}} \int_{\theta-\theta_0}^{\hat{\theta_2}-\theta_0} \frac{\diff{y}}{y} = \sqrt{\frac{2 \left(\beta_1+\beta_2\right)}{\beta_1 \beta_2}} \log{\left(\frac{\hat{\theta_2}-\theta_0}{\theta-\theta_0} \right)} \notag\\
	\implies \sqrt{\frac{2 \left(\beta_1+\beta_2\right)}{\beta_1 \beta_2}} \log{\left(\frac{\hat{\theta_2}-\theta_0}{\theta-\theta_0} \right)} &= \sqrt{\frac{2 z_c}{\delta \epsilon^2}} \left( \zeta - \hat{\zeta}\right) \notag\\
	\theta &= \theta_0 + \left(\hat{\theta_2} - \theta_0 \right) e^{-\sqrt{\frac{z_c \beta_1 \beta_2}{\left(\beta_1+\beta_2\right) \delta \epsilon^2}} \left(\zeta-\hat{\zeta}\right)},
	\end{align}
	We observe that this agrees with the intermediate solution found in \cref{main-eq:theta_sol_rightIL}.
	
	\subsubsection{Right matching}
	We proceed similarly as in case 1 and 2 in \cref{section:left_match_int}.
	We first enforce continuity where $\hat{\zeta}=1-\hat{z}_2$
	\begin{equation}
	\theta\left(\hat{\zeta}\right) = \delta^{-1} \left( \frac{z_c \beta_2}{2} \left(\frac{\hat{\zeta}}{\epsilon}\right)^2 - \sqrt{2 z_c \beta_2 c-2 z_c\delta C_0 }\left(\frac{\hat{\zeta}}{\epsilon}\right) + c\right) =\hat{\theta_2}.
	\end{equation}
	We will define 
	\begin{align}
	F_2 &= \deriv{\theta}{\zeta}\rvert_{\hat{\zeta} \text {case 3}} - \deriv{\theta}{\zeta} \rvert_{\hat{\zeta} \text {case 4}}  \notag \\
	&= 
	\frac{1}{\epsilon \delta} \left( z_c \beta_2 \left(\frac{\hat{\zeta}}{\epsilon}\right) - \sqrt{2 z_c \beta_2 c - 2 z_c \delta  C_0} \right) + \frac{a}{\epsilon \sqrt{\delta}} \left( \hat{\theta_2}-\theta_0\right).
	\end{align}
	Similar to the left hand side matching we find that $F_2'=0$ for
	\begin{equation}     \hat{\zeta}=\frac{\left(a \sqrt{2 z_c \beta_2 c - 2 z_c \delta C_0} -z_c \beta_2 \sqrt{\delta}\right)}{a z_c \beta_2} \epsilon.
	\end{equation}
	Which corresponds to 
	\begin{equation}
	\theta\left(\hat{\zeta}\right) = \left( \frac{\beta_1 + \beta_2}{2 \beta_1} +\frac{C_0}{\beta_2}\right).
	\end{equation}

	\begin{figure}[htb]
		\begin{subfigure}{0.45\linewidth}
			\includegraphics[width=\linewidth]{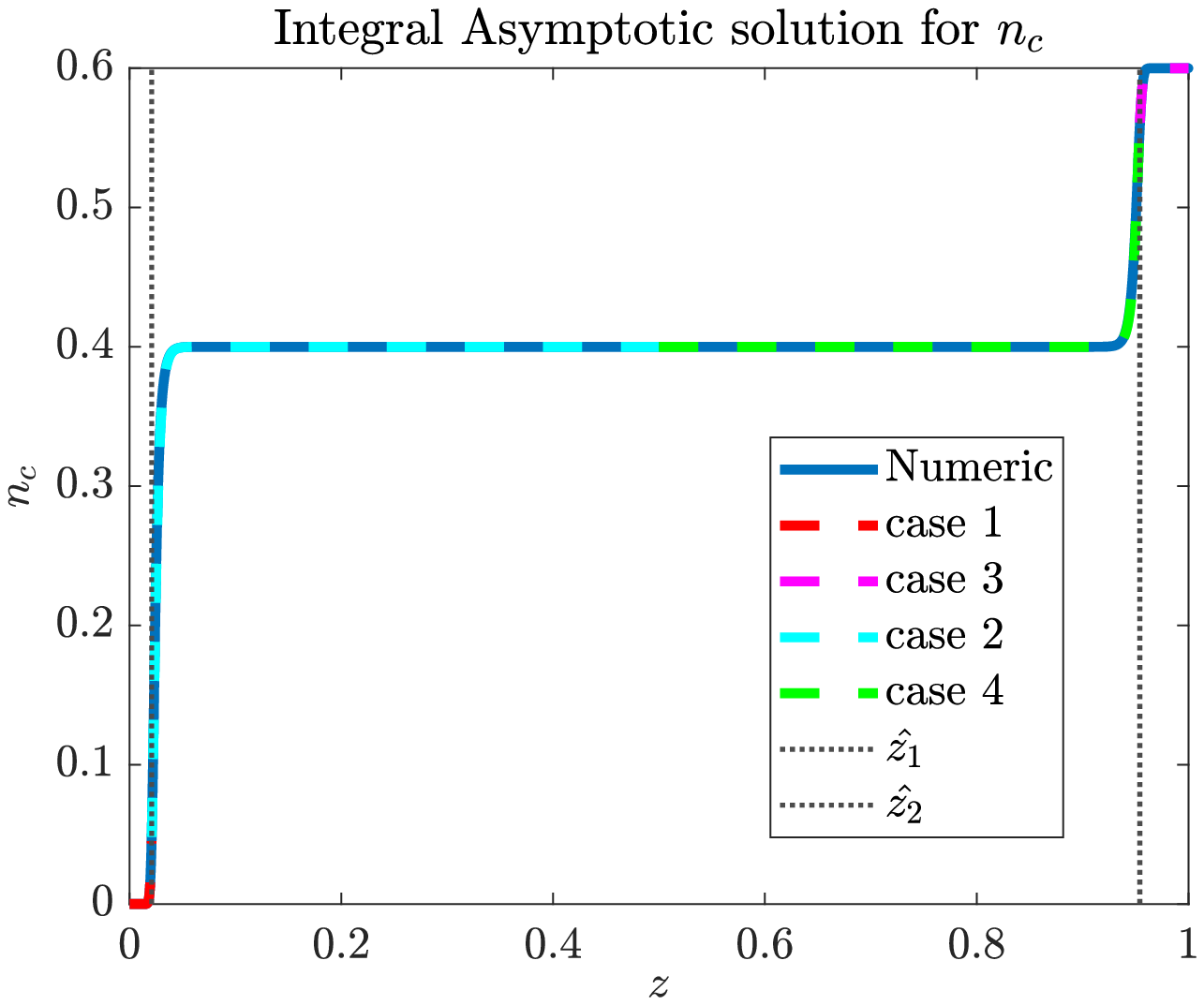}
			\caption{Lithium concentration asymptotics versus numerics}
			\label{fig:asym_v_numerics_int_a}
		\end{subfigure}
		\begin{subfigure}{0.45\linewidth}
			\includegraphics[width=\linewidth]{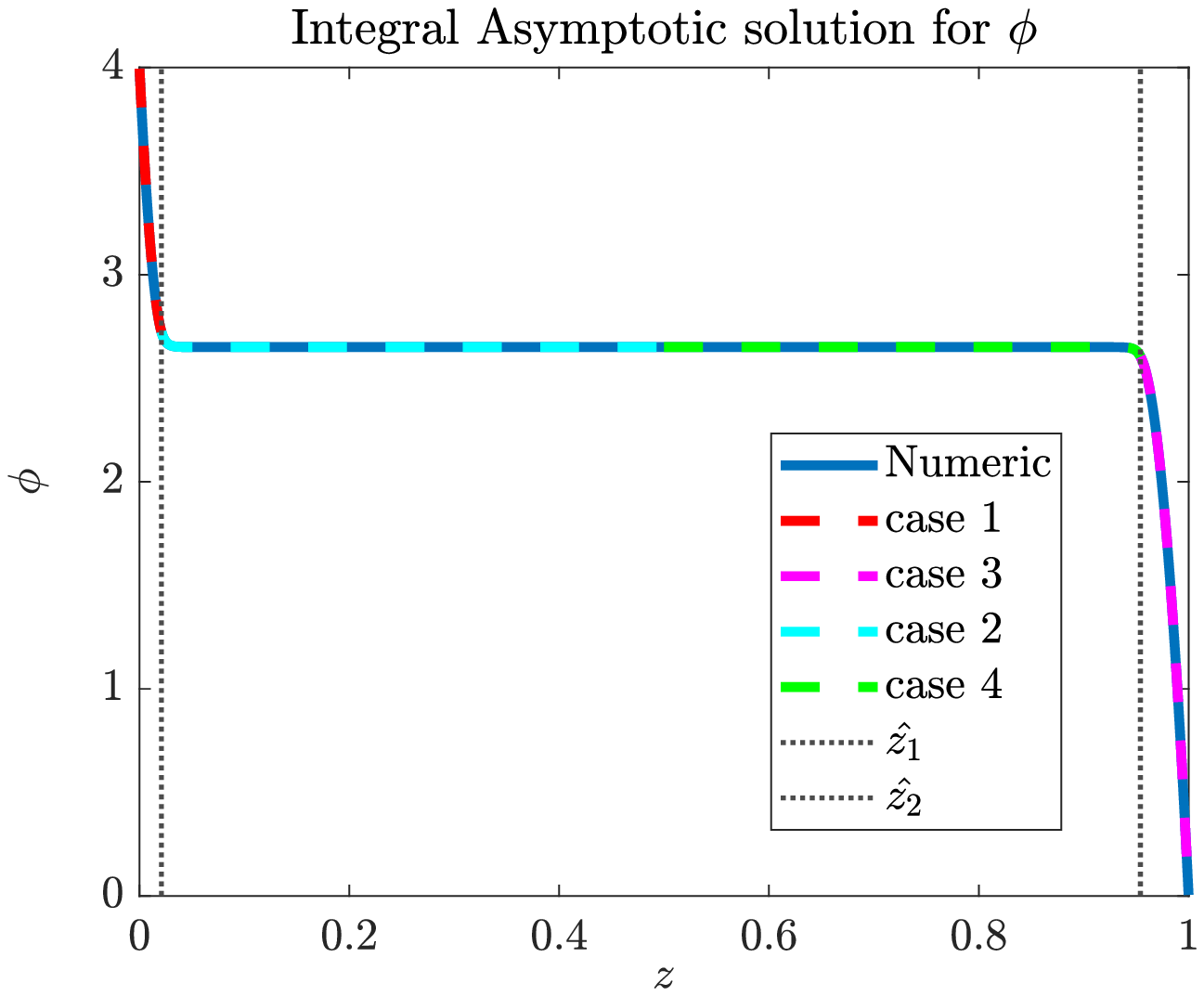}
			\caption{Electric potential asymptotics versus numerics}
			\label{fig:asym_v_numerics_int_b}
		\end{subfigure}
		\caption{Integral Asymptotic versus numerical solutions for lithium concentration and electric potential. Parameter values are as follows: $z_c=1, \quad z_a=-1, \quad m_c=0.3, \quad m_a=0.7, \quad n_a=0.4,\quad \nu=0.6$, with $\frac{\beta_1}{\beta_2}=2 < \beta^*, \quad \frac{\beta_2}{\beta_1}=\frac{1}{2} < \beta^*$. We also have $\lambda=1.5 \times 10^{-3}$, $\delta^{-1}=170$, and $\Delta V = 4$. BL, BR refer to the left and right boundary layer solutions respectfully, similarly IL, IR refer to the left and right intermediate layer solutions respectfully. $z_L, z_R$ refer to the leading order transition points from the boundary to the intermediate layers.}
		\label{fig:asym_v_numerics_int}
	\end{figure}

	\subsection{Comparisons}
\begin{figure}[htb]
		\begin{subfigure}{0.45\linewidth}
			\includegraphics[width=\linewidth]{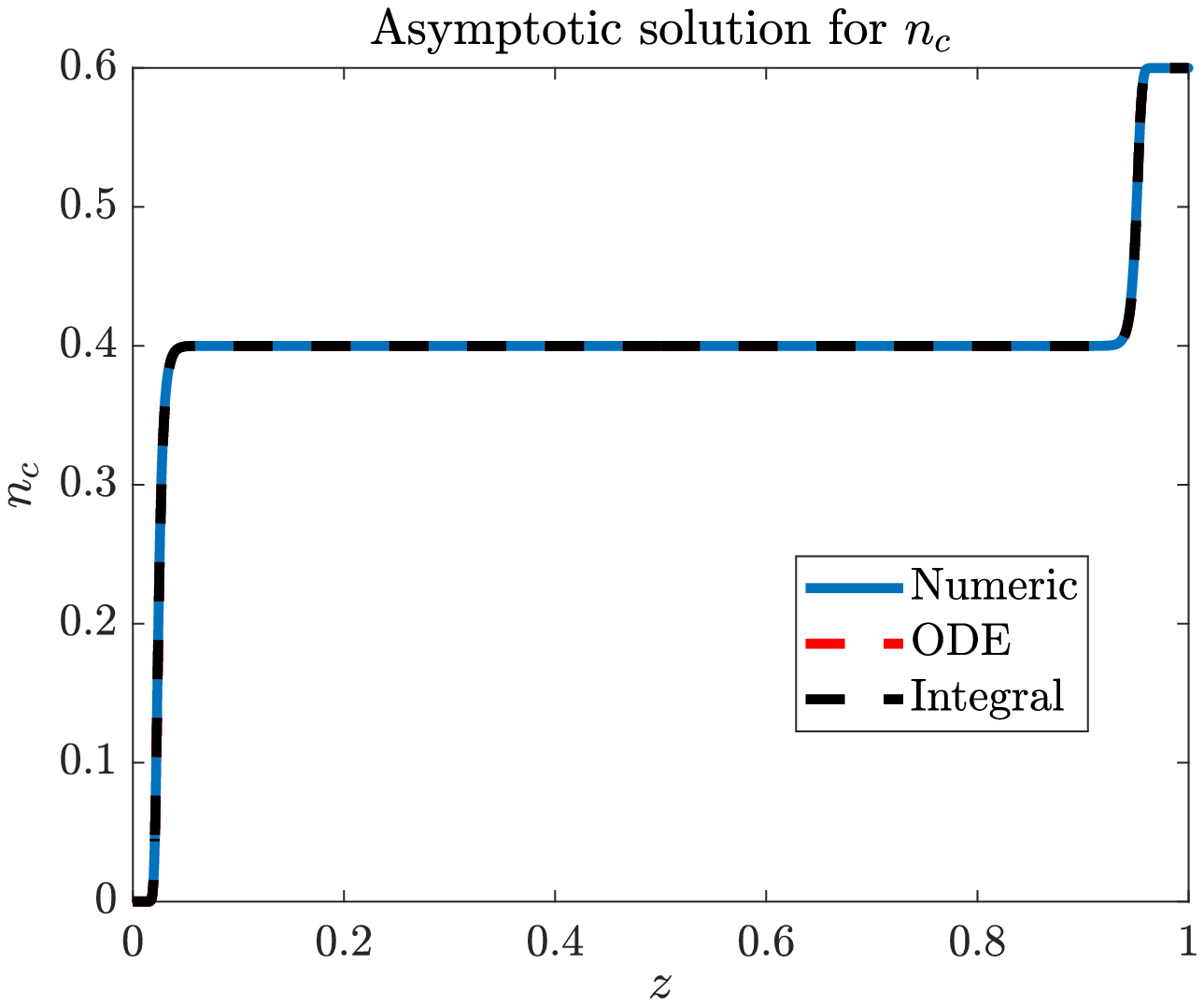}
			\caption{Lithium concentration ODE versus integral asymptotics versus numerics}
			\label{fig:ode_v_int_v_numerics_a}
		\end{subfigure}
		\begin{subfigure}{0.45\linewidth}
			\includegraphics[width=\linewidth]{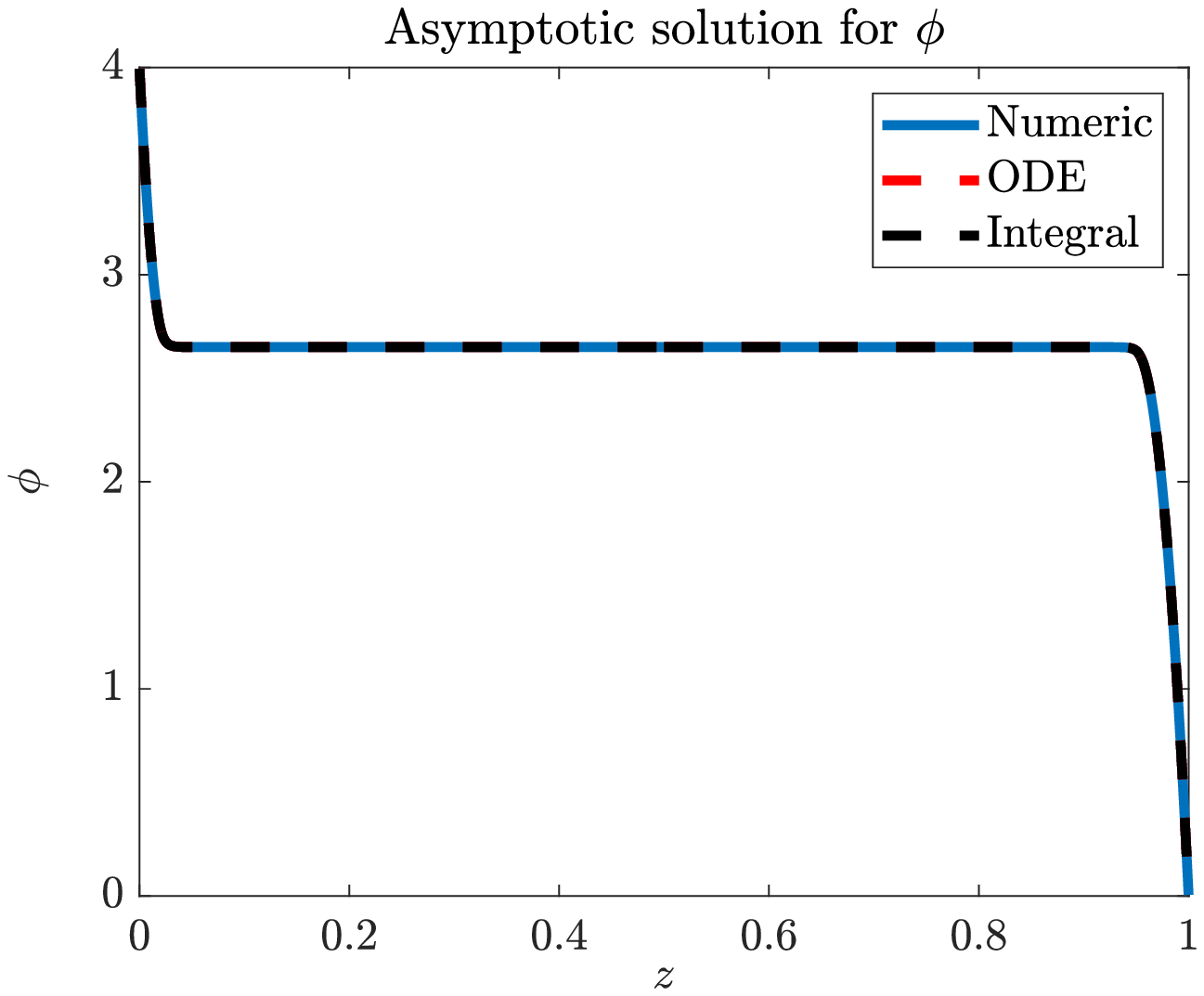}
			\caption{Electric potential ODE versus integral asymptotics versus numerics}
			\label{fig:ode_v_int_v_numerics_b}
		\end{subfigure}
		\caption{ODE versus integral asymptotic versus numerical solutions for lithium concentration and electric potential. Parameter values are as follows: $z_c=1, \quad z_a=-1, \quad m_c=0.3, \quad m_a=0.7, \quad n_a=0.4,\quad \nu=0.6$, with $\frac{\beta_1}{\beta_2}=2 < \beta^*, \quad \frac{\beta_2}{\beta_1}=\frac{1}{2} < \beta^*$, $\lambda=1.5 \times 10^{-3}$, $\delta^{-1}=170$, and $\Delta V = 4$. }
		\label{fig:ode_v_int_v_numerics}
	\end{figure}

	In this section we show that both asymptotic solutions are equivalent and we show the agreement between the integral asymptotic solution and the numerical solution.
	In \cref{fig:asym_v_numerics_int} we plot the integral asymptotic solution versus the numerical solution for both $n_c$ in \cref{fig:asym_v_numerics_int_a} and $\phi$ in \cref{fig:asym_v_numerics_int_b}. As with the ODE asymptotics, we see very good agreement between the two solutions.
	Figure \ref{fig:ode_v_int_v_numerics_a} shows the solution for $n_c$ obtained via all three methods: the numerical solution is shown in blue, the ODE asymptotics are shown in red and the integral asymptotics are shown in black. We observe that the two asymptotic solutions are indistinguishable, and both show excellent agreement with the numerics. We see similar results for the $\phi$ profile as shown in \cref{fig:ode_v_int_v_numerics_b}.

	\bibliographystyle{siamplain}
	\bibliography{References}
	\makeatletter\@input{xx.tex}\makeatother

%% file: solid_electrolyte_shared.tex
\usepackage{lipsum}
\usepackage{amsmath}
\usepackage{amssymb}
\usepackage{amsfonts}
\usepackage{graphicx}
\usepackage{graphics} 
\usepackage{mathtools}	
\usepackage{epstopdf}
\usepackage{subcaption}
\usepackage[version=4]{mhchem}
\usepackage[scientific-notation=true]{siunitx}
\usepackage{hhline} 
\usepackage{titleps}
\usepackage{algorithmic}
\ifpdf
  \DeclareGraphicsExtensions{.eps,.pdf,.png,.jpg}
\else
  \DeclareGraphicsExtensions{.eps}
\fi

\newsiamremark{remark}{Remark}
\newsiamremark{hypothesis}{Hypothesis}
\crefname{hypothesis}{Hypothesis}{Hypotheses}
\newsiamthm{claim}{Claim}

\newcommand{\ie}{\emph{i.e.}~}

\newcommand{\etal}{\emph{et al.}~}

\newcommand{\psibl}{\psi_{\rm{BL}}}
\newcommand{\psibr}{\psi_{\rm{BR}}}
\newcommand{\xl}{x_{\rm{L}}}
\newcommand{\xr}{x_{\rm{R}}}
\newcommand{\yl}{y_{\rm{L}}}
\newcommand{\yr}{y_{\rm{R}}}
\newcommand{\xil}{\xi_{\rm{L}}}
\newcommand{\xir}{\xi_{\rm{R}}}
\newcommand{\thl}{\Theta_{\rm{L}}}
\newcommand{\thr}{\Theta_{\rm{R}}}

\makeatletter
\providecommand*{\diff}%
{\@ifnextchar^{\DIfF}{\DIfF^{}}}
\def\DIfF^#1{%
	\mathop{\mathrm{\mathstrut d}}%
	\nolimits^{#1}\gobblespace}
\def\gobblespace{%
	\futurelet\diffarg\opspace}
\def\opspace{%
	\let\DiffSpace\!%
	\ifx\diffarg(%
	\let\DiffSpace\relax
	\else
	\ifx\diffarg[%
	\let\DiffSpace\relax
	\else
	\ifx\diffarg\{%
	\let\DiffSpace\relax
	\fi\fi\fi\DiffSpace}

\providecommand*{\deriv}[3][]{%
	\frac{\diff^{#1}#2}{\diff #3^{#1}}}

\providecommand*{\pderiv}[3][]{%
	\frac{\partial^{#1}#2}%
	{\partial #3^{#1}}}

\usepackage{verbatim}
\headers{Asymptotic Analysis of Space Charge Layers in a Solid Electrolyte}{Laura M. Keane and Iain R. Moyles}

\title{An Asymptotic Analysis of Space Charge Layers in a Mathematical Model of a Solid Electrolyte \thanks{Submitted to the editors DATE.
\funding{This work was funded by the NSERC Vanier Canada Graduate scholarship Grant No. 434051. I.R.M. acknowledges The Natural Sciences and Engineering Research
Council of Canada Discovery Grant 2019-06337}}}

\author{Laura M. Keane\thanks{Department of Mathematics and Statistics, York University, Toronto, Canada 
  (\email{lkeane@yorku.ca}).}
\and Iain R. Moyles\thanks{Department of Mathematics and Statistics, York University, Toronto, CA 
  (\email{imoyles@yorku.ca}).}
}

\usepackage{amsopn}


%% file: solid_electrolyte.bbl
\begin{thebibliography}{10}

\bibitem{ahmad2022chemomechanics}
{\sc Z.~Ahmad, V.~Venturi, S.~Sripad, and V.~Viswanathan}, {\em Chemomechanics:
  Friend or foe of the “and problem” of solid-state batteries?}, Current
  Opinion in Solid State and Materials Science, 26 (2022), p.~101002.

\bibitem{becker2021statics}
{\sc K.~Becker-Steinberger, S.~Schardt, B.~Horstmann, and A.~Latz}, {\em
  Statics and dynamics of space-charge-layers in polarized inorganic solid
  electrolytes}, arXiv preprint arXiv:2101.10294,  (2021).

\bibitem{biesheuvel2011diffuse}
{\sc P.~Biesheuvel, Y.~Fu, and M.~Z. Bazant}, {\em Diffuse charge and faradaic
  reactions in porous electrodes}, Physical Review E, 83 (2011), p.~061507.

\bibitem{braun2015thermodynamically}
{\sc S.~Braun, C.~Yada, and A.~Latz}, {\em Thermodynamically consistent model
  for space-charge-layer formation in a solid electrolyte}, The Journal of
  Physical Chemistry C, 119 (2015), pp.~22281--22288.

\bibitem{cheng2020revealing}
{\sc Z.~Cheng, M.~Liu, S.~Ganapathy, C.~Li, Z.~Li, X.~Zhang, P.~He, H.~Zhou,
  and M.~Wagemaker}, {\em Revealing the impact of space-charge layers on the
  li-ion transport in all-solid-state batteries}, Joule, 4 (2020),
  pp.~1311--1323.

\bibitem{de2018space}
{\sc N.~J. de~Klerk and M.~Wagemaker}, {\em Space-charge layers in
  all-solid-state batteries; important or negligible?}, ACS applied energy
  materials, 1 (2018), pp.~5609--5618.

\bibitem{dixit2021polymorphism}
{\sc M.~Dixit, B.~Vishugopi, W.~Zaman, P.~Kenesei, J.-S. Park, J.~Almer,
  P.~Mukherjee, and K.~Hatzell}, {\em Polymorphism of garnet solid electrolytes
  and its implications on grain level chemo-mechanics},  (2021).

\bibitem{doyle1993modeling}
{\sc M.~Doyle, T.~F. Fuller, and J.~Newman}, {\em Modeling of galvanostatic
  charge and discharge of the lithium/polymer/insertion cell}, Journal of the
  Electrochemical society, 140 (1993), p.~1526.

\bibitem{farrell2000primary}
{\sc T.~W. Farrell, C.~P. Please, D.~McElwain, and D.~Swinkels}, {\em Primary
  alkaline battery cathodes a three-scale model}, Journal of the
  Electrochemical Society, 147 (2000), p.~4034.

\bibitem{foster2017mathematical}
{\sc J.~M. Foster, S.~J. Chapman, G.~Richardson, and B.~Protas}, {\em A
  mathematical model for mechanically-induced deterioration of the binder in
  lithium-ion electrodes}, SIAM Journal on Applied Mathematics, 77 (2017),
  pp.~2172--2198.

\bibitem{fuller1994simulation}
{\sc T.~F. Fuller, M.~Doyle, and J.~Newman}, {\em Simulation and optimization
  of the dual lithium ion insertion cell}, Journal of the electrochemical
  society, 141 (1994), p.~1.

\bibitem{gokcen1996gibbs}
{\sc N.~Gokcen}, {\em Gibbs-duhem-margules laws}, Journal of phase equilibria,
  17 (1996), pp.~50--51.

\bibitem{gross2019modelling}
{\sc A.~Gro{\ss} and S.~Sakong}, {\em Modelling the electric double layer at
  electrode/electrolyte interfaces}, Current Opinion in Electrochemistry, 14
  (2019), pp.~1--6.

\bibitem{he2018theory}
{\sc F.~He, P.~Biesheuvel, M.~Z. Bazant, and T.~A. Hatton}, {\em Theory of
  water treatment by capacitive deionization with redox active porous
  electrodes}, Water research, 132 (2018), pp.~282--291.

\bibitem{hennessy2020asymptotic}
{\sc M.~G. Hennessy and I.~R. Moyles}, {\em Asymptotic reduction and
  homogenization of a thermo-electrochemical model for a lithium-ion battery},
  Applied Mathematical Modelling, 80 (2020), pp.~724--754.

\bibitem{howey2020free}
{\sc D.~A. Howey, S.~A. Roberts, V.~Viswanathan, A.~Mistry, M.~Beuse, E.~Khoo,
  S.~C. DeCaluwe, and V.~Sulzer}, {\em Free radicals: making a case for battery
  modeling}, The Electrochemical Society Interface, 29 (2020), p.~30.

\bibitem{hu2011linear}
{\sc Y.~Hu and S.~Yurkovich}, {\em Linear parameter varying battery model
  identification using subspace methods}, Journal of Power Sources, 196 (2011),
  pp.~2913--2923.

\bibitem{hu2011electro}
{\sc Y.~Hu, S.~Yurkovich, Y.~Guezennec, and B.~Yurkovich}, {\em Electro-thermal
  battery model identification for automotive applications}, Journal of Power
  Sources, 196 (2011), pp.~449--457.

\bibitem{huang2020editors}
{\sc J.~Huang, Y.~Gao, J.~Luo, S.~Wang, C.~Li, S.~Chen, and J.~Zhang}, {\em
  Editors’ choice—review—impedance response of porous electrodes:
  theoretical framework, physical models and applications}, Journal of the
  Electrochemical Society, 167 (2020), p.~166503.

\bibitem{janek2016solid}
{\sc J.~Janek and W.~G. Zeier}, {\em A solid future for battery development},
  Nature Energy, 1 (2016), pp.~1--4.

\bibitem{katzenmeier2022modeling}
{\sc L.~Katzenmeier, M.~Goosswein, A.~Gagliardi, and A.~S. Bandarenka}, {\em
  Modeling of space-charge layers in solid-state electrolytes: A kinetic monte
  carlo approach and its validation}, The Journal of Physical Chemistry C, 126
  (2022), pp.~10900--10909.

\bibitem{katzenmeier2022nature}
{\sc L.~M. Katzenmeier}, {\em Nature of Space Charge Layers in Li+ Conducting
  Glass Ceramics}, PhD thesis, Technische Universitat Munchen, 2022.

\bibitem{kennedy2017understanding}
{\sc T.~Kennedy, M.~Brandon, F.~Laffir, and K.~M. Ryan}, {\em Understanding the
  influence of electrolyte additives on the electrochemical performance and
  morphology evolution of silicon nanowire based lithium-ion battery anodes},
  Journal of Power Sources, 359 (2017), pp.~601--610.

\bibitem{kennedy2016advances}
{\sc T.~Kennedy, M.~Brandon, and K.~M. Ryan}, {\em Advances in the application
  of silicon and germanium nanowires for high-performance lithium-ion
  batteries}, Advanced Materials, 28 (2016), pp.~5696--5704.

\bibitem{kennedy2014high}
{\sc T.~Kennedy, E.~Mullane, H.~Geaney, M.~Osiak, C.~O’Dwyer, and K.~M.
  Ryan}, {\em High-performance germanium nanowire-based lithium-ion battery
  anodes extending over 1000 cycles through in situ formation of a continuous
  porous network}, Nano letters, 14 (2014), pp.~716--723.

\bibitem{kim2022transport}
{\sc H.-K. Kim, P.~Barai, K.~Chavan, and V.~Srinivasan}, {\em Transport and
  mechanical behavior in peo-llzo composite electrolytes}, Journal of Solid
  State Electrochemistry, 26 (2022), pp.~2059--2075.

\bibitem{kittel2005introduction}
{\sc C.~Kittel}, {\em Introduction to solid state physics}, John Wiley \& sons,
  inc, 2005.

\bibitem{knauth2009inorganic}
{\sc P.~Knauth}, {\em Inorganic solid li ion conductors: An overview}, Solid
  State Ionics, 180 (2009), pp.~911--916.

\bibitem{landstorfer2011advanced}
{\sc M.~Landstorfer, S.~Funken, and T.~Jacob}, {\em An advanced model framework
  for solid electrolyte intercalation batteries}, Physical Chemistry Chemical
  Physics, 13 (2011), pp.~12817--12825.

\bibitem{li2019dendrite}
{\sc G.~Li and C.~W. Monroe}, {\em Dendrite nucleation in lithium-conductive
  ceramics}, Physical Chemistry Chemical Physics, 21 (2019), pp.~20354--20359.

\bibitem{li2021transport}
{\sc G.~Li and C.~W. Monroe}, {\em Transport of secondary carriers in a solid
  lithium-ion conductor}, Electrochimica Acta, 389 (2021), p.~138563.

\bibitem{liu2022unlocking}
{\sc J.~Liu, H.~Yuan, H.~Liu, C.-Z. Zhao, Y.~Lu, X.-B. Cheng, J.-Q. Huang, and
  Q.~Zhang}, {\em Unlocking the failure mechanism of solid state lithium metal
  batteries}, Advanced Energy Materials, 12 (2022), p.~2100748.

\bibitem{liu2011reversible}
{\sc X.~H. Liu, S.~Huang, S.~T. Picraux, J.~Li, T.~Zhu, and J.~Y. Huang}, {\em
  Reversible nanopore formation in ge nanowires during lithiation--delithiation
  cycling: An in situ transmission electron microscopy study}, Nano letters, 11
  (2011), pp.~3991--3997.

\bibitem{magnussen2019toward}
{\sc O.~M. Magnussen and A.~Gro{\ss}}, {\em Toward an atomic-scale
  understanding of electrochemical interface structure and dynamics}, Journal
  of the American Chemical Society, 141 (2019), pp.~4777--4790.

\bibitem{manthiram2017lithium}
{\sc A.~Manthiram, X.~Yu, and S.~Wang}, {\em Lithium battery chemistries
  enabled by solid-state electrolytes}, Nature Reviews Materials, 2 (2017),
  pp.~1--16.

\bibitem{marcicki2013design}
{\sc J.~Marcicki, M.~Canova, A.~T. Conlisk, and G.~Rizzoni}, {\em Design and
  parametrization analysis of a reduced-order electrochemical model of
  graphite/lifepo4 cells for soc/soh estimation}, Journal of Power Sources, 237
  (2013), pp.~310--324.

\bibitem{margules1895zusammensetzung}
{\sc M.~Margules}, {\em {\"U}ber die zusammensetzung der ges{\"a}ttigten
  d{\"a}mpfe von mischungen}, Sitzungsber. Akad. Wiss. Wien, math.-naturwiss.
  Klasse, 104 (1895), pp.~1243--1278.

\bibitem{marquis2019asymptotic}
{\sc S.~G. Marquis, V.~Sulzer, R.~Timms, C.~P. Please, and S.~J. Chapman}, {\em
  An asymptotic derivation of a single particle model with electrolyte},
  Journal of The Electrochemical Society, 166 (2019), p.~A3693.

\bibitem{mistry2020molar}
{\sc A.~Mistry and P.~P. Mukherjee}, {\em Molar volume mismatch: A malefactor
  for irregular metallic electrodeposition with solid electrolytes}, Journal of
  the Electrochemical Society, 167 (2020), p.~082510.

\bibitem{moyles2019asymptotic}
{\sc I.~R. Moyles, M.~G. Hennessy, T.~G. Myers, and B.~R. Wetton}, {\em
  Asymptotic reduction of a porous electrode model for lithium-ion batteries},
  SIAM Journal on Applied Mathematics, 79 (2019), pp.~1528--1549.

\bibitem{mozhzhukhina2020direct}
{\sc N.~Mozhzhukhina, E.~Flores, R.~Lundstrom, V.~Nystrom, P.~G. Kitz,
  K.~Edstrom, and E.~J. Berg}, {\em Direct operando observation of double layer
  charging and early solid electrolyte interphase formation in li-ion battery
  electrolytes}, The journal of physical chemistry letters, 11 (2020),
  pp.~4119--4123.

\bibitem{newman2012electrochemical}
{\sc J.~Newman and K.~E. Thomas-Alyea}, {\em Electrochemical systems}, John
  Wiley \& Sons, 2012.

\bibitem{newman1975porous}
{\sc J.~Newman and W.~Tiedemann}, {\em Porous-electrode theory with battery
  applications}, AIChE Journal, 21 (1975), pp.~25--41.

\bibitem{newman1962theoretical}
{\sc J.~S. Newman and C.~W. Tobias}, {\em Theoretical analysis of current
  distribution in porous electrodes}, Journal of The Electrochemical Society,
  109 (1962), p.~1183.

\bibitem{planella2022continuum}
{\sc F.~B. Planella, W.~Ai, A.~Boyce, A.~Ghosh, I.~Korotkin, S.~Sahu,
  V.~Sulzer, R.~Timms, T.~Tranter, M.~Zyskin, et~al.}, {\em A continuum of
  physics-based lithium-ion battery models reviewed}, Progress in Energy,
  (2022).

\bibitem{plett2015battery}
{\sc G.~L. Plett}, {\em Battery management systems, Volume I: Battery
  modeling}, Artech House, 2015.

\bibitem{randau2020benchmarking}
{\sc S.~Randau, D.~A. Weber, O.~K{\"o}tz, R.~Koerver, P.~Braun, A.~Weber,
  E.~Ivers-Tiff{\'e}e, T.~Adermann, J.~Kulisch, W.~G. Zeier, et~al.}, {\em
  Benchmarking the performance of all-solid-state lithium batteries}, Nature
  Energy, 5 (2020), pp.~259--270.

\bibitem{richardson2020generalised}
{\sc G.~Richardson, I.~Korotkin, R.~Ranom, M.~Castle, and J.~Foster}, {\em
  Generalised single particle models for high-rate operation of graded
  lithium-ion electrodes: systematic derivation and validation}, Electrochimica
  Acta, 339 (2020), p.~135862.

\bibitem{richardson2020charge}
{\sc G.~W. Richardson, J.~M. Foster, R.~Ranom, C.~P. Please, and A.~M. Ramos},
  {\em Charge transport modelling of lithium ion batteries}, arXiv preprint
  arXiv:2002.00806,  (2020).

\bibitem{safari2011modeling}
{\sc M.~Safari and C.~Delacourt}, {\em Modeling of a commercial
  graphite/lifepo4 cell}, Journal of The Electrochemical Society, 158 (2011),
  p.~A562.

\bibitem{sakong2022structure}
{\sc S.~Sakong, J.~Huang, M.~Eikerling, and A.~Gro{\ss}}, {\em The structure of
  the electric double layer: Atomistic vs. continuum approaches}, Current
  Opinion in Electrochemistry,  (2022), p.~100953.

\bibitem{shen2018unlocking}
{\sc Y.~Shen, Y.~Zhang, S.~Han, J.~Wang, Z.~Peng, and L.~Chen}, {\em Unlocking
  the energy capabilities of lithium metal electrode with solid-state
  electrolytes}, Joule, 2 (2018), pp.~1674--1689.

\bibitem{singh2018theory}
{\sc K.~Singh, H.~Bouwmeester, L.~De~Smet, M.~Bazant, and P.~Biesheuvel}, {\em
  Theory of water desalination with intercalation materials}, Physical review
  applied, 9 (2018), p.~064036.

\bibitem{sinzig2023finite}
{\sc S.~Sinzig, T.~Hollweck, C.~P. Schmidt, and W.~A. Wall}, {\em A finite
  element formulation to three-dimensionally resolve space-charge layers in
  solid electrolytes}, Journal of The Electrochemical Society,  (2023).

\bibitem{smith2017multiphase}
{\sc R.~B. Smith and M.~Z. Bazant}, {\em Multiphase porous electrode theory},
  Journal of The Electrochemical Society, 164 (2017), p.~E3291.

\bibitem{stokes2017direct}
{\sc K.~Stokes, H.~Geaney, G.~Flynn, M.~Sheehan, T.~Kennedy, and K.~M. Ryan},
  {\em Direct synthesis of alloyed si1--x ge x nanowires for
  performance-tunable lithium ion battery anodes}, ACS nano, 11 (2017),
  pp.~10088--10096.

\bibitem{sulzer2019faster}
{\sc V.~Sulzer, S.~J. Chapman, C.~P. Please, D.~A. Howey, and C.~W. Monroe},
  {\em Faster lead-acid battery simulations from porous-electrode theory: Part
  ii. asymptotic analysis}, Journal of The Electrochemical Society, 166 (2019),
  p.~A2372.

\bibitem{swift2021modeling}
{\sc M.~W. Swift, J.~W. Swift, and Y.~Qi}, {\em Modeling the electrical double
  layer at solid-state electrochemical interfaces}, Nature Computational
  Science, 1 (2021), pp.~212--220.

\bibitem{takada2013progress}
{\sc K.~Takada}, {\em Progress and prospective of solid-state lithium
  batteries}, Acta Materialia, 61 (2013), pp.~759--770.

\bibitem{tarascon2001issues}
{\sc J.~Tarascon and M.~Armand}, {\em Issues and challenges facing rechargeable
  lithium batteries}, Nature, 414 (2001), pp.~359--367.

\bibitem{tasaki2003decomposition}
{\sc K.~Tasaki, K.~Kanda, S.~Nakamura, and M.~Ue}, {\em Decomposition of
  lipf6and stability of pf 5 in li-ion battery electrolytes: Density functional
  theory and molecular dynamics studies}, Journal of The Electrochemical
  Society, 150 (2003), p.~A1628.

\bibitem{wu2022solid}
{\sc F.~Wu, L.~Liu, S.~Wang, J.~Xu, P.~Lu, W.~Yan, J.~Peng, D.~Wu, and H.~Li},
  {\em Solid state ionics-selected topics and new directions}, Progress in
  Materials Science,  (2022), p.~100921.

\bibitem{xia2019practical}
{\sc S.~Xia, X.~Wu, Z.~Zhang, Y.~Cui, and W.~Liu}, {\em Practical challenges
  and future perspectives of all-solid-state lithium-metal batteries}, Chem, 5
  (2019), pp.~753--785.

\bibitem{xu2021competitive}
{\sc R.~Xu, C.~Yan, and J.-Q. Huang}, {\em Competitive solid-electrolyte
  interphase formation on working lithium anodes}, Trends in Chemistry, 3
  (2021), pp.~5--14.

\bibitem{yamada2013lithium}
{\sc H.~Yamada, K.~Suzuki, Y.~Oga, I.~Saruwatari, and I.~Moriguchi}, {\em
  Lithium depletion in the solid electrolyte adjacent to cathode materials},
  ECS Transactions, 50 (2013), p.~1.

\bibitem{zhang2022research}
{\sc Q.~Zhang, Y.~Kong, K.~Gao, Y.~Wen, Q.~Zhang, H.~Fang, C.~Ma, and Y.~Du},
  {\em Research progress on space charge layer effect in lithium-ion
  solid-state battery}, Science China Technological Sciences,  (2022),
  pp.~1--13.

\bibitem{zhang2023designing}
{\sc S.~Zhang, J.~Ma, S.~Dong, and G.~Cui}, {\em Designing all-solid-state
  batteries by theoretical computation: A review}, Electrochemical Energy
  Reviews, 6 (2023), p.~4.

\bibitem{zhao2019solid}
{\sc W.~Zhao, J.~Yi, P.~He, and H.~Zhou}, {\em Solid-state electrolytes for
  lithium-ion batteries: fundamentals, challenges and perspectives},
  Electrochemical Energy Reviews, 2 (2019), pp.~574--605.

\end{thebibliography}


\begin{thebibliography}{1}

\bibitem{braun2015thermodynamically}
{\sc S.~Braun, C.~Yada, and A.~Latz}, {\em Thermodynamically consistent model
  for space-charge-layer formation in a solid electrolyte}, The Journal of
  Physical Chemistry C, 119 (2015), pp.~22281--22288.

\bibitem{de2013non}
{\sc S.~R. De~Groot and P.~Mazur}, {\em Non-equilibrium thermodynamics},
  Courier Corporation, 2013.

\bibitem{gokcen1996gibbs}
{\sc N.~Gokcen}, {\em Gibbs-duhem-margules laws}, Journal of phase equilibria,
  17 (1996), pp.~50--51.

\bibitem{margules1895zusammensetzung}
{\sc M.~Margules}, {\em {\"U}ber die zusammensetzung der ges{\"a}ttigten
  d{\"a}mpfe von mischungen}, Sitzungsber. Akad. Wiss. Wien, math.-naturwiss.
  Klasse, 104 (1895), pp.~1243--1278.

\bibitem{onsager1931reciprocal}
{\sc L.~Onsager}, {\em Reciprocal relations in irreversible processes. i.},
  Physical review, 37 (1931), p.~405.

\bibitem{prigogine1963introduction}
{\sc I.~Prigogine and P.~Van~Rysselberghe}, {\em Introduction to thermodynamics
  of irreversible processes}, Journal of The Electrochemical Society, 110
  (1963), p.~97C.

\bibitem{rajagopal2004thermomechanical}
{\sc K.~R. Rajagopal and A.~R. Srinivasa}, {\em On thermomechanical
  restrictions of continua}, Proceedings of the Royal Society of London. Series
  A: Mathematical, Physical and Engineering Sciences, 460 (2004), pp.~631--651.

\bibitem{taylor1993multicomponent}
{\sc R.~Taylor and R.~Krishna}, {\em Multicomponent mass transfer}, vol.~2,
  John Wiley \& Sons, 1993.

\end{thebibliography}
